\pdfoutput=1
\documentclass[usenatbib]{mn2e}
\usepackage{apjfonts}
\usepackage{amssymb}
\usepackage{amsmath}
\usepackage{ctable}
\usepackage{url}
\usepackage{fixltx2e} 

\newcommand{\be}{\begin{equation}}
\newcommand{\ee}{\end{equation}}

\newcommand{\etal}{et al.}

\newcommand{\msun}{M_{\sun}}
\newcommand{\lsun}{L_{\sun}}

\newcommand{\qeos}{q_{\rm eos}}

\newcommand{\paperone}{Paper {\small I}}
\newcommand{\papertwo}{Paper {\small II}}
\newcommand{\paperthree}{Paper {\small III}}

\newcommand{\movieurl}{\url{https://www.cfa.harvard.edu/~phopkins/Site/Movies_sbw_mgr.html}}

\newcommand{\scaleup}{}

\newcommand\plotonesize[2]
 {\centering \leavevmode \includegraphics[width={#2\columnwidth}]{#1}}
\newcommand{\plotsidesize}[2]
 {\centering \leavevmode \includegraphics[width={#2\textwidth}]{#1}}
\newcommand{\acknowledgments}{\begin{small}\section*{Acknowledgments}\end{small}}
\newcommand\altaffilmark[1]{$^{#1}$}
\newcommand\altaffiltext[1]{$^{#1}$}
\title[Mergers with GMC Formation \&\ Destruction]{
Star Formation in Galaxy Mergers with Realistic Models of Stellar Feedback \&\ the 
Interstellar Medium\vspace{-0.6cm}}

\author[Hopkins \etal]{
\parbox[t]{\textwidth}{ 
Philip F. Hopkins\altaffilmark{1}\thanks{E-mail:phopkins@astro.berkeley.edu},
Thomas J.\ Cox\altaffilmark{2}, 
Lars Hernquist\altaffilmark{3},
Desika Narayanan\altaffilmark{4}, 
Christopher C.~Hayward\altaffilmark{5}, \&\ 
Norman Murray\altaffilmark{6,7} 
} 
\vspace*{6pt} \\
\altaffiltext{1}{Department of Astronomy, University of California
  Berkeley, Berkeley, CA 94720} \\
\altaffiltext{2}{Carnegie Observatories, 813 Santa Barbara Street, Pasadena, CA 91101} \\ 
\altaffiltext{3}{Harvard-Smithsonian Center for Astrophysics, 60 
Garden Street, Cambridge, MA 02138} \\ 
\altaffiltext{4}{Steward Observatory, University of Arizona, 933 
N Cherry Ave, Tucson, Az, 85721} \\ 
\altaffiltext{5}{Heidelberger Institut f\"ur Theoretische Studien, Schlo\ss-Wolfsbrunnenweg 35, 69118 Heidelberg, Germany} \\
\altaffiltext{6}{Canadian Institute for Theoretical Astrophysics, 
60 St.\ George Street, University of Toronto, ON M5S 3H8, Canada} \\
\altaffiltext{7}{Canada Research Chair in Astrophysics} 
\vspace{-0.5cm}
}

\date{Submitted to MNRAS, June, 2012\vspace{-0.6cm}}
\begin{document}
\maketitle
\label{firstpage}

\begin{abstract}
We use hydrodynamic simulations with detailed, explicit models for stellar
feedback to study galaxy mergers. These high resolution ($\sim 1\,{\rm
pc}$) simulations follow the formation and destruction of individual
giant molecular clouds (GMCs) and star clusters. 
%
%
We find that the final starburst is
dominated by in situ star formation, fueled by gas which flows inwards
due to global torques. The resulting high gas density results in rapid
star formation.  The gas is self gravitating, and forms massive
($\lesssim10^{10}\,M_\odot$) GMCs and subsequently super-starclusters
(with masses up to $10^8\,M_\odot$). However, in contrast to some
recent simulations, the bulk of new stars which eventually form the
central bulge are not born in superclusters which then sink to the
center of the galaxy. This is because feedback efficiently disperses
GMCs after they turn several percent of their mass into stars. In
other words, most of the mass that reaches the nucleus does so in the
form of gas. The Kennicutt-Schmidt law emerges naturally as a
consequence of feedback balancing gravitational collapse, independent
of the small-scale star formation microphysics. The same mechanisms
that drive this relation in isolated galaxies, in particular radiation
pressure from IR photons, extend, with no fine-tuning, over seven
decades in star formation rate (SFR) to regulate star formation in the
most extreme starburst systems with densities $\gtrsim
10^{4}\,\msun\,{\rm pc^{-2}}$. This feedback also drives super-winds
with large mass loss rates; however, a significant fraction of the
wind material falls back onto the disks at later times, leading to
higher post-starburst SFRs in the presence of stellar feedback. This
suggests that strong AGN feedback may be required to explain the sharp
cutoffs in star formation rate that are observed in post-merger
galaxies.

We compare the results to those from simulations with no explicit
resolution of GMCs or feedback (``effective equation of state'' [EOS]
models). We find that global galaxy properties are similar between EOS
and resolved-feedback models. The relic structure and mass profile,
and the total mass of stars formed in the nuclear starburst are quite
similar, as is the morphological structure during and after merger
(tails, bridges, etc.). Disk survival in sufficiently
gas-rich mergers is similar in the two cases, and the new models
follow the same scalings derived for the efficiency of disk
re-formation after a merger as derived from previous work with the
simplified EOS models. While the global galaxy properties are similar
between EOS and feedback models, sub-galaxy scale properties and the
star formation rates can be quite different: the more detailed models
exhibit significantly higher star formation in tails and bridges
(especially in shocks), and allow us to resolve the formation of super
star-clusters. In the new models, the star formation is more strongly
time variable and drops more sharply between close passages. The
instantaneous burst enhancement can be higher or lower, depending on
the details of the orbit and initial structural properties of the
galaxies; first passage bursts are more sensitive to these details
than those at final coalescence.

\end{abstract}


\vspace{-1.0cm}
\section{Introduction}
\label{sec:intro}

A wide range of observed phenomena indicate that gas-rich mergers are important to 
galaxy evolution and star formation. 
In the local Universe, the population of star-forming galaxies appears 
to transition from ``quiescent'' (non-disturbed) disks (which dominate the 
{\em total} star formation rate/IR luminosity density) at the luminous 
infrared galaxy (LIRG) threshold $10^{11}\,\lsun$ ($\dot{M}_{\ast}\sim 10-20\,\msun\,{\rm yr^{-1}}$)  
to violently disturbed systems
at a few times this luminosity. The most intense
starbursts at $z=0$, ultraluminous infrared galaxies (ULIRGs; $L_{\rm IR}>10^{12}\,\lsun$), are 
invariably
associated with mergers \citep[e.g.][]{joseph85,sanders96:ulirgs.mergers,
evans:ulirgs.are.mergers}, and are fueled by compact, central concentrations of 
gas \citep{scoville86,sargent87} which provide material to feed black hole (BH)
growth \citep{sanders88:quasars}, and to boost the concentration and central phase space density
of merging spirals to match those of ellipticals
\citep{hernquist:phasespace,robertson:fp}. With central densities as large as $\sim1000$ times 
those in Milky Way giant molecular clouds (GMCs), these systems provide a laboratory 
for studying star formation 
under the most extreme conditions.

In addition, various studies have shown that the mass involved in these starburst events 
is critical for explaining the relations between spirals, mergers, and ellipticals, 
and has a dramatic impact on the properties of merger 
remnants \citep[e.g.,][]{LakeDressler86,Doyon94,ShierFischer98,James99,
Genzel01,tacconi:ulirgs.sb.profiles,dasyra:mass.ratio.conditions,dasyra:pg.qso.dynamics,
rj:profiles,rothberg.joseph:kinematics,hopkins:cusps.ell,hopkins:cores}. 
Even a small mass fraction of a few percent formed in these nuclear starbursts 
can have dramatic implications for the mass profile structure \citep{mihos:cusps}, 
phase space densities \citep{hernquist:phasespace}, rotation and higher-order kinematics 
\citep{cox:kinematics}, kinematically decoupled components 
\citep{hoffman:dissipation.and.gal.kinematics,hoffman:mgr.orbit.structure.vs.fgas}, stellar population gradients \citep{soto:ssp.grad.in.ulirgs,
kewley:2010.gal.pair.metal.grad.evol,torrey:2011.metallicity.evol.merger}, and growth of the central BH 
\citep{dimatteo:msigma,hopkins:qso.all,hopkins:seyfert.limits}. 

At higher redshifts ($z\sim1-3$), galaxies with the luminosities of LIRGs and ULIRGs may be more ``normal'' galaxies, in the sense that they are relatively undisturbed disks rather than mergers
\citep{yan:z2.sf.seds,sajina:pah.qso.vs.sf,
dey:2008.dog.population,melbourne:2008.dog.morph.smooth,
dasyra:highz.ulirg.imaging.not.major}. 
However, at those same redshifts, yet more luminous systems 
appear, including large populations of Hyper-LIRGs (HyLIRGs; 
$L_{\rm IR}>10^{13}\,\lsun$) and bright sub-millimeter galaxies 
\citep[e.g.][]{chapman:submm.lfs,younger:highz.smgs,
younger:sma.hylirg.obs,casey:highz.ulirg.pops}. 
These hyperluminous objects exhibit many of the traits 
associated with merger-driven starbursts, including morphological 
disturbances, and may be linked to the emergence of 
massive, quenched (non star-forming), compact ellipticals 
at times as early as $z\sim2-4$ 
\citep{papovich:highz.sb.gal.timescales,
younger:smg.sizes,tacconi:smg.maximal.sb.sizes,
schinnerer:submm.merger.w.compact.mol.gas,
chapman:submm.halo.clustering,tacconi:smg.mgr.lifetime.to.quiescent}. 
Reproducing their abundance 
and luminosities remains a challenge for current models of 
galaxy formation \citep{baugh:sam,
swinbank:smg.counts.vs.durham,
narayanan:smg.modeling,
younger:warm.ulirg.evol,hayward:2010.smg.counts,hayward:2011.smg.merger.rt}.

Modeling the consequences of these mergers for star formation and the nuclear structure of galaxies requires following the highly non-linear, resonant, and chaotic interplay between gas (with shocks, cooling, star formation, and feedback), stars, and dark matter over a very large dynamic range. High-resolution numerical hydrodynamic simulations are the method of choice. However, until recently, computational limitations have made it impossible to resolve the $\sim$pc (and $<1000\,$yr) scales of structure in the ISM relevant for star formation and stellar feedback while following the four or five orders-of-magnitude larger global evolution of a galaxy or merger. Perforce, most previous models have adopted a variety of sub-grid approaches to describe ``effective'' ISM properties below the resolution scales of the simulations. 

All numerical galaxy simulations require some sort of ``sub-grid'' model; for example, even with box sizes of order one parsec, current star formation simulations treat radiative feedback in a very approximate manner, while it is clear that such feedback is crucial to determining, e.g., the inital mass function. Unfortunately, without a physically well founded sub-grid model, the predictive power of numerical simulations is limited. If the average star formation properties are put in by hand (and they are in most galaxy scale simultions to date) the global star formation rate clearly cannot be predicted. Similarly, if the effects of stellar feedback are put in by hand, e.g., by invoking an ``effective pressure'' or turbulent dispersion, in order to suppress runaway cooling and clumping, the detailed physics and implications for star formation cannot be determined. Moreover, it is by no means clear if commonly used prescriptions capture the key physics, nor whether they can be extrapolated from ``quiescent'' systems (which are typically used to calibrate the models) to the very different ISM conditions in mergers. Typically subgrid models also have several adjustable parameters, which further limit their predictive power; it is often unclear whether observational constraints imply differences in the merger dynamics or more subtle tweaks in the sub-grid ISM assumptions. This has occasionally led to contradictory conclusions in the literature.

Numerical simulations of isolated galaxies and galaxy mergers can now achieve the dynamic range required to resolve the formation of GMCs and ISM structure, $\sim 1-10$\,pc \citep[see e.g.][]{saitoh:2008.highres.disks.high.sf.thold,tasker:2009.gmc.form.evol.gravalone,bournaud:2010.grav.turbulence.lmc,dobbs:2011.why.gmcs.unbound}. However, improving resolution alone -- beyond the implicit ``averaging'' scales for the ISM model -- has no clear meaning if the  physics that govern star formation and ISM structure on these scales are not included. Generally, models have not attempted to include these physics or have included a very limited sub-set of the relevant processes. In fact, a large number of feedback mechanisms may drive turbulence in the ISM and help disrupt GMCs, including: photo-ionization, stellar winds, radiation pressure from UV and IR photons, proto-stellar jets, cosmic rays, supernovae, and gravitational cascades from large scales \citep[e.g.][and references therein]{mac-low:2004.turb.sf.review}.  

In \citet{hopkins:rad.pressure.sf.fb} (\paperone) and \citet{hopkins:fb.ism.prop} (\papertwo) we developed a new set of numerical models to incorporate feedback on small scales in GMCs and star-forming regions, into simulations with pc-scale resolution.\footnote{\label{foot:url}Movies of these simulations are available at \movieurl} These simulations include the momentum imparted locally (on sub-GMC scales) from stellar radiation pressure, radiation pressure on larger scales via the light that escapes star-forming regions, HII photoionization heating, as well as the heating, momentum deposition, and mass loss by SNe (Type-I and Type-II)  and stellar winds (from O and AGB stars).  The feedback is tied to the young stars, with the energetics and time-dependence taken directly from stellar evolution models. Our method also includes cooling to temperatures $<100\,$K, and a treatment of the molecular/atomic transition in gas and its effect on star formation. 

We showed in Papers I \& II that in isolated disk galaxies these feedback mechanisms produce a quasi-steady ISM  in which giant molecular clouds form and disperse rapidly, after turning just a few percent of their mass into stars.   This leads to an ISM with phase structure, turbulent velocity dispersions, scale heights, and GMC properties (mass functions, sizes, scaling laws) in reasonable agreement with observations. In \citet{hopkins:stellar.fb.winds} (\paperthree), we showed that these same models of stellar feedback  {\em predict} the elusive winds invoked in almost all galaxy formation models; the {\em combination} of multiple feedback mechanisms is critical to give rise to massive, multi-phase winds having a broad distribution of velocities, with material both stirred in local fountains and unbound from the disk. 

Papers I \& II showed that the global star formation rate found in the simulations did not depend on the sub-grid model for star formation, over a broad range of sub-grid model parameters and even model types. The conclusion drawn was that the rate of star formation in the simulations was controlled by the amount of feedback required to maintain the gas disk in hydrostatic equilibrium, and with a Toomre Q parameter of order unity. If this is true in real galaxies, it will simplify the task of understanding galaxy formation greatly. 

In this paper, we extend these models to idealized major mergers between galaxies. In this first exploration, we compare some of the global properties of the remnants -- particularly those tied to star formation and ISM physics -- to the results of simulations using more simplified sub-grid ISM/feedback models.  Specifically, we investigate the consequences of a more detailed and explicit treatment of the ISM for the star formation histories, spatial distribution 
of stellar mass and starburst stars, survival and re-formation of gas disks, and origins of the global Kennicutt-Schmidt law. All these properties can be compared to the results from previous sub-grid models in a straightforward manner. We also test the proposition that the star formation rate is controlled by feedback, in mergers as well as in quiescent galaxies, since we do not adjust our star formation prescription.

In companion papers, we examine in more detail some of the phenomena that could not be predicted with previous models: the phase structure of the ISM and properties of starburst ``super-winds'' driven by stellar feedback, as well as the physics of star cluster formation in tidally shocked and starburst regions.

\begin{figure}
    \centering
    \scaleup
    \plotonesize{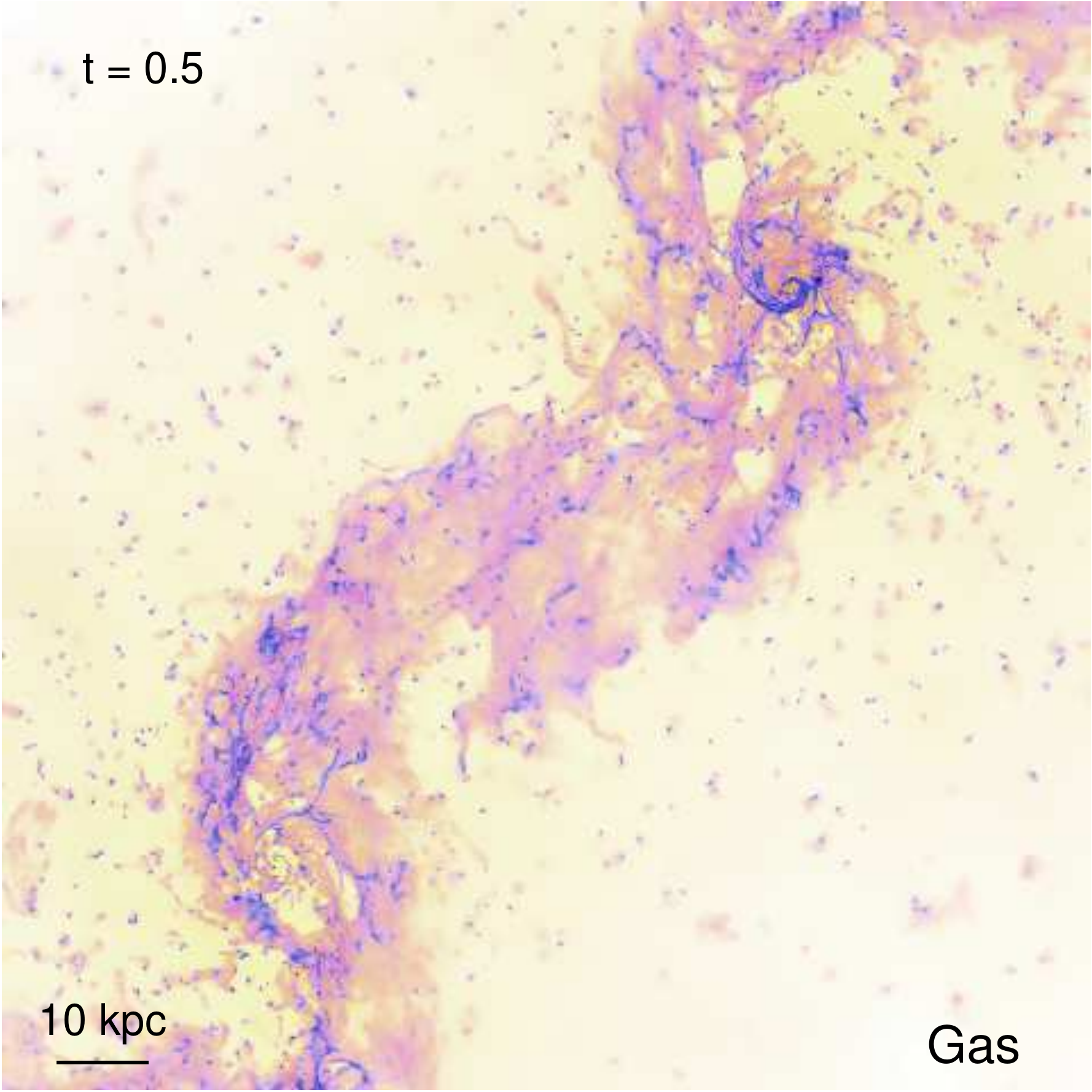}{1}
    \caption{Morphology of the gas in a standard simulation (high-resolution with all 
    explicit feedback mechanisms included) 
    of a merger of the HiZ disk model: a massive, 
    $z\sim2-4$ starburst disk merger.  
    The time is near apocenter after first passage. 
    This color projection emphasizes the cold, star-forming gas. 
    Brightness encodes projected gas density (light-to-dark with increasing density; logarithmically 
    scaled with a $\approx4\,$dex stretch); color encodes gas temperature 
    with the blue/violet material being $T\lesssim1000\,$K molecular and atomic gas, 
    pink/red $\sim10^{4}-10^{5}$\,K 
    warm ionized gas, and yellow/white $\gtrsim10^{6}\,$K hot gas. 
    Gravitational collapse forms giant molecular clouds and proto-star cluster complexes 
    throughout the gas. The outflows present include a component in dense clumps.
    Images of the other simulations at various times are in Appendix~\ref{sec:appendix:images}.
    \label{fig:morph.1}}
\end{figure}
\begin{figure}
    \centering
    \plotonesize{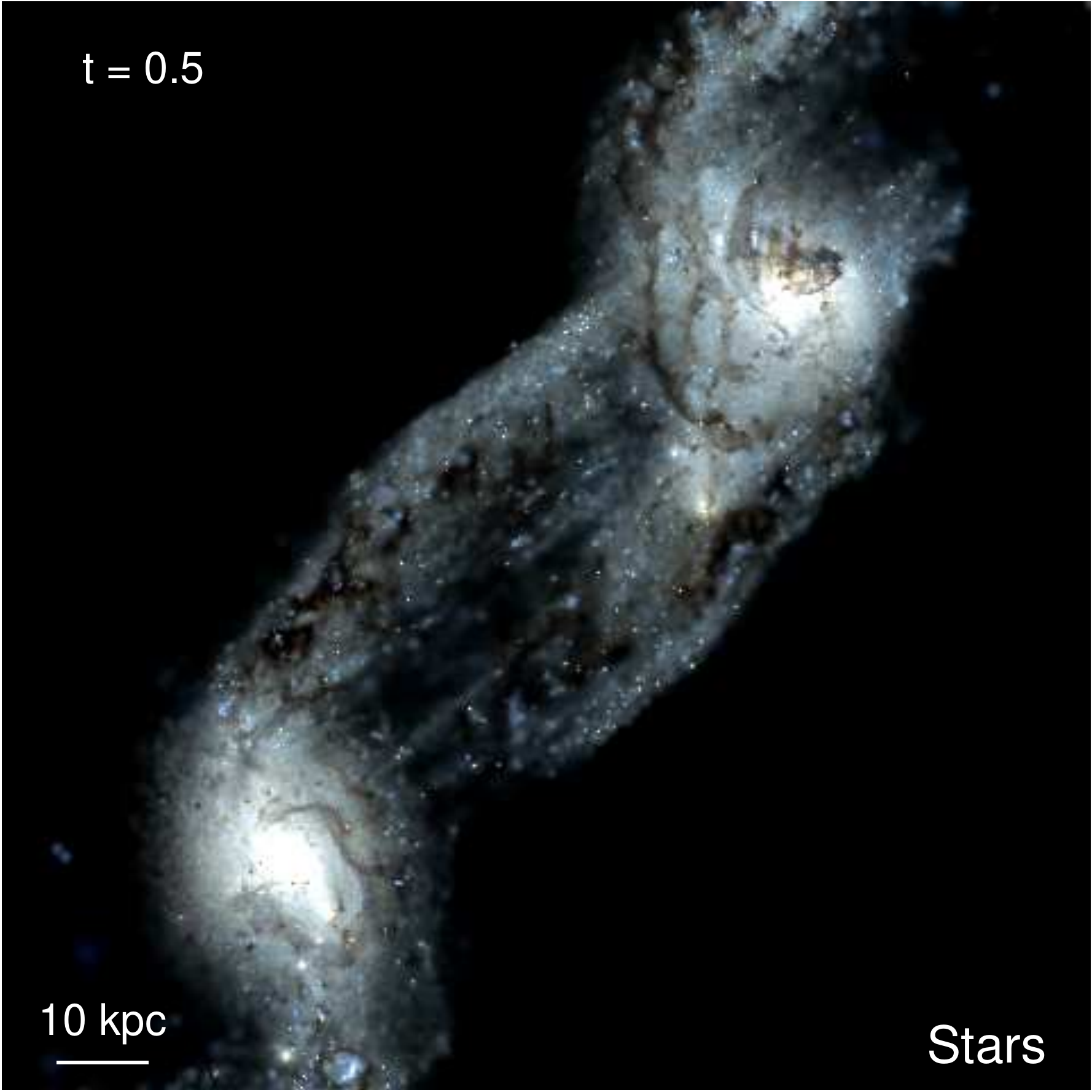}{1}
    \caption{The stars at the same time as Fig.~\ref{fig:morph.1}.
    The image is a mock $ugr$ (SDSS-band) composite, 
    with the spectrum of all stars calculated from their known age and metallicity, 
    and dust extinction/reddening accounted for from the line-of-sight dust mass. 
    The brightness follows a logarithmic scale with a stretch of $\approx2\,$dex. Young 
    star clusters are visible throughout the system as bright white pixels. 
    The nuclei contain most of the star formation, evident in their saturated brightness. 
    Fine structure in the dust gives rise to complicated filaments, dust lanes, and patchy obscuration of star-forming regions. A few super-starclusters are apparent as the brightest young stellar concentrations.     Images of the other simulations at various times are in Appendix~\ref{sec:appendix:images}.
    \label{fig:morph.3}}
\end{figure}

\begin{figure*}
    \centering
    \scaleup
\begin{tabular}{cc}
    \includegraphics[width={1.01\columnwidth}]{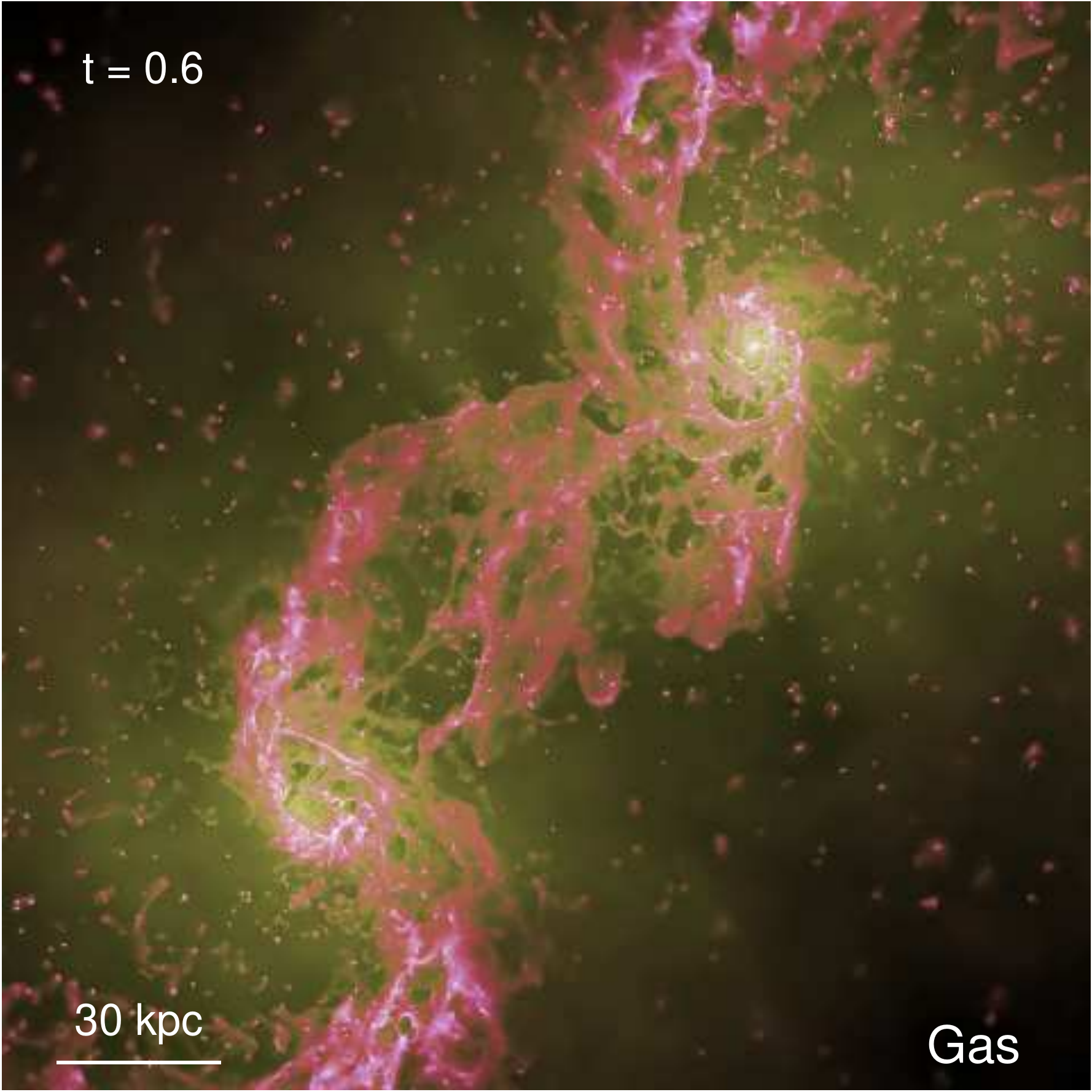} &
    \includegraphics[width={0.95\columnwidth}]{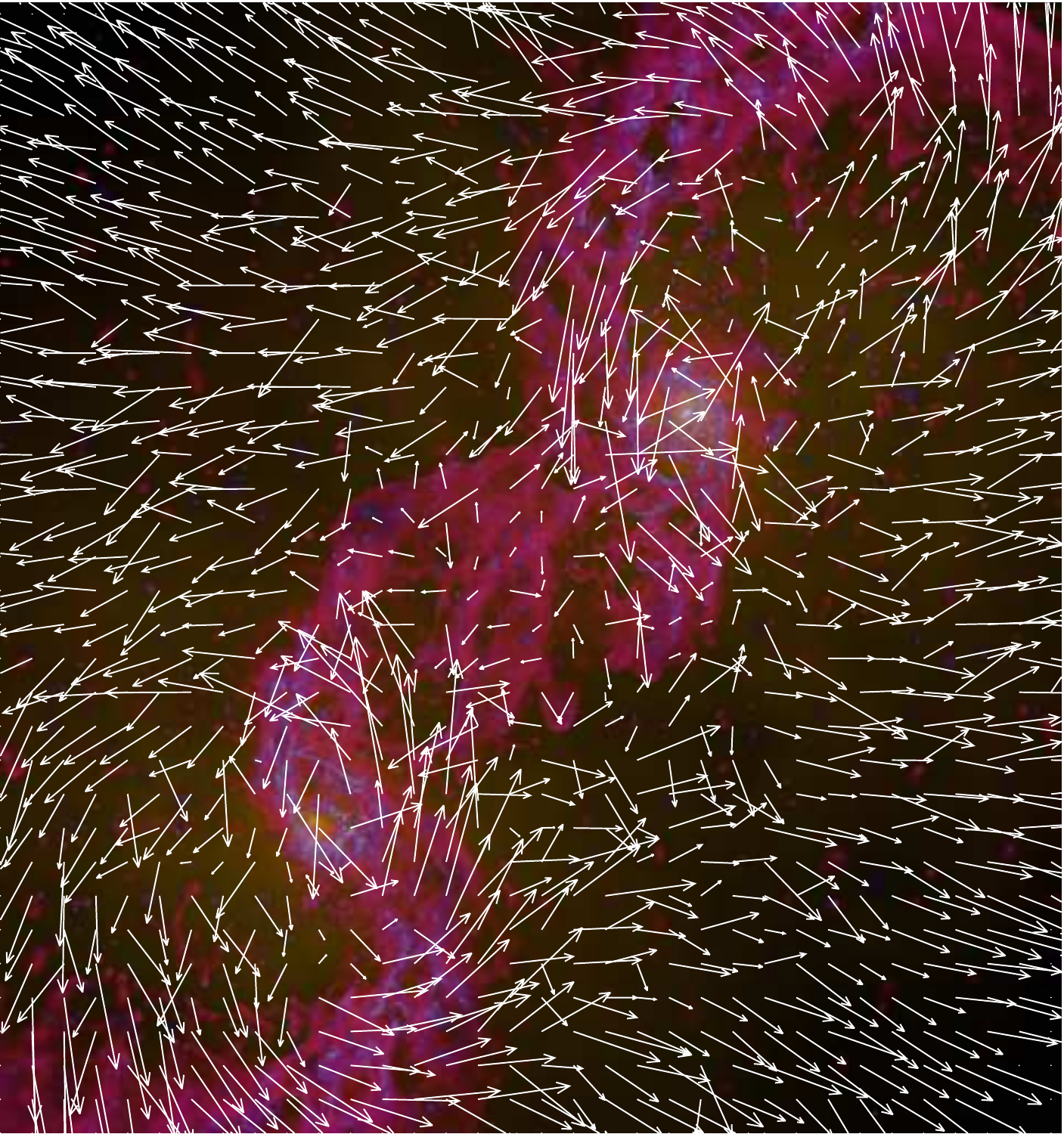}
\end{tabular}
    \caption{{\em Left:} Gas, as Fig.~\ref{fig:morph.1}, at a slightly later time, with a color projection here 
    chosen to emphasize the ionized and hot gas. Brightness now increases with surface density; color 
    encodes temperature in a similar manner (blue/pink/yellow representing cold/warm/hot phases). 
    Violent outflows are present, emerging both from the complexes and 
    the system as a whole, driven by the massive starburst. The volume-filling ``hot'' component 
    is now visible, with multiple ``bubbles'' driven by separate local events. 
    {\em Right:} Same, but with the (projected in-plane) velocity vectors plotted. The vectors interpolate the gas velocities evenly over the image; their length is linearly proportional to the magnitude of the local velocity with the longest plotted corresponding to $\approx500\,{\rm km\,s^{-1}}$. 
    \label{fig:morph.2}}
\end{figure*}


\begin{footnotesize}
\ctable[
  caption={{\normalsize Galaxy Models}\label{tbl:sim.ics}},center,star
  ]{lcccccccccccccc}{
\tnote[ ]{Parameters describing our (isolated) galaxy models, used as the initial conditions for all of the mergers: \\ 
{\bf (1)} Model name: shorthand for models of an isolated SMC-mass dwarf ({\bf SMC}); local gas-rich galaxy ({\bf Sbc}); MW-analogue ({\bf MW}); and high-redshift massive starburst ({\bf HiZ}).
{\bf (2)} $\epsilon_{g}$: gravitational force softening. Higher-resolution tests of the isolated galaxies use half this value (\paperone).
{\bf (3)} $M_{\rm halo}$: halo mass. 
{\bf (4)} $c$: halo concentration. Values lie on the halo mass-concentration 
relation at each redshift ($z=0$ for SMC, Sbc, and MW; $z=2$ for HiZ). 
{\bf (5)} $V_{\rm max}$: halo maximum circular velocity. 
{\bf (6)} $M_{\rm bary}$: total baryonic mass.
{\bf (7)} $M_{\rm b}$: bulge mass.
{\bf (8)} $a$: \citet{hernquist:profile} profile scale-length for bulge. 
{\bf (9)} $M_{d}$: stellar disk mass.
{\bf (10)} $r_{d}$: stellar disk scale length. 
{\bf (11)} $h$: stellar disk scale-height.
{\bf (12)} $M_{g}$: gas disk mass.
{\bf (13)} $r_{g}$: gas disk scale length (gas scale-height determined so that $Q=1$).
{\bf (14)} $f_{\rm gas}$: average gas fraction of the disk.
inside of the stellar $R_{e}$ ($M_{\rm g}[<R_{e}]/(M_{\rm g}[<R_{e}]+M_{\rm d}[<R_{e}])$).
The total gas fraction, including the extended disk, is $\sim50\%$ larger.
{\bf (15)} $Z$: initial metallicity (in solar units) of the gas and stars.
}
}{
\hline\hline
\multicolumn{1}{c}{Model} &
\multicolumn{1}{c}{$\epsilon_{\rm g}$} &
\multicolumn{1}{c}{$M_{\rm halo}$} & 
\multicolumn{1}{c}{$c$} & 
\multicolumn{1}{c}{$V_{\rm max}$} & 
\multicolumn{1}{c}{$M_{\rm bary}$} & 
\multicolumn{1}{c}{$M_{\rm b}$} & 
\multicolumn{1}{c}{$a$} & 
\multicolumn{1}{c}{$M_{\rm d}$} & 
\multicolumn{1}{c}{$r_{d}$} & 
\multicolumn{1}{c}{$h$} & 
\multicolumn{1}{c}{$M_{\rm g}$} & 
\multicolumn{1}{c}{$r_{g}$} &
\multicolumn{1}{c}{$f_{\rm gas}$} &
\multicolumn{1}{c}{$Z$} \\
\multicolumn{1}{c}{\,} &
\multicolumn{1}{c}{[pc]} &
\multicolumn{1}{c}{[$\msun$]} & 
\multicolumn{1}{c}{\,} & 
\multicolumn{1}{c}{[${\rm km\,s^{-1}}$]} & 
\multicolumn{1}{c}{[$\msun$]} & 
\multicolumn{1}{c}{[$\msun$]} & 
\multicolumn{1}{c}{[kpc]} & 
\multicolumn{1}{c}{[$\msun$]} & 
\multicolumn{1}{c}{[kpc]} & 
\multicolumn{1}{c}{[pc]} & 
\multicolumn{1}{c}{[$\msun$]} & 
\multicolumn{1}{c}{[kpc]} &
\multicolumn{1}{c}{\,} &
\multicolumn{1}{c}{[$Z_{\sun}$]} \\
\hline
{\bf SMC} & 1.0 & 2.0e10 & 15 & 46 & 8.9e8 & 1e7 & 0.25 & 1.3e8 & 0.7 & 140 & 7.5e8 & 2.1 & 0.56 & 0.1 \\ 
{\bf Sbc} & 3.1 & 1.5e11 & 11 & 86 & 1.05e10 & 1e9 & 0.35 & 4e9 & 1.3 & 320 & 5.5e9 & 2.6 & 0.36 & 0.3 \\ 
{\bf MW} & 4.0 & 1.6e12 & 12 & 190 & 7.13e10 & 1.5e10 & 1.0 & 4.73e10 & 3.0 & 300 & 0.9e10 & 6.0 & 0.09  & 1.0 \\ 
{\bf HiZ} & 7.0 & 1.4e12 & 3.5 & 230 & 1.07e11 & 7e9 & 1.2 & 3e10 & 1.6 & 130 & 7e10 & 3.2 & 0.49 & 0.5 \\ 
\hline\hline\\
}
\end{footnotesize}

\begin{figure*}
    \centering
    \scaleup
    \plotsidesize{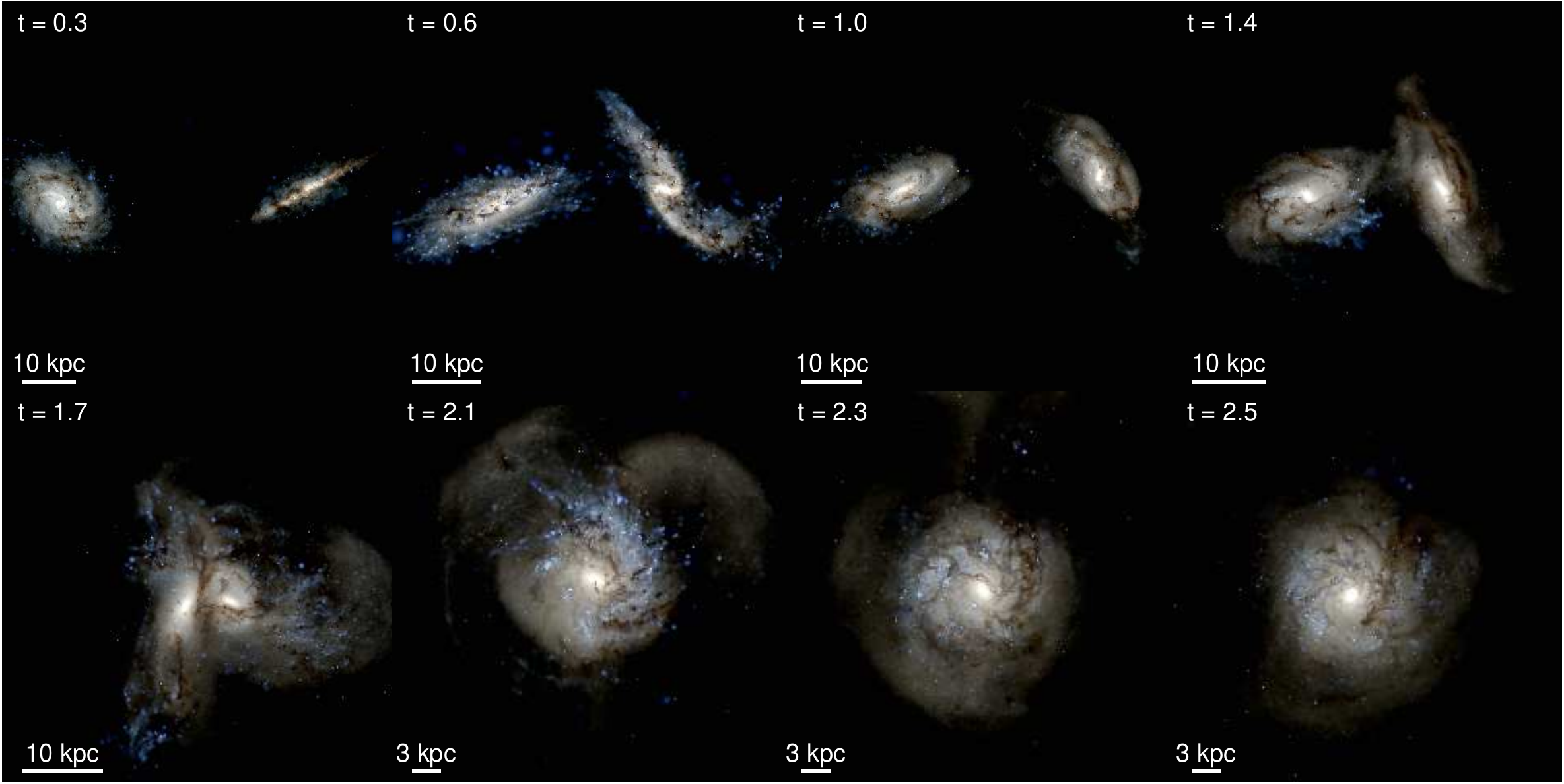}{1.0}
    \caption{Morphology of the {\bf f} (retrograde) merger of the Sbc galaxy (a dwarf starburst); each shows the optical as Fig.~\ref{fig:morph.3} at different times during the merger.
    \label{fig:stile.sbc.e}}
\end{figure*}

\begin{figure*}
    \centering
    \scaleup
    \plotsidesize{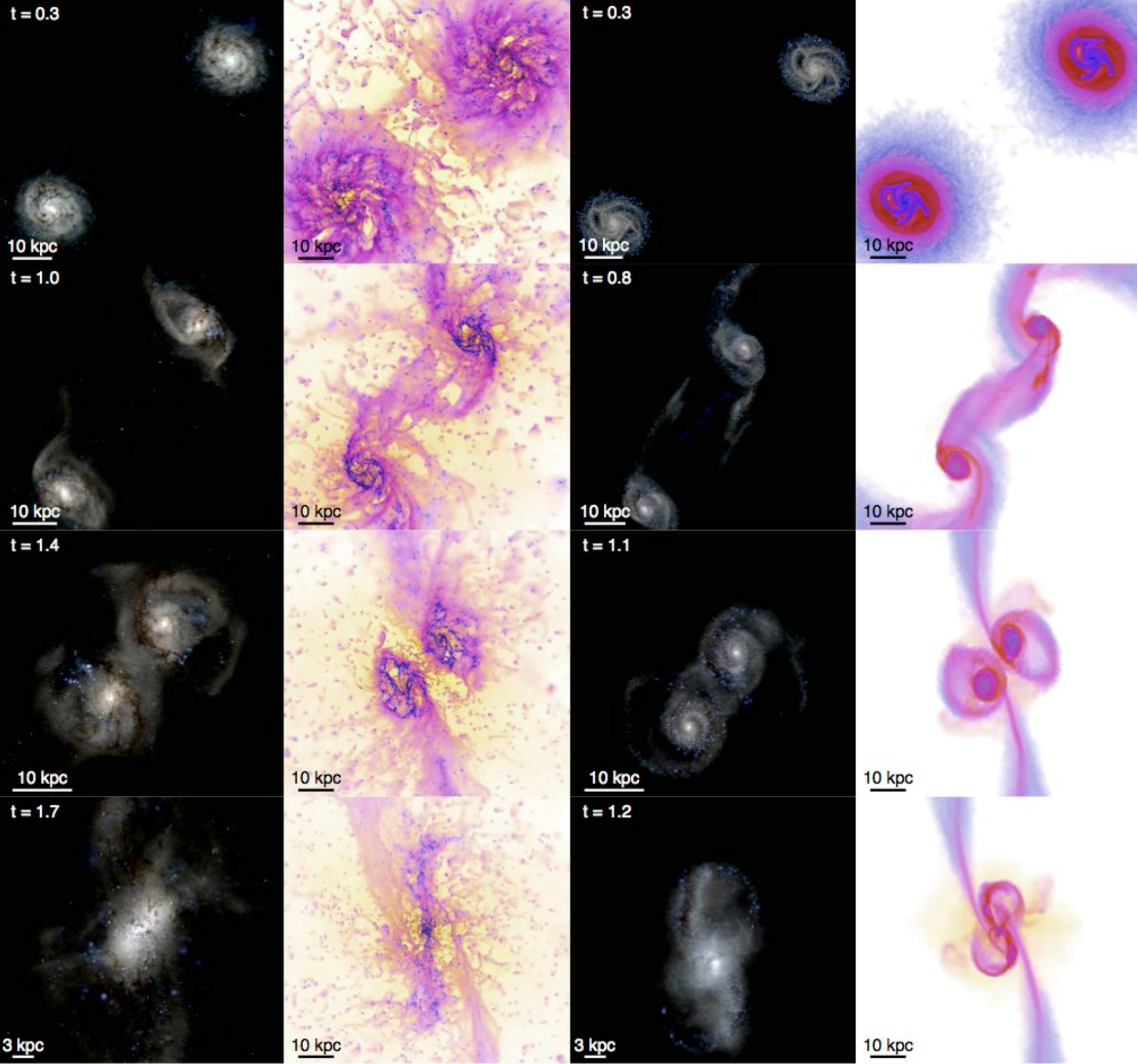}{1.0}
    \caption{Comparison of the morphology in gas and stars in the {\bf e} (prograde) merger of the Sbc (dwarf starburst) galaxy; each shows the optical and gas as Figs.~\ref{fig:morph.1}-\ref{fig:stile.sbc.e} at different times during the merger (top to bottom: before first passage, after first passage, before final coalescence, and after coalescence, respectively). {\em Left:} Simulations with explicit stellar feedback models (stars and gas). {\em Right:} Simulations with simplified sub-grid treatment of the ISM and stellar feedback; the lack of small-scale structure (molecular clouds, star clusters, and dust lanes) and galactic winds is apparent. A complete set of images for all the simulations (with additional panels) is given in Appendix~\ref{sec:appendix:images}. 
    \label{fig:morph.explicit.vs.subgrid}}
\end{figure*}

\vspace{-0.5cm}
\section{Methods}
\label{sec:sims}

The simulation techniques and galaxy models used here are described in detail in 
\paperone\ (Sec.~2 \&\ Tables~1-3) and \papertwo\ (Sec.~2).
We briefly summarize the most important properties here. 
The simulations were performed with the parallel TreeSPH code {\small 
GADGET-3} \citep{springel:gadget}, in its fully
conservative formulation \citep{springel:entropy}.
They include stars, dark matter, and gas, 
with cooling, shocks, star formation, and stellar feedback.

\vspace{-0.5cm}
\subsection{Disk Models}

We implement the ISM
prescription in four distinct initial disk models spanning a range of galaxy types (summarized in Table~\ref{tbl:sim.ics}). 
Each has a bulge, stellar and gaseous disk, halo, and central BH (although to isolate the 
role of stellar feedback, models for BH growth and feedback are disabled). 
At our standard resolution, each model has $\approx 0.3-1\times10^{8}$ total particles, 
giving particle masses of $500-1000\,\msun$ and $1-5$\,pc smoothing lengths. 
A couple of ultra-high resolution runs (of isolated versions of these disks) for 
convergence tests employ $\approx 10^{9}$ particles with sub-pc resolution.\footnote{These tests are described in \paperone\ and \papertwo, and used to check convergence in small-scale ISM properties. We have also run every simulation described in this paper with $10$ times fewer particles ($2$ times larger softening); although the small-scale properties such as the structure of individual star clusters differ, all  properties plotted in this paper (more global values in general) are found to be indistinguishable.}
The disk models include: 

(1) SMC: an SMC-like dwarf, with baryonic mass $M_{\rm bar}=8.9\times10^{8}\,\msun$ 
and halo mass $M_{\rm halo}=2\times10^{10}\,\msun$ (concentration $c=15$), 
a \citet{hernquist:profile} profile bulge with a mass $m_{b}=10^{7}\,\msun$, and exponential 
stellar ($m_{d}=1.3\times10^{8}\,\msun$) and gas disks ($m_{g}=7.5\times10^{8}\,\msun$) 
with scale-lengths $h_{d}=0.7$ and $h_{g}=2.1$\,kpc, respectively. 
The initial stellar scale-height is $z_{0}=140$\,pc and the stellar disk is initialized such that the 
Toomre $Q=1$ everywhere. The gas and stars are initialized with uniform metallicity $Z=0.1\,Z_{\sun}$.

(2) MW: a MW-like galaxy, with halo and baryonic properties of $(M_{\rm halo},\,c)=(1.6\times10^{12}\,\msun,\,12)$ and $(M_{\rm bar},\,m_{b},\,m_{d},\,m_{g})=(7.1,\,1.5,\,4.7,\,0.9)\times10^{10}\,\msun$, $Z=Z_{\sun}$, and scale-lengths $(h_{d},\,h_{g},\,z_{0})=(3.0,\,6.0,\,0.3)\,{\rm kpc}$. 

(3) Sbc: a LIRG-like galaxy (i.e.\ a more gas-rich spiral than is characteristic 
of those observed at low redshifts)
with $(M_{\rm halo},\,c)=(1.5\times10^{11}\,\msun,\,11)$,  
$(M_{\rm bar},\,m_{b},\,m_{d},\,m_{g})=(10.5,\,1.0,\,4.0,\,5.5)\times10^{9}\,\msun$, 
$Z=0.3\,Z_{\sun}$, 
and $(h_{d},\,h_{g},\,z_{0})=(1.3,\,2.6,\,0.13)\,{\rm kpc}$. 

(4) HiZ: a high-redshift massive starburst disk, chosen to match the 
properties of the observed non-merging but rapidly star-forming SMG 
population, with 
$(M_{\rm halo},\,c)=(1.4\times10^{12}\,\msun,\,3.5)$ and a 
virial radius appropriately rescaled for a halo at $z=2$ rather than $z=0$,  $(M_{\rm bar},\,m_{b},\,m_{d},\,m_{g})=(10.7,\,0.7,\,3,\,7)\times10^{10}\,\msun$, $Z=0.5\,Z_{\sun}$, and 
$(h_{d},\,h_{g},\,z_{0})=(1.6,\,3.2,\,0.32)\,{\rm kpc}$.

\vspace{-0.5cm}
\subsection{Merger Parameters}
\label{sec:sims:mergers}

The purpose of this paper is to contrast models with and without 
explicit stellar feedback, not to present a systematic study of all 
merger parameters -- extensive studies of this nature can be found 
in the literature \citep{barneshernquist92,barnes02:gasdisks.after.mergers,
naab:boxy.disky.massratio,cox:kinematics,cox:massratio.starbursts,
robertson:fp,younger:minor.mergers,dimatteo:merger.induced.sb.sims,
burkert:anisotropy,hoffman:dissipation.and.gal.kinematics,
jesseit:merger.rem.spin.vs.gas}. We therefore focus on a small but representative 
subset of possible mergers.

We consider equal-mass mergers (merging identical copies of galaxies (1)-(4)). 
The pairs are placed on parabolic orbits 
\citep[motivated by cosmological simulations; see][]{benson:cosmo.orbits,
khochfar:cosmo.orbits} with the spin axis of each disk specified by $\theta$ and $\phi$ 
in spherical coordinates. We choose two representative orientations from among those
which have been studied in a number of other works \citep{cox:kinematics,
barnes:disk.halo.mergers,hernquist:bulgeless.mergers,hernquist:bulge.mergers,naab:minor.mergers}.
The first (orbit {\bf e} in \citealt{cox:kinematics}) is near-prograde 
with $(\theta_{1},\,\phi_{1},\,\theta_{2},\,\phi_{2})=(30,\,60,\,-30,\,45)$; 
the second (orbit {\bf f}) is near-retrograde (or polar-retrograde) 
with  $(\theta_{1},\,\phi_{1},\,\theta_{2},\,\phi_{2})=(60,\,60,\,150,\,0)$. 
This spans both a strong resonant interaction (prograde) and weak, out-of-resonance 
(retrograde) interaction. 
We choose these rather than perfectly prograde 
$(\theta_{1},\,\phi_{1},\,\theta_{2},\,\phi_{2})=(0,\,0,\,0,\,0)$ 
and retrograde $(\theta_{1},\,\phi_{1},\,\theta_{2},\,\phi_{2})=(180,\,0,\,180,\,0)$ orbits 
because there are known pathological behaviors 
associated with the narrow phase space of perfectly resonant orbits.

We have compared a subset of low-resolution simulations that vary orbital 
inclinations and pericentric passage distances; the variation in remnant properties 
is consistent with previous studies \citep[e.g.][]{cox:kinematics}. 
But this variation is largely a dependence on the strength of gravitational torques and 
total angular momentum; the {\em differences} between models with/without feedback 
do not dramatically depend on these properties. In any case our {\bf e} and {\bf f} orbits 
reasonably bracket the range from most to least violent orbits.

\vspace{-0.5cm}
\subsection{Explicit Feedback Models}

The most important physics in these simulations is the model of 
stellar feedback.   
We include feedback from a variety of mechanisms, each of which we briefly describe below.  More details about our implementations of this physics are given in \paperone\ and \papertwo.
 We use a  \citet{kroupa:imf} initial mass function (IMF) throughout and use STARBURST99 \citep{starburst99} to calculate the stellar luminosity, mass return from stellar winds, supernova rate, etc. as a function of the age and metallicity of each star particle.

(1) {\bf Local Momentum Deposition} from Radiation Pressure, 
Supernovae, \&\ Stellar Winds: In \paperone, we present the radiation pressure aspect of this 
model for feedback from young star clusters in detail. At each timestep, gas particles identify the nearest density peak representing the center of the nearest star-forming ``clump'' or GMC-analog. 
We calculate the total luminosity of the star particles inside the sphere defined by the distance from the center of this star-forming region to the gas particle of interest;  the incident flux on the gas is then determined assuming that the local star forming region is optically thick to the UV radiation.

The rate of momentum deposition from radiation pressure is then $\dot{P}_{\rm rad}\approx (1+\tau_{\rm IR})\,L_{\rm incident}/c$ 
where the term 
$1+\tau_{\rm IR}$ accounts for the fact that most of the initial optical/UV radiation is 
absorbed and re-radiated in the IR; $\tau_{\rm IR}=\Sigma_{\rm gas}\,\kappa_{\rm IR}$ 
is the optical depth in the IR, which allows for the fact that the momentum is boosted 
by multiple scatterings in optically thick regions. 
Here $\Sigma_{\rm gas}$ is calculated self-consistently as the average surface density of the identified clump, with $\kappa_{\rm IR}\approx5\,(Z/Z_{\sun})\,{\rm g^{-1}\,cm^{2}}$ approximately 
constant over the relevant physical range of dust temperatures. 
The imparted acceleration is directed along the flux vector.   In \paperone\ we discuss numerous technical aspects of this implementation -- such as the effects of resolution, photon leakage, 
and how the momentum is discretized -- and show that essentially all our conclusions are robust to uncertainties in these choices. 

The {\em direct} momentum of SNe ejecta and stellar winds $\dot{P}_{\rm SNe}$ 
and $\dot{P}_{\rm w}$ are similarly tabulated from STARBURST99 and injected as an appropriate function of age and metallicity to the gas within a smoothing length of each star. This source of turbulent energy is almost always smaller than that due to radiation pressure discussed above.  In some cases, however, in particular in dwarf galaxies, the work done by bubbles of gas shock-{\em heated}  by supernovae and/or stellar winds is dynamically important; this is discussed below. 

(2) {\bf Supernova and Stellar Wind Shock-Heating}: The gas shocked by 
supernovae and stellar winds can be heated to high temperatures, generating bubbles and 
filaments of hot gas. We tabulate the Type-I and Type-II SNe rates from \citet{mannucci:2006.snIa.rates}  and STARBURST99, respectively, as a function of age and 
metallicity for all star particles, and stochastically determine at 
each timestep if a SN occurs.  The SNe are resolved 
discretely in time (as opposed to continuous energy injection). For each SN, the appropriate thermal energy is injected into the gas within a smoothing length of the star particle. Similarly, stellar winds are assumed to shock locally and so we inject the appropriate tabulated mechanical 
power $L(t,\,Z)$ as a continuous function of age and metallicity into the gas within a smoothing length of the star particles.   The specific energy of these stellar winds is large for young stellar populations in which fast winds from massive stars dominate, but declines rapidly at later times when slower, AGB winds dominate. 

(3) {\bf Gas Recycling:} Gas mass is returned continuously
to the ISM from stellar evolution, at a rate tabulated from SNe and stellar mass 
loss in STARBURST99. The integrated mass fraction recycled 
is $\sim0.3$.  

(4) {\bf Photo-Heating of HII Regions}: For each star particle, we tabulate the 
rate of production of ionizing photons; starting from the nearest gas particle and moving radially outwards, we then ionize each non-ionized gas particle (using the gas and stellar properties to determine the photon production rate needed to maintain the particle as fully ionized). Gas which is ionized is immediately heated to $\sim10^{4}$ K, unless is already above this temperature; moreover, the gas is not allowed to cool below $10^4$ K until it is no longer in an HII region. This method allows for overlapping, non-spherical HII regions that can extend to radii $\sim$kpc. 

(5) {\bf Long-Range Radiation Pressure:} Radiation pressure from photons absorbed 
in the immediate vicinity of stars is captured in mechanism (1). However, photons that 
escape these regions can still be absorbed at larger radii. For each star particle, we 
construct the intrinsic SED ($L_{\nu}$) as a function of age and metallicity; 
we then use the local density and density gradients to estimate the integrated 
column density and attenuation of the SED using $\tau_{\nu} = \kappa_{\nu}\,\Sigma \approx \kappa_\nu \, \rho\,(h_{\rm sml} + |\nabla\ln\rho|^{-1})$, where $h_{\rm sml}$ is the smoothing length and $\kappa_\nu$ is the frequency-dependent opacity (assuming dust opacities that 
scale with metallicity, as in (1) above).   The resulting ``escaped'' SED gives a frequency-dependent 
flux ${\bf F}_{\nu}$ that is propagated to large distances. We construct a force tree for 
this long range force in an identical fashion to the gravity tree, since after attenuating the flux near the star particle, the stellar flux is assumed to decrease $\propto r^{-2}$. Each gas particle then sees an incident net flux vector  ${\bf F}_{\nu}^{i}$, integrated over all stars in the galaxy. 

Extensive numerical tests of the feedback models are presented in \paperone, \papertwo, 
and \paperthree. 

In these models, gas follows an atomic cooling curve with additional fine-structure 
cooling to $<100\,$K, with no cooling floor imposed. 
Star formation follows the model in \citet{hopkins:binding.sf.prescription}. 
Regions which are locally self-gravitating on the smallest resolved scale, 
i.e.\ which have $\alpha\equiv \delta v^{2}\,\delta r/G\,m_{\rm gas}(<\delta r) \rightarrow 
({|\nabla\cdot {\bf v}|^{2} + |\nabla\times {\bf v}|^{2}})/({2\,G\,\rho}) < 1$ 
should collapse to much higher densities in a free-fall time absent feedback; 
we therefore assign them an instantaneous SFR of $\dot{\rho}_{\ast}=\rho/t_{\rm ff}(\rho)$. 
We stress that because feedback is present and can regulate against further collapse once stars form, the average efficiency in the dense gas is much lower (typically $\sim1\%$).
We further follow \citet{krumholz:2011.molecular.prescription} and calculate the molecular 
fraction within the dense gas as a function of the local column density and 
metallicity, and allow star formation only from that gas; however as shown in 
\papertwo\ and \citet{hopkins:binding.sf.prescription} this has essentially no effect on our results.

As noted in the introduction, we showed in \paperone\ and \papertwo\ that the galaxy structure and SFR are basically independent of the small-scale SF law, density threshold (provided it is high), and treatment of molecular chemistry, because the structure and SFR are feedback-regulated. As a further test of this result, we have re-run several of the models described here with the more simplified (and frequently used) star formation prescription in which star formation is restricted to occur only in gas above a density $n>1000\,{\rm cm^{-3}}$, with a rate $\dot{\rho}_{\ast}=0.015\,\rho/t_{\rm ff}$ corresponding to observed average efficiencies ``enforced.'' This has almost no effect on global properties or the galaxy-average SFRs, but it does tend to ``smear out'' the SFR in dense regions such as the merging nuclei, where {\em all} the gas is well above this threshold.

\vspace{-0.5cm}
\subsection{Models with an ``Effective'' Equation of State}
\label{sec:sims:qeos}

In previous work on galaxy mergers, it was not possible to explicitly resolve 
the processes described above. Instead, simulations used a variety of sub-grid 
approaches to model some ``effective'' ISM properties on the resolution scales of the simulations (often a few hundred pc). 

We want to assess the accuracy of such subgrid models.
We therefore compare otherwise identical disk merger models using the effective equation of state (EOS) approach in \citet{springel:multiphase}. Since high-density GMCs are not resolved, the SF threshold is lower, chosen to represent densities where multi-phase structure should appear (here we adopt $n>1\,{\rm cm^{-3}}$); the efficiency must be correspondingly tuned to match the global Kennicutt relation (here $\dot{\rho}_{\ast}=0.03\,\rho/t_{\rm ff}$). The model attempts to account for the complex sub-grid feedback physics described above by simply assigning the star-forming gas an effective sound speed $c_{\rm eff}$ representing e.g.\ a sub-grid turbulent ISM pressure. 

In the model of \citet{springel:multiphase}, the effective pressure is motivated by the ISM model in \citet{mckee.ostriker:ism}. Because the assumed sub-grid ISM properties are uncertain, the model allows for interpolation in the EOS with the parameter $\qeos$ \citep[see][]{springel:models,springel:spiral.in.merger,robertson:disk.formation,
robertson:msigma.evolution}. The standard ``maximal feedback'' case ($\qeos=1$) assumes $100\%$ of the 
energy from supernovae couples to an inter-cloud diffuse phase with long cooling times; this leads to a fairly ``stiff'' equation of state, $c_{\rm eff}\sim50-200\,{\rm km\,s^{-1}}$.
The opposite ``weak feedback'' case ($\qeos=0$) has $c_{\rm eff}\sim10\,{\rm km\,s^{-1}}$, motivated by the minimum dispersion observed in gas in most galaxies. This is quite similar to the effective EOS used in 
\citet{teyssier:2010.clumpy.sb.in.mergers}. Intermediate values interpolate 
as $c_{\rm eff}^{2}=\qeos\,c_{\rm eff}^{2}[q=1]+(1-\qeos)\,c_{\rm eff}^{2}[q=0]$. 
In \citet{hopkins:zoom.sims}, various choices are compared to observations of star-forming systems as a function of density; they suggest that $\qeos\sim0.1-0.3$ gives a reasonable representation of the observations. We therefore adopt $\qeos=0.25$ as our standard choice; but we have also re-run comparison models with $\qeos=1$ and our conclusions are nearly identical.

\begin{figure}
    \centering
    \scaleup
    \plotonesize{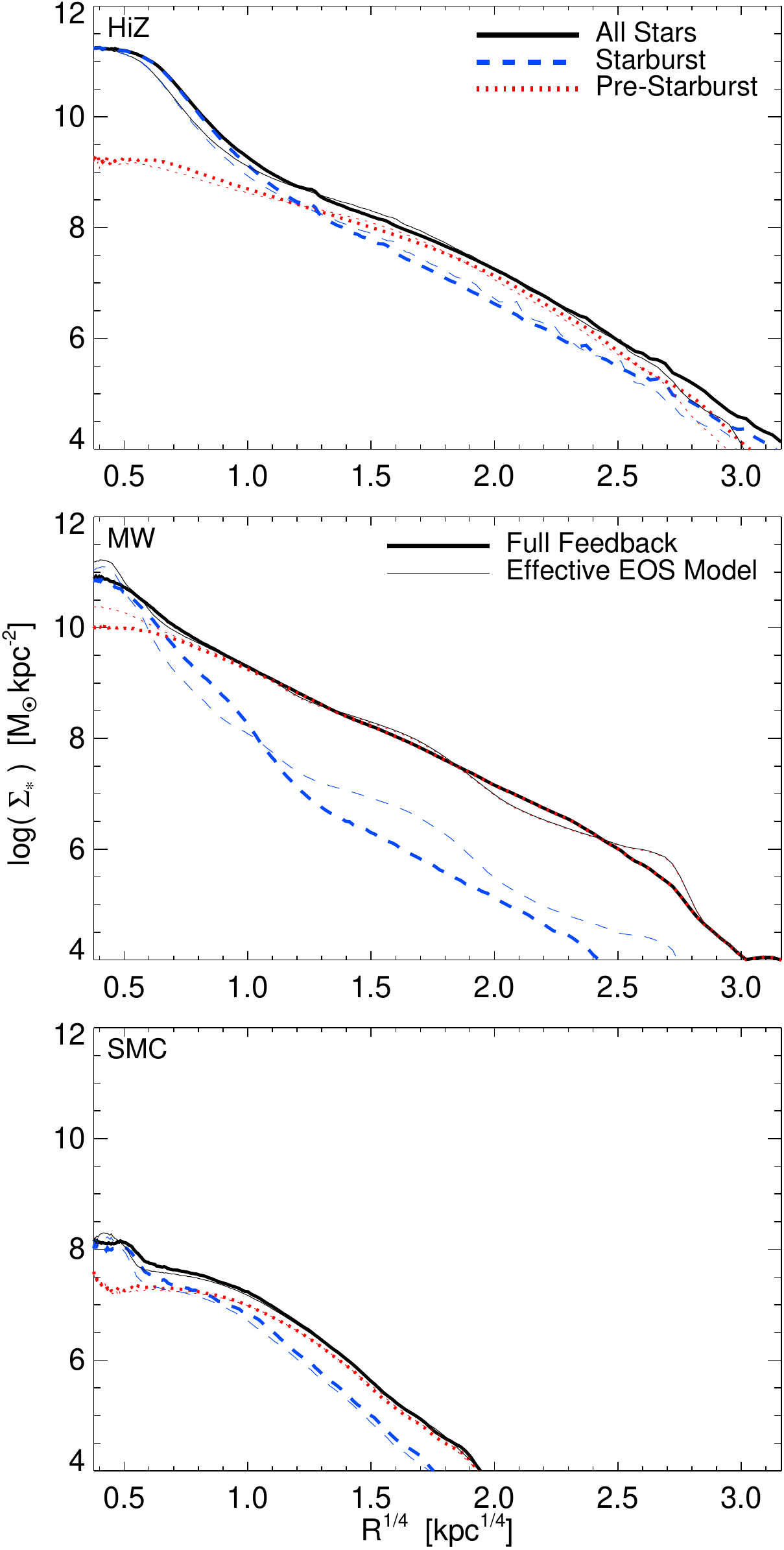}{0.9}
    \caption{Merger remnant stellar mass profiles, with explicit (thick lines) and 
    implicit/sub-resolution (thin lines) 
    stellar feedback prescriptions.
    {\em Top:} High-redshift gas-rich starburst merger.
    {\em Middle:} MW-like galaxy merger.
    {\em Bottom:} SMC-mass dwarf merger.
    For each simulation, we plot the total stellar surface density (black solid), 
    and the separate dissipationless/violently relaxed component (stars formed before 
    the merger simulation; red dotted) and dissipation/starburst 
    component (stars formed in the merger-induced starburst; blue dashed). 
    The profiles are averaged over $\approx 100$ viewing angles. 
    In all cases, the profiles agree very well independent of the details 
    of the feedback prescription. 
    Despite the inner structure, dynamics, and winds associated with the 
    starburst being modified considerably, the  
    ``total starburst mass'' and characteristic radii where the post-starburst 
    light dominates the profile appear robust. This is expected if 
    these properties are set by large-scale gravitational torques.
    \label{fig:massprofile.compare}}
\end{figure}

\vspace{-0.5cm}
\section{Morphologies}
\label{sec:morph}

\vspace{-0.1cm}
\subsection{Merger Morphologies}
\label{sec:morph:merger}
 
Figures~\ref{fig:morph.1}-\ref{fig:morph.3} show the gas and stars in some representative 
stages of the merger simulations, for our standard case (``explicit'' feedback 
model) where all feedback mechanisms are present. Figure~\ref{fig:stile.sbc.e}-\ref{fig:morph.explicit.vs.subgrid} and Figures~\ref{fig:tile.sbc.e}-\ref{fig:tile.smc.f} (in Appendix~\ref{sec:appendix:images}) extend this to show the gas and stars in a number of merger stages, from different viewing angles, for each merger. Figure~\ref{fig:morph.explicit.vs.subgrid} specifically compares this in a simulation with the explicit feedback model and one with identical initial conditions using the sub-grid treatment of the ISM and star formation. 

In each image set, the gas maps show the projected gas density (intensity) 
and temperature (color, with blue representing cold molecular gas at 
$T\lesssim 1000\,$K, pink representing the warm ionized gas at $\sim10^{4}-10^{5}$\,K, 
and yellow representing the hot, X-ray emitting gas at $\gtrsim10^{6}\,$K).\footnote{The projected temperatures are logarithmically-averaged and surface-density weighted, so reflect the temperature of most of the line-of-sight gas mass, rather than the temperature that contains most of the thermal energy.}
The stellar maps show a mock three-color observed
image, specifically a $u/g/r$ composite. 
The stellar luminosity in each band is calculated from each star particle 
according to the STARBURST99 model given its age, mass, and metallicity 
(and smoothed over the appropriate kernel). We then attenuate the stars 
following the method of \citet{hopkins:lifetimes.letter}: we calculate the total dust 
column (from the simulated gas) along the line-of-sight to each star particle 
for the chosen viewing angle (assuming a constant 
dust-to-metals ratio, i.e.\, dust-to-gas equal to the MW value 
times $Z/Z_{\sun}$), and apply a MW-like extinction and 
reddening curve \citep[as tabulated in][]{pei92:reddening.curves}.\footnote{Accounting for dust scattering changes the detailed spectrum, but the colors and morphologies in the images are not significantly altered \citep{jonsson:sunrise.attenuation,wuyts:model.numbers.and.colors.vs.obs,wuyts:2010.highz.sizes.vs.models}.}

As seen in \papertwo\ for isolated galaxies, gravitational collapse forms 
GMC complexes in the gas, which are individually 
dispersed rapidly by feedback once they turn a few percent of their mass into stars, 
but are replaced by continuously forming new clumps. The ISM as a whole is a 
a highly supersonically turbulent multiphase medium, with volume-filling hot gas (heated 
by SNe and O-star winds) making up a few percent of the mass, a ``warm'' phase of gas maintained by HII photoheating at $\sim 10^{4}\,$K, which constitutes a large filling factor and mass fraction at large radii from the galaxy nuclei, and GMC complexes in dense molecular filaments (and volume-filling molecular gas near the galaxy centers), with most of the gas mass. 

Violent outflows are clearly evident, both emerging from individual GMCs and from 
the merging systems as a whole. The composition of these outflows is discussed in 
detail in \paperthree. Briefly, the volume filling hot gas is primarily venting SNe and O-star 
wind heated gas; there is some contribution from gravitational shocks, but it is much 
smaller (see \citealt{cox:xray.gas} and discussion below). Occasionally the early stages of these outflows are evident as ``bubbles'' breaking out of the warm phase gas. The warm/cold clumps embedded in the 
outflow predominantly come from material directly accelerated by radiation pressure 
(which is much more efficiently at launching cold gas out of the disk, as opposed to 
the entrainment of that cold material by hot, tenuous outflows). 

In the stellar images, young stars clusters are visible throughout the galaxies. 
Most of the stars form in resolved clusters; in a companion paper we discuss the nature and properties of these clusters in detail. Most of them, however, are open clusters that correspond to GMCs which have converted $\sim5-10\%$ of their mass into stars (see \papertwo), and so are unbound once the much larger (marginally bound) GMC gas mass is dispersed by feedback. The bright nuclei contain most of the star formation, as discussed below. The massive gas inflows give rise to large dust column densities along many sightlines; face-on, we see heavily extincted dust lanes with complex structure (filaments, feathering, etc.). The patchy obscuration is typical of local merging systems. A few super-star clusters are visible as the brightest young stellar ``knots,'' often outside the nuclei in tidally shocked arms. 

\vspace{-0.5cm}
\subsection{Merger Remnant Morphologies}
\label{sec:morph:remnant}

As expected for the remnants of very major (equal-mass) mergers, 
the relic stars typically form a bulge-dominated galaxy once the merger has 
completed. 
In Figure~\ref{fig:massprofile.compare}, we show this quantitatively by 
comparing the surface stellar mass density profiles of the 
remnant stars. 
We plot the median profile averaged over $1000$ lines-of-sight 
uniformly sampling the unit sphere, but the sightline-to-sightline 
differences are small.
Despite being initially disk-dominated, the HiZ and MW models 
form clear $r^{1/4}$-law bulges; the SMC and Sbc cases may still have significant disks (discussed below), 
evident in the curvature (lower Sersic $n_{s}$) here.

The final mass profiles of models with explicit feedback or with an effective EOS are quite similar. This is, of course, expected where the profile is dominated by violent relaxation of pre-existing stars. However, in previous EOS models, it has generally been found that the central $\sim$kpc is dominated by stars formed in situ in the nuclear starburst \citep[from gas driven to the center by strong torques; see][]{mihos:cusps,hopkins:cusps.mergers}. We therefore examine the mass profile of two sub-components: the stars formed in the starburst,\footnote{For simplicity, we define this as all stars formed within $\pm150\,$Myr of the peak in the SFR near coalescence, but small changes in the definition have no effect on our conclusions.}
and the violently relaxed ``envelope'' of stars formed before this starburst/coalescence (all stars formed previous to this cut). In all cases, the explicit feedback models are nearly identical to the ``effective'' equation of state models. In particular, the starburst component, not just the dissipationless envelope, appears to have similar mass, shape, and radius.

There are some systematic second-order differences. The full feedback models tend to produce a somewhat more massive $\sim$kpc-scale core (at the tens of percent level), because dissipation is more efficient. There are subtle differences at large radii ($\sim10\,$kpc) owing to extended star formation (discussed below), although the total stellar mass in young stars involved at these radii is very small.

\begin{figure}
    \centering
    \scaleup
    \plotonesize{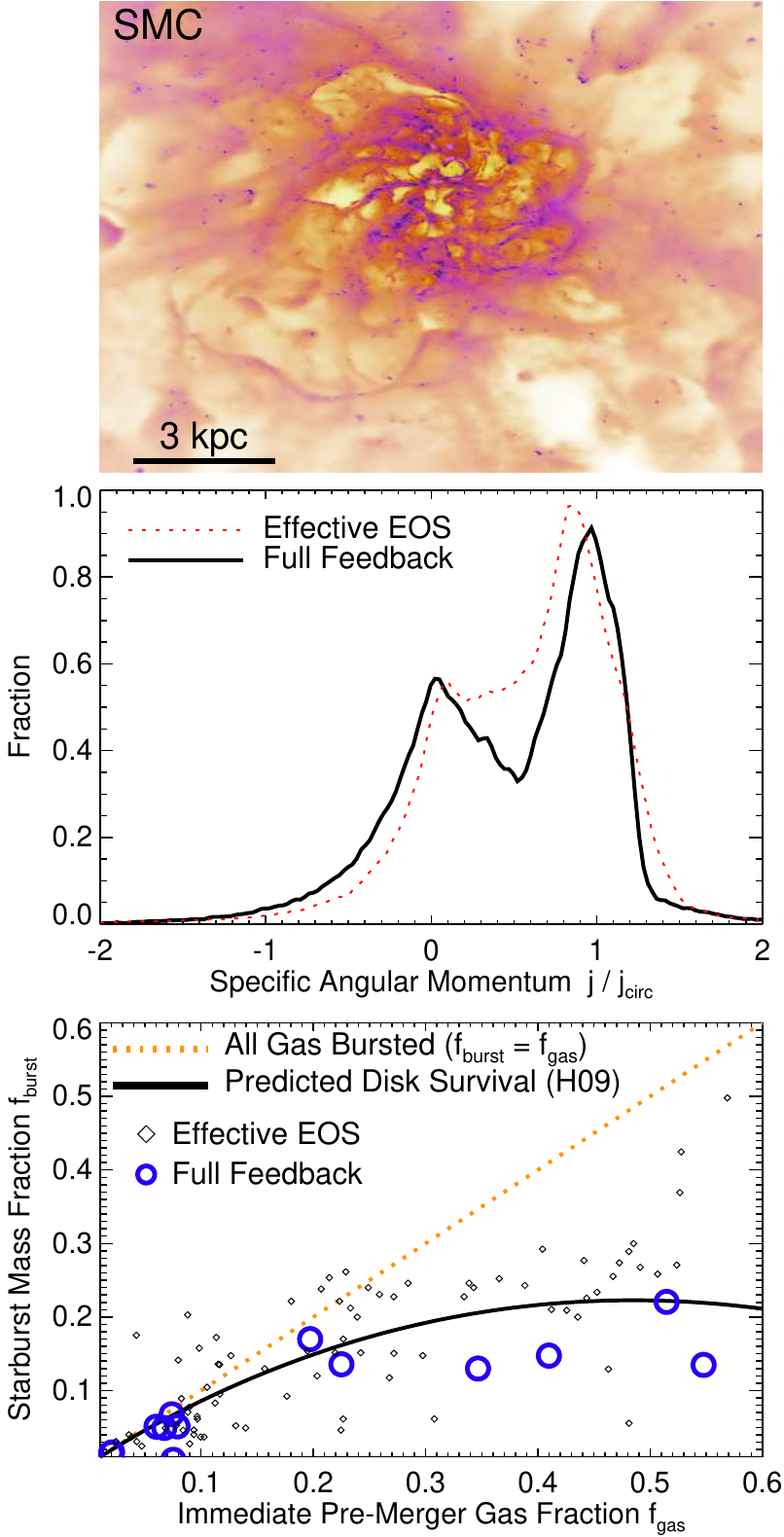}{1}
    \caption{Disk survival in mergers. 
    {\em Top:} Gas image of the SMC {\bf f} merger remnant (as Fig.~\ref{fig:morph.1}), with our full feedback model, 
    after final coalescence. The gas has clearly re-formed a large, rotating disk (face-on here), 
    very similar to the isolated progenitor(s). 
    {\em Middle:} Distribution of specific angular momentum $j$ of the baryons within $5$\,kpc of the center of the same model, after the merger; values are compared to the value $j_{\rm circ}$ of a pure circular orbit at the same radius. The distribution is clearly bimodal, with a peak containing $\approx30\%$ of the mass near $j/j_{\rm circ}=0$ (bulge), and $\approx70\%$ in a peak near $j/j_{\rm circ}=1$ (disk). The disk lies on the same Tully-Fisher relation as its progenitors.
    {\em Bottom:} Mass fraction of the disks which is consumed in the final coalescence starburst ($f_{\rm burst}$), as a function of the gas mass fraction just before the burst begins ($f_{\rm gas}$; here $200\,$Myr before the SFR peaks). Dotted line shows the result if all the gas were consumed, $f_{\rm burst}=f_{\rm gas}$. Solid line shows the approximate model for the efficiency of disk survival in gas-rich mergers from \citet{hopkins:disk.survival}, roughly $f_{\rm burst}\approx f_{\rm gas}\,(1-f_{\rm gas})$. Small diamonds compare the effective EOS simulations used to analyze disk survival therein. Large circles compare the simulations here. Values $f_{\rm burst}<f_{\rm gas}$ indicate disk survival and re-formation. Disk survival is comparably efficient in full feedback and EOS models.
    \label{fig:disk.survival}}
\end{figure}

\begin{figure}
    \centering
    \plotonesize{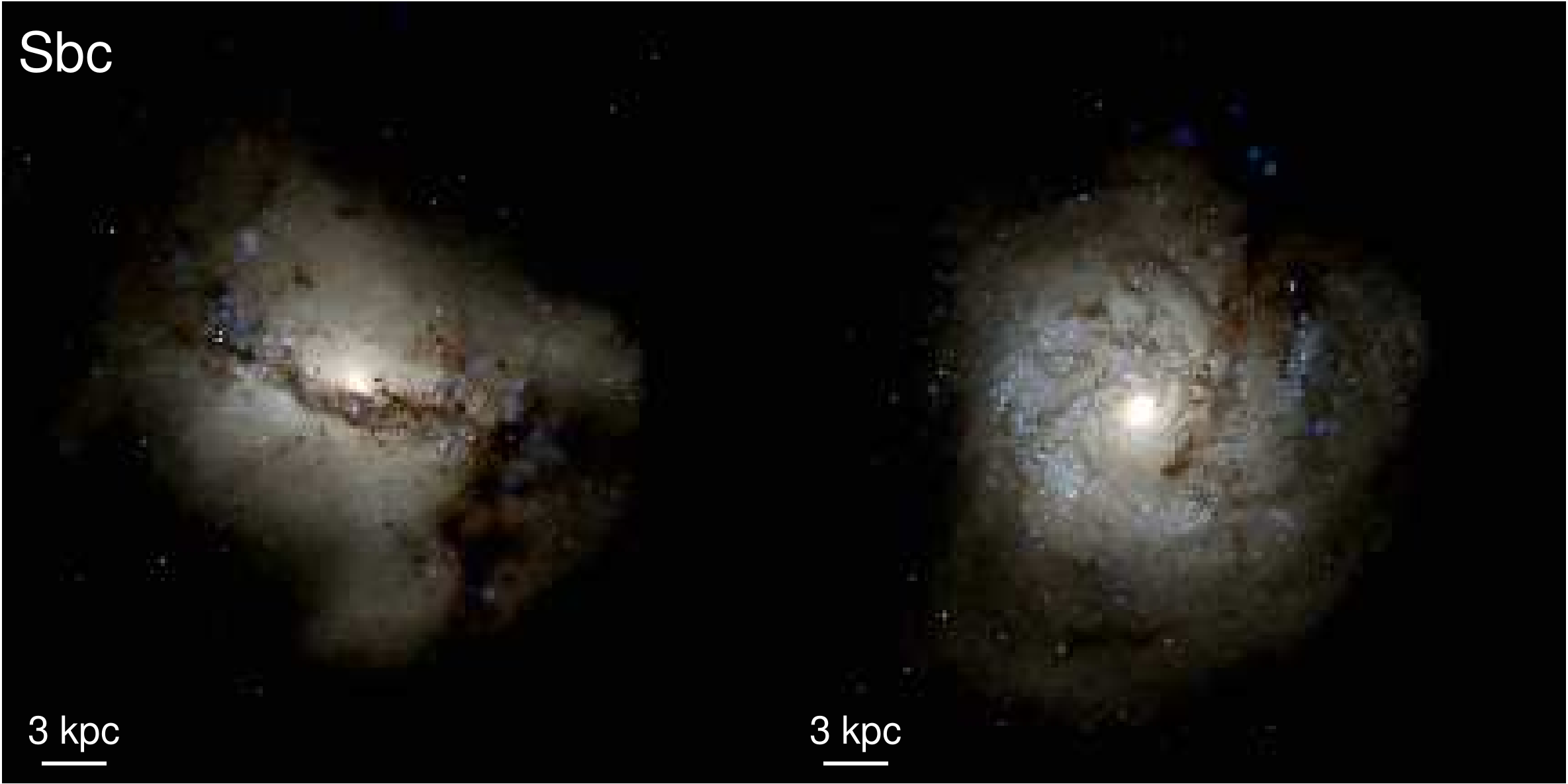}{1.0}
    \caption{Optical image (as Fig.~\ref{fig:morph.3}) of the stars in the relaxed merger remnant 
    of the Sbc {\bf f} merger (viewed edge-on and face-on). The violently relaxed stars form a very extended halo, but this is mostly at surface densities 
    $\sim100$ times smaller than the disk. The merger produces an S0/a (depending on viewing angle), despite the fact that it just experienced an {\em equal-mass} (1:1) merger with no new gas accretion: most situations will give even larger disks.
    \label{fig:disk.tile}}
\end{figure}

\vspace{-0.5cm}
\subsection{Disk Survival}
\label{sec:morph:disksurv}

A controversial subject in recent years has been the question of disk ``survival'' or ``re-formation'' after mergers. Simulations with EOS models predicted that sufficiently gas-rich mergers should leave large relic disks and even disk-dominated galaxies \citep{springel:spiral.in.merger,robertson:disk.formation,hopkins:disk.survival}. \citet{hopkins:disk.survival} showed that in those models, this is governed by global gravitational torques being inefficient at removing angular momentum from gas at large radii in disks, and so is independent of the details of the EOS implementation or introduction of an ad-hoc stellar wind model. However, other simulations, with either no hydrodynamics (``sticky particle'' codes) or weak stellar feedback but efficient cooling, predict much less efficient disk survival \citep{bournaud10}. We therefore examine this question for the first time in simulations with more detailed stellar feedback models. 

Figure~\ref{fig:disk.survival} shows the morphology of one merger remnant (the SMC-SMC retrograde {\bf f} merger) which clearly re-forms a prominent, rotationally supported disk. 
We quantify the disk mass and rotational support following \citet{hopkins:disk.survival}, by plotting the distribution of baryonic mass as a function of its angular momentum $j/j_{c}$, where $j=|{\bf r}\times {\bf v}| $ is the specific angular momentum and $j_{c} = r\,(G\,M_{\rm tot}(<r)/r)^{1/2}$ is the angular momentum of a circular orbit for each particle. The distribution is strongly bimodal, with a peak near zero net angular momentum (bulge stars), and a peak near $j/j_{c}=1$, i.e.\ in a circular orbit. Fitting to this distribution (either cutting at the minimum between the two peaks or assuming the ``bulge'' peak is symmetric about $j=0$ and taking the ``disk'' material as the residual material; see \citealt{hopkins:disk.survival}) we can identify the disk material as having an approximate rotational velocity of $\sim40\,{\rm km\,s^{-1}}$ and mass of $\sim5\times10^{8}\,\msun$, in good agreement with the observed baryonic Tully-Fisher relation \citep{belldejong:tf,mcgaugh:tf}. This is not to say our simulations uniquely predict this relation -- it follows from the initial disks lying on the observed relation by construction -- but it does say that the disk is a ``real'' disk in as much as its progenitors were. Fig.~\ref{fig:disk.tile} shows the optical image of the Sbc {\bf f} merger relic, which also re-forms a prominent disk -- based on the $B/T$ ratio and presence of spiral arms, the face-on image is a typical Sa/b galaxy (edge-on S0/a).

The key question is the quantitative efficiency of survival. Since the behavior of the dissipationless stars (i.e.\ how much they do or do not avoid violent relaxation) is identical in all models, we focus on the amount of gas which is able to survive the merger to re-form a disk. 
\citet{hopkins:disk.survival} parameterize this by examining the efficiency of gas consumption in the merger-induced starbursts. Gas which is not consumed (and not completely expelled by winds) will, inevitably, re-form a disk with whatever net angular momentum it carries (since it can cool). 
If all the gas were efficiently torqued (or locally forced to collapse in the merger), it would all contribute to the starburst, $f_{\rm burst}=f_{\rm gas}$ -- there would be no ``residual'' disk gas survival. \citet{hopkins:disk.survival} argue, however, that at high gas fractions these torques become inefficient so the burst fraction scales sub-linearly with gas fraction as $\sim f_{\rm gas}(1-f_{\rm gas})$; this is the origin of the enhanced survival in gas-rich mergers. We compare the explicit feedback models to the EOS models used to calibrate this relation, and find they are statistically nearly identical. The starburst efficiency is clearly well below $f_{\rm burst}=f_{\rm gas}$; i.e.\ disk survival is efficient. If anything there is slightly {\em more} disk survival in these runs, compared to the EOS runs, owing to the contribution of winds expelling some of the mass that would have contributed to the starburst bulge.

\vspace{-0.5cm}
\section{Star Formation Histories}
\label{sec:sfh}

\begin{figure}
    \centering
    \scaleup
    \plotonesize{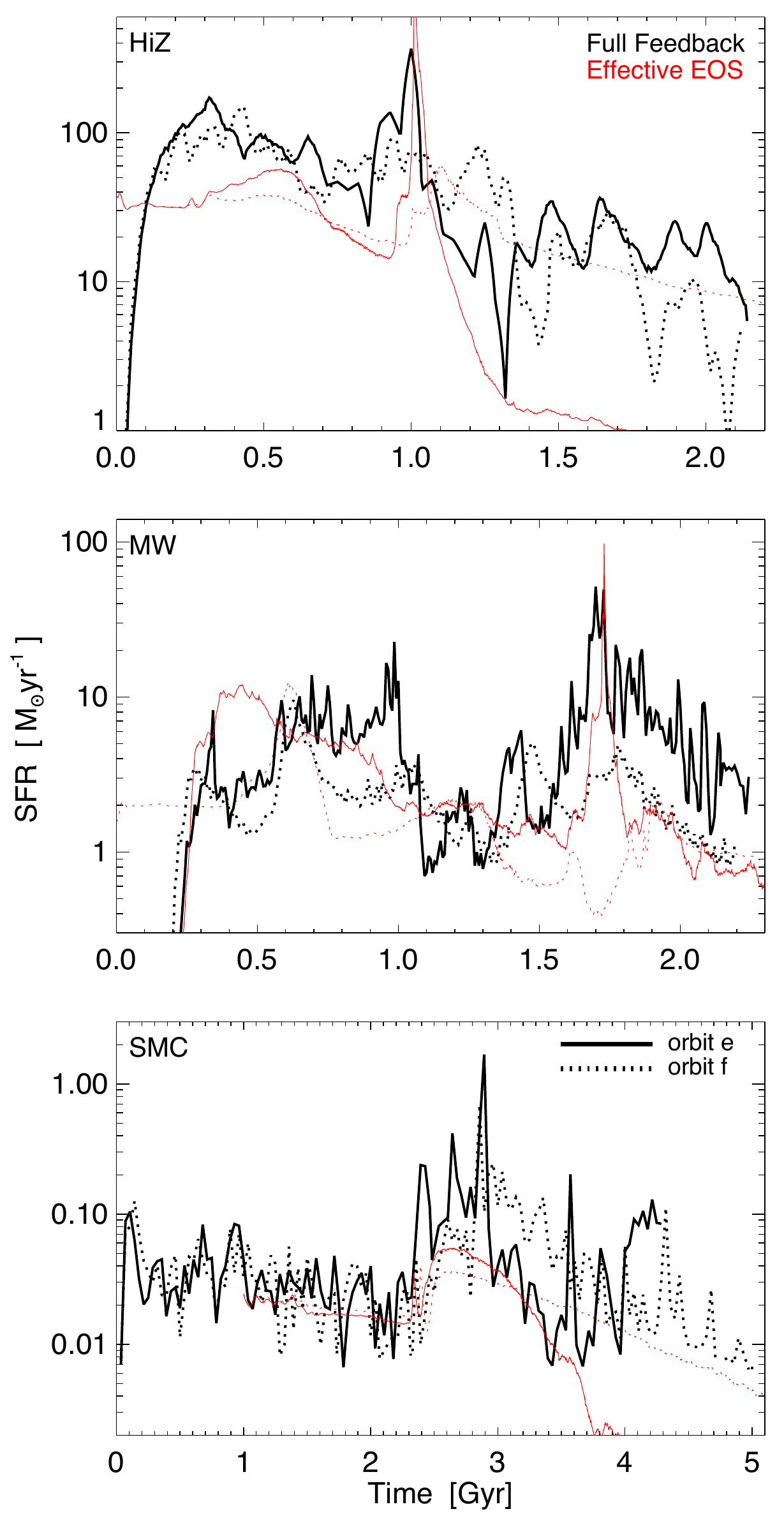}{1.0}
    \caption{Star formation histories in some representative merger simulations, in EOS and explicit feedback models. In all cases, prograde ({\bf e}) models produce more pronounced starbursts; the total mass consumed in the bursts is similar as well. The explicit feedback models predict more variable/bursty SFRs, because the ISM is less homogeneous. In low-mass systems, the bursts are much stronger with explicit feedback models, because the EOS models heavily pressurize the gas relative to the disk $V_{c}$ (making it less compressible). At higher masses, the effect on the peak/duration of the starburst is more sensitive to other details. The ``tail'' of post-starburst SF is significantly enhanced in the explicit feedback models, especially in prograde mergers: in the EOS cases the starburst gas is all exhausted efficiently, but with winds present, some of this is expelled into a fountain that continuously recycles gas for the next couple Gyr. Starburst winds therefore make merger-induced quenching/gas exhaustion (at least {without} AGN) {\em less}, not more, efficient.
    \label{fig:sfh}}
\end{figure}

\vspace{-0.1cm}
\subsection{Starbursts}
\label{sec:sfh:sb}

Figure~\ref{fig:sfh} shows the galaxy-integrated star formation histories of 
each simulation (averaged over $\approx10\,$Myr intervals). The qualitative character of the models is similar, regardless of 
feedback details. In prograde ({\bf e}) orbits, resonant interactions lead to starbursts 
on first passage; 
however since the models here have some pre-existing bulge, these are not always strong \citep{barnes.hernquist.91,barneshernquist96}; they are less prominent in the 
retrograde ({\bf f}) case because the systems are out-of-resonance. In all cases 
there is a starburst on final coalescence, as the gas is channeled to 
the galaxy center.

The integrated mass in stars formed in total and 
in the major starburst(s) are similar; as suggested by Fig.~\ref{fig:massprofile.compare}. 
However, the exact maximum SFR and shape of the SFH can differ significantly. 
It is already known that the duration, amplitude, and variability within the starburst are quite sensitive to 
the initial conditions, and treatment of sub-grid 
star formation recipe applied to the simulations \citep[see][]{cox:feedback}, so this is not surprising.

In all explicit feedback models, there is increasing variability in the SFR, owing to SF being concentrated in resolved GMCs and clusters, and feedback being associated with individual star clusters (and SNe) hence more stochastic in time and space. Interestingly, this suppresses some of the differences between prograde and retrograde encounters in the weaker first-passage bursts; although quantifying this effect in detail requires a larger sample of orbits. 

In the low-mass (SMC and Sbc) cases, the starbursts are much more pronounced with explicit feedback models; this is because the EOS models, even with effective sound speed $\sim10\,{\rm km\,s^{-1}}$, are sufficient to highly pressurize the systems and suppress a strong burst. In this case the allowance for molecular cooling makes the gas more compressible. 

However, especially in the higher-mass cases, expulsion of gas from the nuclei in feedback-driven winds (in the explicit feedback models) tends to ``spread'' the starburst in time. In particular, it gives rise to a long tail of SF as much of the material is not entirely unbound, but kicked into a fountain or stirred up within the disk and then re-condenses. The difference between this and the EOS case is especially clear in the prograde ({\bf e}) models; there, with an EOS model, the gas is efficiently consumed in the starburst, leaving very little continuing SF afterwards. However, in the explicit feedback models, the SFR declines from a similar peak much more slowly, remaining at more than an order-of-magnitude higher SFR for $\sim$Gyrs.

This has important implications for ``quenching'' of star formation in massive galaxies. If quenching were possible without the presence of some additional feedback source -- say, from an AGN -- then the simulations here are the most optimal case for this. They are isolated galaxies, so there is zero new accretion; moreover, an equal-mass merger represents the most efficient means to exhaust a large amount of gas quickly via star formation, much moreso than an isolated disk \citep[see e.g.][]{hopkins:groups.qso,hopkins:groups.ell}. But we find that with the presence of stellar winds, many of our models -- including the already gas-poor MW-like system, maintain post-merger SFRs nearly as large as their steady-state pre-merger SFR. The systems simulated here would take several Gyr to cross the ``green valley'' and turn red, much longer than the $<$Gyr quenching timescale required by observations \citep[see][]{martin:mass.flux,snyder:2011.ka.gal.sims}. Far from resolving this by gas expulsion, stellar feedback makes the ``quenching problem'' {\em harder}. 

As shown in \citet{moster:2011.gas.halo.merger.fx}, addition of gas halos around the merging galaxies (even without continuous accretion) only further enhances the post-merger SFR. Also, as shown in Fig.~\ref{fig:sfh}, the magnitude of the differences between models with and without the explicit stellar feedback models does depend significantly on the merger orbital parameters, so it may be possible that some orbits are ``more efficient'' at exhausting gas than the limited pair we explore here. However, at least in these simulations, the smaller differences are seen in the retrograde orbits, i.e.\ the case where both models preserve significant gas throughout the merger, so it does not appear likely to resolve this issue. 

Of course, the simulations here are not fully cosmological, so there may be other scenarios, such as a series of rapid mergers rather than a single major merger, which can exhaust gas and so terminate star formation more efficiently. But if gas is to be swept out of galaxies efficiently after a merger, our models imply that some other form of feedback -- perhaps from bright quasars -- may be necessary. This is also suggested by observations of late-stage mergers, which find that in the AGN-dominated systems at quasar luminosities, outflow masses are enhanced and the outflow velocities reach $\sim1000\,{\rm km\,s^{-1}}$, larger than those we find driven by stellar feedback \citep{feruglio:2010.mrk231.agn.fb,tremonti:postsb.outflows,sturm:2011.ulirg.herschel.outflows,rupke:2011.outflow.mrk231}.

\begin{figure*}
    \centering
    \scaleup
    \plotsidesize{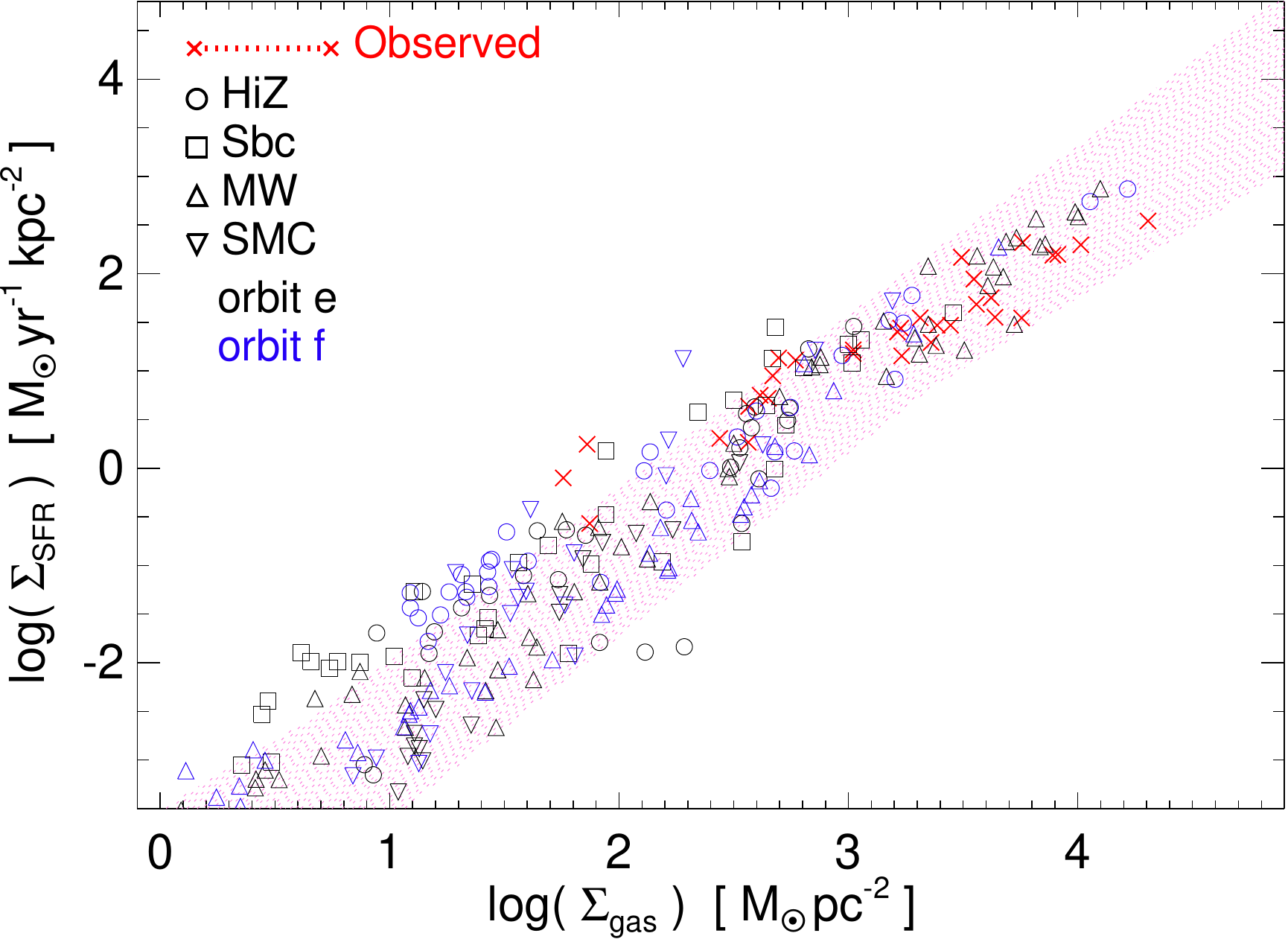}{0.8}
    \caption{Kennicutt-Schmidt relation for the merger simulations with explicit feedback models. We compare the observed starburst galaxies in \citet{kennicutt98} and updated with high-redshift galaxies by \citet{genzel:2010.ks.law} (red $\times$'s); shaded region shows the $90\%$ range at each $\Sigma_{\rm gas}$ from the compilations in those works and \citet{bigiel:2008.mol.kennicutt.on.sub.kpc.scales,daddi:2010.ks.law.highz}. Simulations are plotted at uniformly sampled times and with values measured at the half-SFR radius (averaged over $100$ projections); before coalescence each merging galaxy is treated separately (populating the lower range). EOS models are constructed to lie on the relation by design, but the explicit-feedback merger models have a {\em local} star formation efficiency of unity in self-gravitating dense gas. Without feedback, the mergers lie a factor of $\sim20-100$ above the observed relation (\paperone). With feedback -- with no adjusted parameters -- the KS-law is well-matched. As shown in \paperone\ and \papertwo\ this emerges generically from feedback self-regulating the SFR (independent of the small-scale SF law); with the mechanisms included here, this extends even to extreme starbursts. There is no bimodality in the relation, though it may shift towards the upper envelope at high SFR.
    \label{fig:ks.law}}
\end{figure*}

\vspace{-0.5cm}
\subsection{The Kennicutt-Schmidt Law}
\label{sec:sfh:kslaw}

Figure~\ref{fig:ks.law} plots the positions of the simulations with explicit feedback on the 
Schmidt-Kennicutt relation during the merger (the EOS models are placed on the relation by construction, and so are not interesting here). The agreement between these simulations and the observed relation for both isolated galaxies (the low-$\Sigma_{\rm gas}$ regime, corresponding to when the galaxies are well-separated disks) and observed merging ULIRG/starburst systems is excellent.

We remind the reader that there is no ``tunable'' parameter in these models -- the feedback strength is not adjusted to match the relation, and the instantaneous star formation efficiency we assign in self-gravitating gas is unity, not the $\sim1-2\%$ which is needed to match the Kennicutt relation. As we showed in \paperone\ and \papertwo\ for isolated disks, the 
self-regulation of the SFR in these systems is a consequence of feedback; \paperone\ and \citet{hopkins:binding.sf.prescription} show that their location on the Kennicutt relation has almost nothing to do with the actual star formation prescription, once feedback is included (but {\em is} sensitive to the strength of feedback, if we were to artificially adjust it). Also note that there is no evidence for bimodality in the relation, as suggested by \citet{genzel:2010.ks.law,daddi:2010.ks.law.highz}; we discuss this in \S~\ref{sec:discussion}. In \paperone, we show that the observed ``cutoff'' appears at sufficiently low $\Sigma$, but here it is less obvious since we always average over the half-SFR radius.

\begin{figure}
    \centering
    \scaleup
    \plotonesize{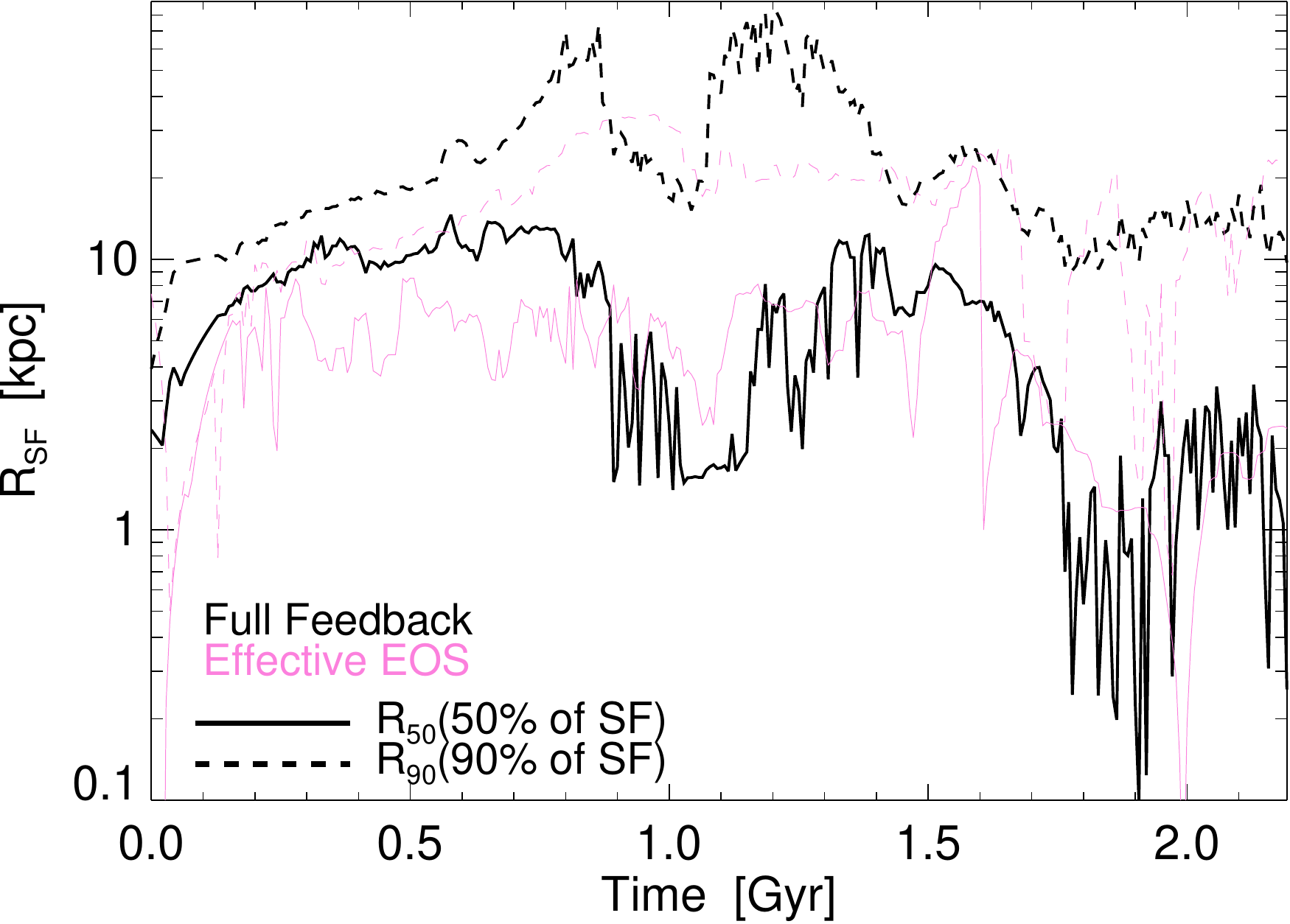}{1}
    \caption{Radial distribution of SF as a function of merger phase. For clarity we focus on one example (MW {\bf e}), but the others are qualitatively similar. We plot the radii enclosing $50\%$ ($R_{50}$) and $90\%$ ($R_{90}$) of the total SFR.$^{\ref{foot:r.half.sfr}}$. 
    In both EOS and full feedback models, the half-SFR radius initially reflects the isolated disks, but drops to $\sim$kpc due to inflows induced by global torques on first passage, then to sub-kpc scales at the peak of the final coalescence starburst. The ``compression'' at the center is slightly larger in the explicit feedback models because of cooling to low temperatures allowing higher densities. 
    The outer extent of SF ($R_{90}$) is generally similar, except leading into first passage and between passages; here, more extended SF in tidal shocks (in arms and the inter-galaxy bridge region) is captured in the full feedback models, contributing $\sim 20-40\%$ of the total SFR.
    \label{fig:r.sf}}
\end{figure}

\begin{figure}
    \centering
    \scaleup
    \plotonesize{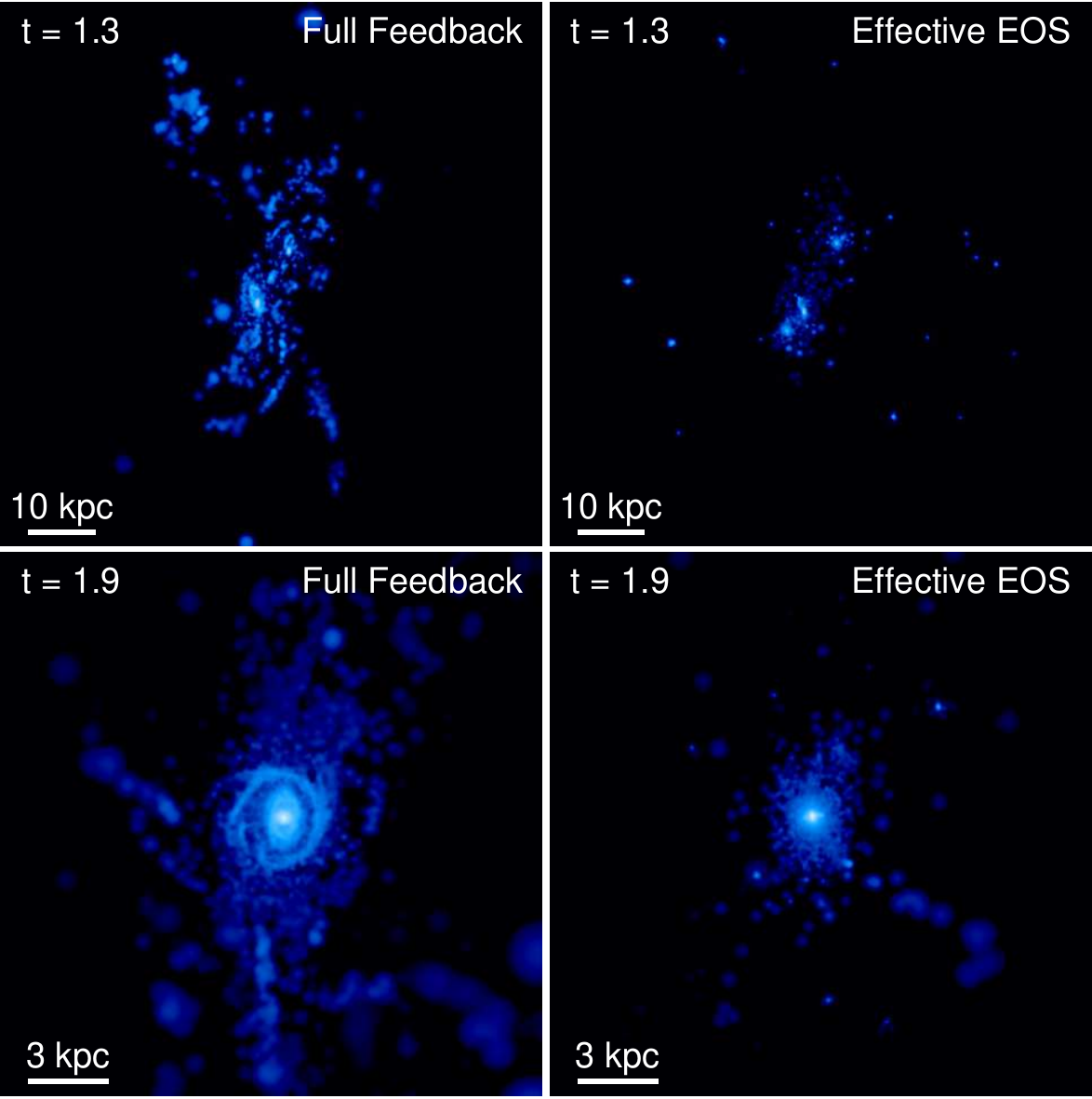}{1}
    \caption{Morphology of the extended SF plotted in Fig.~\ref{fig:r.sf}, for the full feedback ({\em left}) and EOS ({\em right}) models. We plot only the very young ($<10$\,Myr old) stars (projected surface density increasing with brightness with $\approx2\,$dex stretch). We focus on the same example as Fig.~\ref{fig:r.sf} (MW {\bf e}), at two times: just after first passage ({\em top}) and near final coalescence ({\em bottom}). The extended SF is much more prominent in the full feedback models, especially after first passage in the tail/bridge regions. At final coalescence the kpc-scale starburst, containing most of the SF, has similar density and morphology in both runs, but is surrounded by more extended emission and a more prominent surviving star-forming disk in the full feedback model.
    \label{fig:sf.extended.map}}
\end{figure}

\vspace{-0.5cm}
\subsection{Where Do Stars Form?}
\label{sec:sfh:locations}

In Figure~\ref{fig:r.sf}, we plot the radii that enclose 
$50\%$ ($R_{50}$) and $90\%$ ($R_{90}$) 
of the star formation in the merging galaxies, as a function of time.\footnote{\label{foot:r.half.sfr}When the galaxies are well-separated, we should consider the SFR within each separately; because our code is quasi-Lagrangian, we simply separate the particles belonging to each progenitor disk at all times, and take the appropriate radii separately (then plot the average of the two).}
At early stages the radii are large -- basically the values for the isolated disks in seen in \papertwo. At first passage, strong torques lead to shocks, dissipation, and rapid 
inflow; $R_{50}$ drops rapidly as the gas piles up in a $\sim$kpc-scale starburst. After this starburst, $R_{50}$ increases again, until final coalescence when new inflows are driven to small scales and most of the star formation comes from a sub-kpc nuclear starburst. This is the standard scenario of merger-driven starbursts in simplified EOS models as well 
\citep{hernquist.89,barnes.hernquist.91,mihos:starbursts.94,mihos:starbursts.96}.

The evolution of $R_{90}$ is significantly different. At early times, it also reflects the isolated disk sizes. However, going into first passage (especially just before the first-passage starburst), and between this starburst and final coalescence, the radii actually become much {\em larger}, $\sim10-100$\,kpc. This is a consequence of star formation in tidal arms and the bridge region where the gas shocks between the two disks (especially prominent in prograde orbits). This is shown in Fig.~\ref{fig:sf.extended.map}, where we plot the morphology of the young stars for the same simulations. In these stages, $\sim20-50\%$ of the total SFR is contributed by this tidal/shock region. In contrast, it is well-known that in models with a strong effective EOS, large density contrasts in shocks are poorly captured and so the SFR in these regions tends to be small, $\sim10\%$ of the total \citep[see][]{barnes:2004.shock.sf.mice,karl:2010.antennae.model}. Even though strong feedback is present, the gas in these simulations is (at least instantaneously) more compressible, so these shocks give rise to large density jumps that promote rapid star formation, significantly enhancing the contribution of the tidal region to the total SFR \citep[seen in e.g.\ the no-feedback results with varied cooling prescriptions in][]{saitoh:2009.firstpassage.shocks.wcooling}. This may be further enhanced at even higher resolution, with the ability to better resolve strong density gradients in shocks on sub-pc scales; but at that scale, the effect of magnetic fields on shock compression is also important.

\vspace{-0.5cm}
\section{Models with Weak or No Feedback}
\label{sec:nofb}

\begin{figure}
    \centering
    \scaleup
    \plotonesize{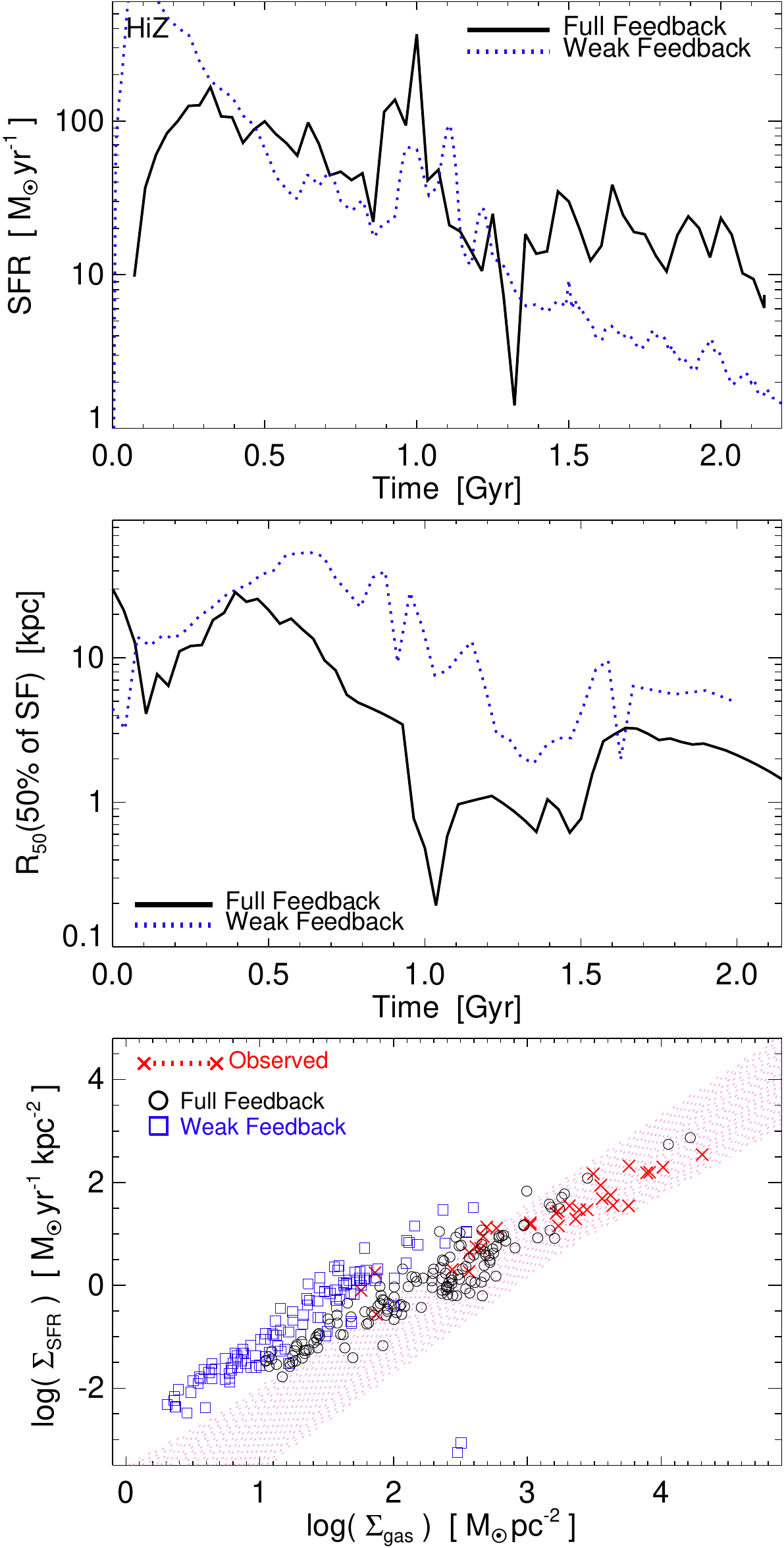}{1}
    \caption{Comparison of a gas-rich merger with our full feedback model to one with artificially weak stellar feedback. We compare our standard HiZ {\bf e} model to a simulation with identical initial conditions but an effective EOS with $\qeos=0$ (cooling floor at $10^{4}\,$K). 
    {\em Top:} SF history as Fig.~\ref{fig:sfh}. Without strong feedback, most of the gas is consumed by a massive burst within the disks when they begin to interact. The burst ``enhancement'' in final coalescence is much weaker. The post-merger SFR declines much more rapidly as the gas is exhausted, with no feedback-driven fountain to maintain gas infall. The surviving disk fraction is lower by a factor $\sim3$ in the weak-feedback model. 
    {\em Middle:} Half-SFR radius as Fig.~\ref{fig:r.sf}. Even on coalescence, most of the SFR in the weak-feedback case comes from large radii $\gtrsim5-10\,$kpc, as inflowing gas undergoes catastrophic fragmentation and turns into stars before it reaches the nucleus. With full feedback present, the dip in $R_{50}$ at coalescence reflects inflows driving a starburst (as Fig.~\ref{fig:r.sf}).
    {\em Bottom:} Kennicutt-Schmidt law as Fig.~\ref{fig:ks.law}. With weak feedback, the highest observed surface densities (seen in merging nuclei) are never reached (gas fragments and turns to stars before reaching the central regions). The separated galaxies, constructed to lie on the KS law when isolated, also jump to much higher SFRs, predicting SFR densities an order of magnitude larger than observed and an apparently bimodal relation. 
    \label{fig:nofb}}
\end{figure}

In \paperone\ and \papertwo, we discuss the properties of isolated simulations with no feedback. The models undergo catastrophic fragmentation with runaway gas collapse to arbitrarily high densities, turning most of the gas into stars in a single dynamical time. We have not focused on these models here, because they are clearly ruled out by e.g.\ the observed \citet{kennicutt98} relation. However, we briefly consider the differences between the predictions of the models here and a model with ``weak'' feedback implemented. Specifically, we consider an effective EOS model with $\qeos=0$, i.e.\ a cooling floor at $10^{4}\,$K or minimum turbulent velocity dispersion of $10\,{\rm km\,s^{-1}}$. This is commonly adopted in historical models. This is not a no-feedback model; so it is not necessarily the case that the gas will experience runaway fragmentation. In the isolated MW or SMC models, this is comparable to the observed velocity dispersion and that predicted from our full feedback model (see \papertwo). However, in more extreme systems -- the mergers here, and even the isolated HiZ models -- this is smaller than the predicted dispersions and the turbulent support needed to resist runaway gas fragmentation. 

Fig.~\ref{fig:nofb} plots the SFR versus time (as Fig.~\ref{fig:sfh}), the half-SFR radius versus time (as Fig.~\ref{fig:r.sf}), and the Kennicutt-Schmidt relation (as Fig.~\ref{fig:ks.law}) for this model and our standard full feedback model. To highlight the differences, we focus on the most extreme case: the HiZ model and {\bf e} orbit. This reaches the highest gas densities, SFRs, and gas densities of our simulations, so is most sensitive to the strength of feedback. With the weak feedback model, runaway collapse cannot be prevented. Early in the simulation (even before the galaxies are fully interacting, and especially on first passage), the gas undergoes catastrophic fragmentation and local clumps collapse to the highest resolvable densities ($\sim10^{6}\,{\rm cm^{-3}}$), turning into stars rapidly. Because of this, the SFR is much higher on early passages than final coalescence. But even in that coalescence, gas collapses rapidly as soon as it collects in filaments, before it can reach the nucleus. The effective radius of SF in this stage is $\sim5-10\,$kpc, rather than the $\lesssim1\,$kpc traditional ``nuclear starburst'' we see in the simulations with full feedback; stellar clumps form during the inflow and then sink collisionlessly into the bulge.\footnote{Note that even during the final starburst, the total extent of the star-forming region in the full feedback models is quite large, $\gtrsim10\,$pc; quantitative comparisons between similar simulations and observations of high-redshift extreme starbursts (sub-millimeter galaxies) shows that their spatial distributions agree well \citep[see e.g.][]{younger:smg.sizes,narayanan:molecular.gas.in.smgs,narayanan:smg.modeling,ivison:smg.is.complex.merger.w.agn}.} As a result the SFR is more flat in the final coalescence (the burst enhancement is suppressed). This also means that the high nuclear gas densities ($\Sigma_{\rm gas}\gg 10^{3}\,\msun\,{\rm pc^{-2}}$) observed in ULIRGs are not actually reached. The predicted SFR densities on the \citet{kennicutt98} relation are uniformly an order of magnitude larger than observed (recall, the observations include high-redshift systems analogous to the models here). Since the initial effective EOS model is tuned to lie on this relation, this produces an apparent sharp bimodality in the relation. Finally, since there is no super-wind, the SFR declines rapidly post-merger (having exhausted all the gas and with little material at large radii to ``rain back in''). At the end of the simulation, we measure the efficiency of disk survival as in \S~\ref{sec:morph:disksurv}; the surviving/re-formed disk mass in the weak-feedback simulation is only $\sim1/3$ of the mass of the re-formed disk in the full feedback simulation.

\vspace{-0.5cm}
\section{Discussion}
\label{sec:discussion}

We have studied major galaxy mergers with explicit models for stellar feedback that can follow the formation and destruction of individual GMCs and star clusters.  Our models include star formation only at extremely high densities inside GMCs, and the mass, momentum, and energy flux from SNe (Types I \&\ II), stellar winds (both ``fast'' O-star winds and ``slow'' AGB winds), and radiation pressure (from UV/optical and multiply-scattered IR photons), as well as HII photoionization heating and molecular cooling. As a first study, we focus on simple, global properties, and compare them to those obtained from previous generations of simulations which did not follow these processes explicitly but instead adopted a simplified ``effective equation of state'' sub-grid model of the ISM. 

We find that most global consequences of mergers are similar, in models with explicit feedback and resolved ISM phase structure, and simplified EOS models. Some details, however, can be rather different.

\vspace{-0.5cm}
\subsection{Galaxy Morphologies}

The relic structure and mass profile, and the presence of tidal tails in the merger, are predominantly determined by global gravitational torques, and so depend little on feedback details. This includes the integral properties (mass and effective radius) of the stellar relic of the kpc-scale starburst. 

\citet{mihos:cusps} argued that merger remnants are really two-component systems, and \citet{hopkins:cusps.mergers,hopkins:cusps.fp,hopkins:cusps.evol,hopkins:cusps.ell} later demonstrated this in comparison of observations (of ellipticals and recent merger remnants) and simulations with sub-grid feedback models. The outer profile is dominated by violently relaxed stars formed before the final galaxy merger, which are dissipationlessly scattered into the ``envelope'' at large radii. The inner portion is dominated by ``starburst'' stars which form via gas that is gravitationally torqued and dissipates, falls to the center, and turns into stars in a central starburst; this dominates the central $\sim$kpc and extends inwards to form the central ``cusps'' in gas-rich merger remnants \citep[see][]{kormendysanders92,hibbard.yun:excess.light,rj:profiles,hopkins:cusp.slopes}; it is often responsible for nuclear kinematically decoupled components \citep{hernquist:kinematic.subsystems,hoffman:mgr.orbit.structure.vs.fgas}. 

We find that this decomposition, and the behavior of the mass profile of each sub-component separately (as well as the total) is robust to the inclusion of explicit or effective EOS models. This is reassuring, since \citet{hopkins:cusps.mergers,hopkins:cusps.ell} showed that the simulated mass distributions agree very well with observed mergers and spheroids, and \citet{cox:feedback} showed these results are robust to substantial changes in sub-grid parameterizations of feedback and star formation. The dissipationless profile should obviously be insensitive to the gas physics. The dissipative profiles are similar because, as shown in \citet{hopkins:cusps.mergers}, their characteristic radii are set not by feedback ``pressurizing'' the material, but by the radius at which the inflowing gas becomes self-gravitating, at which point it is no longer strongly torqued by the external stars and dark matter and can turn into stars relatively rapidly. The slope of the central mass profile is set by equilibrium between gravitationally driven inflow and star formation \citep{hopkins:cusp.slopes}. 

We may however expect differences in very low-mass, gas-rich mergers. In the EOS models, these are ``pressurized'' to the point where the effective sound speed is comparable to the circular velocity, and so tend to result in merger remnants that are larger compared to true ellipticals at $\lesssim10^{9}\,\msun$ \citep{hopkins:cusps.ell}. Allowing the gas to be more compressible (instantaneously), while retaining feedback, may resolve this discrepancy.

\vspace{-0.5cm}
\subsection{Disk Survival}

Disk survival -- the efficiency with which gas avoids consumption in the merger and can rapidly re-form a disk after -- appears to be just as, if not slightly more, efficient in models with explicit feedback (appropriately accounting for their properties at the time of merger). They follow the same scalings for the likelihood of disk survival (and gas angular momentum loss) as derived from simplified models in \citet{hopkins:disk.survival}. 

This is expected: \citet{hopkins:disk.survival} found disk survival did not depend on the sub-grid parameters of the EOS model (unless feedback is very weak), and discuss the reasons for this. Gas is dissipative, so gas that survives the merger without turning into stars or losing most of its angular momentum will eventually re-form a rotating disk. The dominant torques driving angular momentum loss in mergers arise from resonant interactions exciting collisionless perturbations in the stars, which in turn force the gas into strong shocks \citep{barnes.hernquist.91,barneshernquist92,barnes02:gasdisks.after.mergers,hopkins:zoom.sims,hopkins:inflow.analytics}. Angular momentum ``cancellation,'' viscosity, other hydrodynamic torques, and the direct torque from the presence of the secondary galaxy are typically much smaller effects. As a result, one can analytically derive a critical radius and resonant conditions that depend only on merger orbit, mass ratio, and galaxy structural properties (not the gas microphysics), outside of which gas will not be strongly torqued and will thus survive to re-form disks. With few stars, gas rich disks have weak collisionless torques, so give larger surviving disks. 

The same has also been seen in cosmological simulations with sub-grid feedback models \citep{governato:disk.rebuilding,governato:2010.dwarf.gal.form,guedes:2011.cosmo.disk.sim.merger.survival}. 
Disks survive multiple major mergers by virtue of conserving gas which immediately re-forms disks; this is enhanced by the presence of feedback-driven outflows which further suppress the mass turned into stars in the center. This is important for the existence of ``realistic'' disks today \citep[][]{robertson.bullock:disk.merger.rem.vs.obs,hopkins:disk.survival.cosmo,stewart:disk.survival.vs.mergerrates}, and there appear to be a growing number of observed disks which have undergone recent mergers \citep{hammer:obs.disk.rebuilding,hammer:hubble.sequence.vs.mergers,yang:post.merger.disk.obs,puech:z06.disk.postmerger,bundy:merger.fraction.new,bundy:2010.passive.disks}.

\vspace{-0.5cm}
\subsection{Starbursts and Star Formation}

The final starbursts are dominated by {in situ} star formation from gas which is torqued, falls into the nucleus and turns into stars at high densities. This is the ``standard'' scenario of merger-driven nuclear starbursts and star formation predicted in early simulations \citep{barnes.hernquist.91,mihos:starbursts.94,mihos:starbursts.96}, and observed at the centers of nearby ULIRGs \citep[strong, rapid inflows evident in their profiles of molecular gas, star formation, and metallicity; see e.g.][]{tacconi:ulirgs.sb.profiles,sanders96:ulirgs.mergers,kormendysanders92,hibbard.yun:excess.light,rj:profiles,titus:ssp.decomp,reichardt:ssp.decomp,michard:ssp.decomp,foster:metallicity.gradients,sanchezblazquez:ssp.gradients,kewley:2010.gal.pair.metal.grad.evol,rupke:2010.metallicity.grad.merger.vs.obs,soto:ssp.grad.in.ulirgs}. \citet{cox:feedback} showed that this is robust to large variations in the implementation of feedback in sub-grid models. As emphasized in many previous studies, we are not saying that most of the stars form in the starburst: this is typically $\sim10\%$ of the total stellar mass \citep{jogee:merger.density.08,robaina:2009.sf.fraction.from.mergers,hopkins:ir.lfs,hopkins:sb.ir.lfs}, with the rest formed in the ``isolated disk'' mode (including times during the merger outside the ``burst'').

Although gas clumps (into GMCs) as it flows in, feedback efficiently disperses these GMCs after they turn just a few percent of their mass into stars. But the material is not expelled completely (which would shut down the burst entirely); most of the recycled gas is unbound {\em locally} from the GMC, but not globally from the galaxy. The typical velocities are the escape velocities from the GMC or local star cluster ($\sim10\,{\rm km\,s^{-1}}$) and so the gas is ``stirred.'' This allows it to reach the nuclei in a coherent flow, so that gravitational torques will still tend to dominate. 

The more explicit feedback models, however, show significantly higher star formation 
in tails and bridges (especially in shocks). This may resolve a long-standing discrepancy between merger models and observed star formation distributions in local mergers \citep{barnes:2004.shock.sf.mice}; a more quantitative comparison with models intended to mimic these specific systems will be the subject of future work.

In the explicit feedback models, the star formation is more strongly time variable (less 
``smoothed'' by the effective EOS). This owes to the more inhomogeneous nature of the ISM -- in extreme cases, a large fraction of the SFR can come from a few super star clusters. The burst ``enhancement'' (peak SFR) can be higher or lower, depending on the orbit and structural properties; as with sub-grid models, first passage bursts are much more sensitive to these details than final coalescence. In low-mass systems, it tends to be higher, because (as noted above) the EOS models may over-pressurize small disks by enforcing a minimum sound speed of $\sim10\,{\rm km\,s^{-1}}$; with molecular cooling the gas can more efficiently collapse to high densities in the nucleus.

In all the explicit feedback cases, the ``tails'' of the starburst tend to be more broad; this owes to the effect of stellar winds ejecting some material at modest velocities, so it falls back into the disk in a fountain and makes the starburst more extended (discussed below).

\vspace{-0.5cm}
\subsection{The Kennicutt-Schmidt Relation}

The explicit feedback models predict SFRs that agree extremely well with observed merging galaxies on the Kennicutt-Schmidt relation. This is a major success of the models: unlike the EOS models, where the kpc-scale SF law is imposed to match the Kennicutt law, the predicted Kennicutt relation in the explicit feedback cases is a true prediction (both in normalization, scatter, and power-law slope). Recall, star formation in these models occurs only in very dense, locally self-gravitating gas (scales $\lesssim$pc), with an instantaneous local efficiency of unity -- not the global $\sim1\%$ value of the Kennicutt relation. And without feedback (see \paperone\ \&\ \papertwo), the models predict a SFR far larger than observed, $\dot{M}_{\ast}\sim M_{\rm gas}/t_{\rm dyn}$. With feedback included -- with parameters not in any way adjusted to reproduce the observations but taken directly from stellar evolution models, the observed relation naturally emerges. 

This is a consequence of feedback-regulated star formation: turbulent support in the gas dissipates locally on a crossing time, leading to collapse, which proceeds until a sufficient number of young stars form (regardless of {\em how} they form, in detail) to provide enough momentum flux in feedback to balance the dissipation. As such, we showed in \paperone\ that the SFR and Kennicutt relation are direct consequences of feedback, and are {\em independent} of the microphysical star formation law once explicit feedback is included. We find the same here; we have re-run some of our simulations replacing our local self-gravity criterion for star formation with a simple density threshold with a very different efficiency (or changing the small-scale density dependence), as discussed in \S~\ref{sec:sims}, and the predicted Kennicutt law is nearly identical. 

The same feedback mechanisms regulating star formation in isolated galaxies extend to the regime of extreme merger-induced starbursts. This is no mean feat: the typical surface densities and star formation rates are orders of magnitude larger than the isolated disk counterparts of these galaxies.  The gas surface densities in the bursts here extend to $\sim100-1000$ times larger than the mean surface density of local GMCs! We are therefore well into the regime where the medium is entirely molecular and high-density (although it may still exhibit phase structure). In \papertwo, we showed that SNe and other ``thermal'' feedback mechanisms become unimportant in gas above densities $\sim1-10\,{\rm cm^{-3}}$ (typical of the diffuse gas in GMCs), because the cooling times are extremely short compared to the dynamical times. We therefore expect that radiation pressure is the most important feedback mechanism in these starburst regions, as we found there for the most dense regions within ``normal'' GMCs. It is a remarkable success of the models, though, that they interpolate naturally into even this extreme regime, without requiring any adjustment. In contrast, most analytic models for the Kennicutt-Schmidt relation have been forced to assume some ad hoc break or transition in the scaling at high densities.

We see no strong bimodality in the predicted relation. But this is not necessarily inconsistent with recent observational claims along these lines \citep{genzel:2010.ks.law,daddi:2010.ks.law.highz}; our merger and isolated galaxy simulations overlap closely with the observations from those papers. What we see is a continuum between these systems, which is completely consistent with the observations \citep[see e.g.][]{narayanan:2011.xco}. The apparent bimodality stems from a strict choice of two separate CO conversion factors; in future work, we will incorporate detailed models for molecular emission to make a quantitative comparison with these observations. \citet{narayanan:2011.xco.model} find that when applying a functional form for the CO-H2 conversion factor which depends on the physical conditions in the ISM to the observed galaxies in Figure~\ref{fig:ks.law}, a slope of $\sim1.8$ emerges. Our results are quite consistent with those, and we find a slope of $\sim1.7$; this suggests that stars do form somewhat more ``efficiently'' (in terms of consumption time) in systems like mergers which occupy the high-surface density region of the Kennicutt-Schmidt plot, though mergers do not lie on a different track of the relation than disks. The increased efficiency likely owes to a larger fraction of gas in the dense phase, a result which has been noted in observations of local ULIRGs \citep[e.g.][]{juneau:2009.enhanced.dense.gas.ulirgs}.

\vspace{-0.5cm}
\subsection{Comparison to Models with Weak Feedback}

Recently, there have been some different conclusions reached in the literature from the study of high-resolution simulations with very different implementations of stellar feedback. \citet{teyssier:2010.clumpy.sb.in.mergers} argued that merger-induced starbursts might not occur ``in situ'' from inflows, but instead from catastrophic fragmentation of tidal arms, bars, and other instabilities. The fragments would turn a large fraction of their mass into stars very rapidly, and then sink as supermassive stellar clusters to the galaxy center via dynamical friction, building the bulge. They also suggest a sharp bimodality in the Kennicutt-Schmidt relation, with the mergers and isolated disks being well-separated, as a consequence of this induced fragmentation in the mergers (though they do discuss how this might be smoothed out). \citet{bournaud10} argued that disk survival is very inefficient in their simulations, for similar reasons -- the gas undergoes dramatic fragmentation and ``clumps'' scatter off one another, providing new torques. 

We believe that these differences stem from the fact that these simulations include relatively weak stellar feedback, which is unable to disperse dense gas clouds (GMCs and especially clouds in the dense tidal or starburst regions of mergers) after they form. In fact, we reproduce these results if we re-run our mergers with an effective EOS model with much weaker ``feedback strength,'' $\qeos=0$. 
Like the $\qeos=0$ models, the \citet{teyssier:2010.clumpy.sb.in.mergers} models adopt an effective EOS, but one in which the ``median'' ISM temperature is much colder than the sub-grid sound speeds used in the EOS models here, which makes catastrophic fragmentation possible. 
Without strong feedback, the clumps cannot then 
disperse and recycle their mass. Specifically, the models either do not include explicit feedback mechanisms or include only SNe; however, we found in \paperone\ and \papertwo\ that SNe have little effect on the dispersal of dense gas, because the cooling time in that gas is much shorter than the dynamical time. As a result, the SNe can stir up the ``diffuse'' medium but cannot recycle dense gas. In our simulations, the most dense clouds are destroyed by a combination of radiation pressure in both UV and IR, photo-ionization, and momentum from O-star winds. 

In models that allow cooling but do not include these explicit feedback mechanisms, therefore, gas cools and collapses into dense GMC-like objects, which steadily contract and form stars, but are not efficiently ``re-mixed'' into the ISM on a short timescale. Collapse (hence star formation) in the dense regions runs away, giving rise to the dramatically enhanced ``fragmentation mode'' of star formation and apparent bimodality in the Kennicutt relation.  In \papertwo\ we show that even if slow star formation is forced in the dense regions, the lack of mixing allows the dense regions to move ballistically, with relatively small interaction cross-sections and long lifetimes. In this limit, gas is effectively no longer collisional. All the dissipation occurs within individual clumps, while the relative motions between clumps cannot be dissipated, just like relative motions of stars; the gas behaves like a collection of ``super-massive'' star or dark matter particles, and loses orbital angular momentum only via processes much slower than shocks, such as scattering and dynamical friction.

In contrast, observations suggest that GMCs are short-lived, with lifetimes of a few free-fall times (few Myr), and turn just a few percent of their mass into stars before dispersing \citep{zuckerman:1974.gmc.constraints,williams:1997.gmc.prop,evans:1999.sf.gmc.review,evans:2009.sf.efficiencies.lifetimes}. Although the relevant physics in galaxy-scale simulations remains quite uncertain (and it is unclear how robustly this can be generalized to all the cases studied in this paper), short lifetimes do appear to be typical even for the most massive ($10^{8}-10^{9}\,\msun$) GMC complexes observed in local mergers \citep{rand:1990.m51.superclumps,planesas:1991.1068.superclumps,wilson:2003.supergiant.clumps,wilson:2006.arp220.superclumps}. So even if most of the gas at a given instant is in dense sub-units (GMCs), the gas can be recycled through the diffuse ISM on a short timescale, and therefore experience normal hydrodynamic forces. So long as the GMC lifetime is short compared to the total duration of the strong torques in the merger ($\sim500\,$Myr), this will be true. 

Likewise, since inflows in the final coalescence of a major merger require only a couple of dynamical times to reach the center, for fragmentation to ``beat'' inflow as a starburst driver the fragmenting clumps in the inflow would have to have a very high instantaneous star formation efficiency, $\dot{M}_{\ast}\sim M_{\rm gas}/t_{\rm dyn}$. This happens in models with weaker feedback; however, the observed Kennicutt relation in ULIRGs and other mergers (even high-redshift systems) implies an efficiency substantially smaller (by as much as a factor of $\sim50$). And direct observations of ULIRGs do suggest that the final coalescence star formation rate is indeed dominated by a nuclear starburst (see references above).

A key conclusion is that, in any model which resolves the formation of dense clouds, predictions critically depend on the physics that may (or may not) {\em destroy} those clouds.

\vspace{-0.5cm}
\subsection{Caveats \&\ Future Directions}

The results here are a preliminary study of how more detailed and explicit stellar feedback and ISM structure influence the results of merger simulations. We have found that the most basic global integral properties of star formation and galaxy morphology tend to be similar to those inferred in previous studies with a simplified treatment of the gas physics. However, we also find that more detailed structural and kinematic properties are sensitive to the gas physics. We do not expect this to be the case outside of the central $\sim$kpc, where the relic galaxy is dominated by stars formed before the merger and violently relaxed. But kinematic substructures in the central region may be sensitive to how the gas collapses and turns into stars. It will also be particularly interesting to examine the effects of more detailed star formation and enrichment models that can separately follow stellar winds and SNe Types I \&\ II on the age, metallicity, and abundance ($\alpha/{\rm Fe}$) gradients in the galaxies; these can place strong constraints on the merger history and role of dissipation in galaxy formation \citep[see e.g.][and references therein]{torrey:2011.metallicity.evol.merger}. 

In a companion paper, we examine the star clusters formed in these simulations. The mass/luminosity distribution, spatial locations, formation time distribution, and physical properties of these clusters represent a powerful constraint on small-scale models of the ISM and star formation. 

We have also restricted our focus to major mergers. Studies of mergers with varying mass ratios suggest that the qualitative behaviors discussed here should not depend on mass ratio for ratios to about 3:1 or 4:1, and even at lower mass ratios they can be considered similar but with an ``efficiency'' of inducing starbursts and violent relaxation that scales approximately linearly with mass ratio \citep{hernquist.mihos:minor.mergers,cox:massratio.starbursts,naab:minor.mergers,younger:minor.mergers}. But at small mass ratios the dominant role of mergers may be more subtle: inducing resonant disk perturbations \citep{donghia:resonant.stripping.dwarfs,donghia:2010.tidal.resonances} and disk heating \citep{purcell:minor.merger.thindisk.destruction,stewart:mw.minor.accretion,walker:disk.fragility.minor.merger,hopkins:disk.heating,moster:2010.thin.disk.heating.vs.gas}, for which the gas response may depend significantly on its phase structure. 

We note that recent studies comparing cosmological simulations done
with {\small GADGET} and the new moving mesh code {\small AREPO}
\citep{springel:arepo} have called into question the reliability of
smoothed particle hydrodynamics (SPH) for some problems related to
galaxy formation in a cosmological context \citep{vogelsberger:2011.arepo.vs.gadget.cosmo,
sijacki:2011.gadget.arepo.hydro.tests,keres:2011.arepo.gadget.disk.angmom,bauer:2011.sph.vs.arepo.shocks,
torrey:2011.arepo.disks}.  However, we have also
performed idealized simulations of mergers between individual
galaxies and found excellent agreement between {\small GADGET}
and {\small AREPO} for e.g. gas-inflow rates, star formation
histories, and the mass in the ensuing starbursts, for
reasons that will be discussed in
\citep{hayward:arepo.gadget.mergers}.  
The discrepancies above are also minimized when the flows of interest are supersonic (as opposed to sub-sonic), which is very much the case here \citep{kitsionas:2009.grid.sph.compare.turbulence,price:2010.grid.sph.compare.turbulence,bauer:2011.sph.vs.arepo.shocks}. We have also compared a subset of our merger simulations run with an alternative formulation of SPH from \citet{hopkins:lagrangian.pressure.sph}, which produces results much more similar to grid codes; in these tests we confirm all of the qualitative conclusions here.

Our new models allow us to follow the structure of the gas in the central regions of starburst systems at high resolution. This makes them an ideal laboratory for studying feedback physics under extreme conditions in, say, the center of Arp 220 and other very dense galaxies. In another companion paper, we examine the properties of the large starburst winds driven by stellar feedback in the mergers, which can reach $\sim10-500\,\msun\,{\rm yr^{-1}}$ outflow rates and so have potentially major implications for metal enrichment and self-regulation of galaxy growth. We have also here, for clarity, neglected AGN feedback in these models, but we expect it may have a significant effect on the systems and their outflows after the final coalescence. For example, it may strongly suppress the otherwise quite high post-merger SFRs we see in these simulations, which -- without something to rapidly ``quench'' them -- may make it difficult or impossible to explain the observed abundance of high-redshift, low-SFR galaxies. With high-resolution models that include the phase structure of the ISM, it becomes meaningful to include much more explicit physical models for AGN feedback. 

Finally, we stress that these models are still approximations, and the treatment of ISM and star formation physics is necessarily still ``sub-grid'' at some level (just at the GMC-level, instead of the kpc-level). We can follow galaxy-wide phenomena such as mergers, and large-scale processes such as the formation of GMCs and massive star clusters. However, we still require assumptions regarding the behavior of gas at densities above $\sim10^{4}-10^{6}\,{\rm cm^{-3}}$, and make simple approximations to the chemistry of the gas (especially at low temperatures); our feedback models also average over the IMF, owing to the fact that a single stellar particle still represents many individual stars, rather than discriminating individual low and high-mass stars. The treatment of radiative feedback, in particular, is not a full radiative transfer calculation (which, for the infrared, is extremely demanding and may be sensitive to very small-scale unresolved ISM structure). It remains, unfortunately, prohibitively expensive to treat these physics much more explicitly in galaxy-wide simulations. As such, the behavior of any individual molecular cloud or star cluster is at best marginally resolved in our simulations and should only be taken as a first approximation to the behavior that might be obtained if the evolution of that system could be followed in detail. However, it is common to consider both these physics and much smaller spatial and mass resolution of sub-stellar masses and $<0.1\,$pc in simulations of the ISM and star formation within small ``patches'' of the ISM. There is considerable and rapidly growing simulation work in these areas, much of it ``building up'' to larger scales such as GMCs, overlapping with our smallest resolved scales. This suggests that considerable progress might be made by combining these approaches and using  smaller-scale (but more accurate and explicit) star formation and ISM simulations to calibrate and test the approximations in galaxy-scale simulations such as those here, which can themselves, in turn, be used to calibrate the kpc-scale sub-grid approaches still required for fully cosmological simulations.

\vspace{-0.25in}
\acknowledgments 
We thank Eliot Quataert for helpful discussions and contributions motivating this work, and our referee, Matthias Steinmetz, for several useful suggestions. Support for PFH was provided by NASA through Einstein Postdoctoral Fellowship Award Number PF1-120083 issued by the Chandra X-ray Observatory Center, which is operated by the Smithsonian Astrophysical Observatory for and on behalf of the NASA under contract NAS8-03060.
\\

\bibliography{/Users/phopkins/Documents/work/papers/ms}

\begin{thebibliography}{182}
\expandafter\ifx\csname natexlab\endcsname\relax\def\natexlab#1{#1}\fi

\bibitem[{{Barnes}(1988)}]{barnes:disk.halo.mergers}
{Barnes}, J.~E. 1988, \apj, 331, 699

\bibitem[{{Barnes}(2002)}]{barnes02:gasdisks.after.mergers}
---. 2002, \mnras, 333, 481

\bibitem[{{Barnes}(2004)}]{barnes:2004.shock.sf.mice}
---. 2004, \mnras, 350, 798

\bibitem[{{Barnes} \& {Hernquist}(1992)}]{barneshernquist92}
{Barnes}, J.~E., \& {Hernquist}, L. 1992, \araa, 30, 705

\bibitem[{{Barnes} \& {Hernquist}(1996)}]{barneshernquist96}
---. 1996, \apj, 471, 115

\bibitem[{{Barnes} \& {Hernquist}(1991)}]{barnes.hernquist.91}
{Barnes}, J.~E., \& {Hernquist}, L.~E. 1991, \apjl, 370, L65

\bibitem[{{Bauer} \& {Springel}(2012)}]{bauer:2011.sph.vs.arepo.shocks}
{Bauer}, A., \& {Springel}, V. 2012, \mnras, 423, 3102

\bibitem[{{Baugh} {et~al.}(2005){Baugh}, {Lacey}, {Frenk}, {Granato}, {Silva},
  {Bressan}, {Benson}, \& {Cole}}]{baugh:sam}
{Baugh}, C.~M., {Lacey}, C.~G., {Frenk}, C.~S., {Granato}, G.~L., {Silva}, L.,
  {Bressan}, A., {Benson}, A.~J., \& {Cole}, S. 2005, \mnras, 356, 1191

\bibitem[{{Bell} \& {de Jong}(2001)}]{belldejong:tf}
{Bell}, E.~F., \& {de Jong}, R.~S. 2001, \apj, 550, 212

\bibitem[{{Benson}(2005)}]{benson:cosmo.orbits}
{Benson}, A.~J. 2005, \mnras, 358, 551

\bibitem[{{Bigiel} {et~al.}(2008){Bigiel}, {Leroy}, {Walter}, {Brinks}, {de
  Blok}, {Madore}, \& {Thornley}}]{bigiel:2008.mol.kennicutt.on.sub.kpc.scales}
{Bigiel}, F., {Leroy}, A., {Walter}, F., {Brinks}, E., {de Blok}, W.~J.~G.,
  {Madore}, B., \& {Thornley}, M.~D. 2008, \aj, 136, 2846

\bibitem[{{Bournaud} {et~al.}(2010){Bournaud}, {Elmegreen}, {Teyssier},
  {Block}, \& {Puerari}}]{bournaud:2010.grav.turbulence.lmc}
{Bournaud}, F., {Elmegreen}, B.~G., {Teyssier}, R., {Block}, D.~L., \&
  {Puerari}, I. 2010, \mnras, 409, 1088

\bibitem[{{Bournaud} {et~al.}(2011)}]{bournaud10}
{Bournaud}, F., {et~al.} 2011, \apj, 730, 4

\bibitem[{{Bundy} {et~al.}(2009){Bundy}, {Fukugita}, {Ellis}, {Targett},
  {Belli}, \& {Kodama}}]{bundy:merger.fraction.new}
{Bundy}, K., {Fukugita}, M., {Ellis}, R.~S., {Targett}, T.~A., {Belli}, S., \&
  {Kodama}, T. 2009, \apj, 697, 1369

\bibitem[{{Bundy} {et~al.}(2010)}]{bundy:2010.passive.disks}
{Bundy}, K., {et~al.} 2010, \apj, 719, 1969

\bibitem[{{Burkert} {et~al.}(2008){Burkert}, {Naab}, {Johansson}, \&
  {Jesseit}}]{burkert:anisotropy}
{Burkert}, A., {Naab}, T., {Johansson}, P.~H., \& {Jesseit}, R. 2008, \apj,
  685, 897

\bibitem[{{Casey} {et~al.}(2009){Casey}, {Chapman}, {Beswick}, {Biggs},
  {Blain}, {Hainline}, {Ivison}, {Muxlow}, \& {Smail}}]{casey:highz.ulirg.pops}
{Casey}, C.~M., {Chapman}, S.~C., {Beswick}, R.~J., {Biggs}, A.~D., {Blain},
  A.~W., {Hainline}, L.~J., {Ivison}, R.~J., {Muxlow}, T., \& {Smail}, I. 2009,
  \mnras, 399, 121

\bibitem[{Chapman {et~al.}(2009)Chapman, Blain, Ibata, Ivison, Smail, \&
  Morrison}]{chapman:submm.halo.clustering}
Chapman, S.~C., Blain, A., Ibata, R., Ivison, R.~J., Smail, I., \& Morrison, G.
  2009, The Astrophysical Journal, 691, 560

\bibitem[{{Chapman} {et~al.}(2005){Chapman}, {Blain}, {Smail}, \&
  {Ivison}}]{chapman:submm.lfs}
{Chapman}, S.~C., {Blain}, A.~W., {Smail}, I., \& {Ivison}, R.~J. 2005, \apj,
  622, 772

\bibitem[{{Cox} {et~al.}(2006{\natexlab{a}}){Cox}, {Di Matteo}, {Hernquist},
  {Hopkins}, {Robertson}, \& {Springel}}]{cox:xray.gas}
{Cox}, T.~J., {Di Matteo}, T., {Hernquist}, L., {Hopkins}, P.~F., {Robertson},
  B., \& {Springel}, V. 2006{\natexlab{a}}, \apj, 643, 692

\bibitem[{{Cox} {et~al.}(2006{\natexlab{b}}){Cox}, {Dutta}, {Di Matteo},
  {Hernquist}, {Hopkins}, {Robertson}, \& {Springel}}]{cox:kinematics}
{Cox}, T.~J., {Dutta}, S.~N., {Di Matteo}, T., {Hernquist}, L., {Hopkins},
  P.~F., {Robertson}, B., \& {Springel}, V. 2006{\natexlab{b}}, \apj, 650, 791

\bibitem[{{Cox} {et~al.}(2006{\natexlab{c}}){Cox}, {Jonsson}, {Primack}, \&
  {Somerville}}]{cox:feedback}
{Cox}, T.~J., {Jonsson}, P., {Primack}, J.~R., \& {Somerville}, R.~S.
  2006{\natexlab{c}}, \mnras, 373, 1013

\bibitem[{{Cox} {et~al.}(2008){Cox}, {Jonsson}, {Somerville}, {Primack}, \&
  {Dekel}}]{cox:massratio.starbursts}
{Cox}, T.~J., {Jonsson}, P., {Somerville}, R.~S., {Primack}, J.~R., \& {Dekel},
  A. 2008, \mnras, 384, 386

\bibitem[{{Daddi} {et~al.}(2010)}]{daddi:2010.ks.law.highz}
{Daddi}, E., {et~al.} 2010, \apjl, 714, L118

\bibitem[{{Dasyra} {et~al.}(2008){Dasyra}, {Yan}, {Helou}, {Surace}, {Sajina},
  \& {Colbert}}]{dasyra:highz.ulirg.imaging.not.major}
{Dasyra}, K.~M., {Yan}, L., {Helou}, G., {Surace}, J., {Sajina}, A., \&
  {Colbert}, J. 2008, \apj, 680, 232

\bibitem[{{Dasyra} {et~al.}(2006)}]{dasyra:mass.ratio.conditions}
{Dasyra}, K.~M., {et~al.} 2006, \apj, 638, 745

\bibitem[{{Dasyra} {et~al.}(2007)}]{dasyra:pg.qso.dynamics}
---. 2007, \apj, 657, 102

\bibitem[{{Dey} {et~al.}(2008){Dey}, {Soifer}, {Desai}, {Brand}, {Le Floc'h},
  {Brown}, {Jannuzi}, {Armus}, {Bussmann}, {Brodwin}, {Bian}, {Eisenhardt},
  {Higdon}, {Weedman}, \& {Willner}}]{dey:2008.dog.population}
{Dey}, A., {Soifer}, B.~T., {Desai}, V., {Brand}, K., {Le Floc'h}, E., {Brown},
  M.~J.~I., {Jannuzi}, B.~T., {Armus}, L., {Bussmann}, S., {Brodwin}, M.,
  {Bian}, C., {Eisenhardt}, P., {Higdon}, S.~J., {Weedman}, D., \& {Willner},
  S.~P. 2008, \apj, 677, 943

\bibitem[{{di Matteo} {et~al.}(2007){di Matteo}, {Combes}, {Melchior}, \&
  {Semelin}}]{dimatteo:merger.induced.sb.sims}
{di Matteo}, P., {Combes}, F., {Melchior}, A.-L., \& {Semelin}, B. 2007, \aap,
  468, 61

\bibitem[{{Di Matteo} {et~al.}(2005){Di Matteo}, {Springel}, \&
  {Hernquist}}]{dimatteo:msigma}
{Di Matteo}, T., {Springel}, V., \& {Hernquist}, L. 2005, \nat, 433, 604

\bibitem[{{Dobbs} {et~al.}(2011){Dobbs}, {Burkert}, \&
  {Pringle}}]{dobbs:2011.why.gmcs.unbound}
{Dobbs}, C.~L., {Burkert}, A., \& {Pringle}, J.~E. 2011, \mnras, 413, 528

\bibitem[{{D'Onghia} {et~al.}(2009){D'Onghia}, {Besla}, {Cox}, \&
  {Hernquist}}]{donghia:resonant.stripping.dwarfs}
{D'Onghia}, E., {Besla}, G., {Cox}, T.~J., \& {Hernquist}, L. 2009, \nat, 460,
  605

\bibitem[{{D'Onghia} {et~al.}(2010){D'Onghia}, {Vogelsberger},
  {Faucher-Giguere}, \& {Hernquist}}]{donghia:2010.tidal.resonances}
{D'Onghia}, E., {Vogelsberger}, M., {Faucher-Giguere}, C.-A., \& {Hernquist},
  L. 2010, \apj, 725, 353

\bibitem[{{Doyon} {et~al.}(1994){Doyon}, {Wells}, {Wright}, {Joseph}, {Nadeau},
  \& {James}}]{Doyon94}
{Doyon}, R., {Wells}, M., {Wright}, G.~S., {Joseph}, R.~D., {Nadeau}, D., \&
  {James}, P.~A. 1994, \apjl, 437, L23

\bibitem[{{Evans} {et~al.}(2009{\natexlab{a}})}]{evans:ulirgs.are.mergers}
{Evans}, D.~A., {et~al.} 2009{\natexlab{a}}, \apj, submitted

\bibitem[{{Evans}
  {et~al.}(2009{\natexlab{b}})}]{evans:2009.sf.efficiencies.lifetimes}
{Evans}, N.~J., {et~al.} 2009{\natexlab{b}}, \apjs, 181, 321

\bibitem[{{Evans}(1999)}]{evans:1999.sf.gmc.review}
{Evans}, II, N.~J. 1999, \araa, 37, 311

\bibitem[{{Feruglio} {et~al.}(2010){Feruglio}, {Maiolino}, {Piconcelli},
  {Menci}, {Aussel}, {Lamastra}, \& {Fiore}}]{feruglio:2010.mrk231.agn.fb}
{Feruglio}, C., {Maiolino}, R., {Piconcelli}, E., {Menci}, N., {Aussel}, H.,
  {Lamastra}, A., \& {Fiore}, F. 2010, \aap, 518, L155

\bibitem[{{Foster} {et~al.}(2009){Foster}, {Proctor}, {Forbes}, {Spolaor},
  {Hopkins}, \& {Brodie}}]{foster:metallicity.gradients}
{Foster}, C., {Proctor}, R.~N., {Forbes}, D.~A., {Spolaor}, M., {Hopkins},
  P.~F., \& {Brodie}, J.~P. 2009, \mnras, 400, 2135

\bibitem[{{Genzel} {et~al.}(2001){Genzel}, {Tacconi}, {Rigopoulou}, {Lutz}, \&
  {Tecza}}]{Genzel01}
{Genzel}, R., {Tacconi}, L.~J., {Rigopoulou}, D., {Lutz}, D., \& {Tecza}, M.
  2001, \apj, 563, 527

\bibitem[{{Genzel} {et~al.}(2010)}]{genzel:2010.ks.law}
{Genzel}, R., {et~al.} 2010, \mnras, 407, 2091

\bibitem[{{Governato} {et~al.}(2009)}]{governato:disk.rebuilding}
{Governato}, F., {et~al.} 2009, \mnras, 398, 312

\bibitem[{{Governato} {et~al.}(2010)}]{governato:2010.dwarf.gal.form}
---. 2010, \nat, 463, 203

\bibitem[{{Guedes} {et~al.}(2011){Guedes}, {Callegari}, {Madau}, \&
  {Mayer}}]{guedes:2011.cosmo.disk.sim.merger.survival}
{Guedes}, J., {Callegari}, S., {Madau}, P., \& {Mayer}, L. 2011, \apj, 742, 76

\bibitem[{{Hammer} {et~al.}(2009){Hammer}, {Flores}, {Puech}, {Yang},
  {Athanassoula}, {Rodrigues}, \&
  {Delgado}}]{hammer:hubble.sequence.vs.mergers}
{Hammer}, F., {Flores}, H., {Puech}, M., {Yang}, Y.~B., {Athanassoula}, E.,
  {Rodrigues}, M., \& {Delgado}, R. 2009, \aap, 507, 1313

\bibitem[{Hammer {et~al.}(2009)Hammer, Flores, Yang, Athanassoula, Puech,
  Rodrigues, \& Peirani}]{hammer:obs.disk.rebuilding}
Hammer, F., Flores, H., Yang, Y.~B., Athanassoula, E., Puech, M., Rodrigues,
  M., \& Peirani, S. 2009, Astronomy and Astrophysics, 496, 381

\bibitem[{{Hayward} {et~al.}(2011{\natexlab{a}}){Hayward}, {Kere{\v s}},
  {Jonsson}, {Narayanan}, {Cox}, \& {Hernquist}}]{hayward:2011.smg.merger.rt}
{Hayward}, C.~C., {Kere{\v s}}, D., {Jonsson}, P., {Narayanan}, D., {Cox},
  T.~J., \& {Hernquist}, L. 2011{\natexlab{a}}, \apj, 743, 159

\bibitem[{{Hayward} {et~al.}(2010){Hayward}, {Narayanan}, {Jonsson}, {Cox},
  {Kere{\v s}}, {Hopkins}, \& {Hernquist}}]{hayward:2010.smg.counts}
{Hayward}, C.~C., {Narayanan}, D., {Jonsson}, P., {Cox}, T.~J., {Kere{\v s}},
  D., {Hopkins}, P.~F., \& {Hernquist}, L. 2010, ASP Conference Proceedings
  [arXiv:1008.4584]

\bibitem[{{Hayward}
  {et~al.}(2011{\natexlab{b}})}]{hayward:arepo.gadget.mergers}
{Hayward}, C.~C., {et~al.} 2011{\natexlab{b}}, \mnras, in preparation

\bibitem[{{Hernquist}(1989)}]{hernquist.89}
{Hernquist}, L. 1989, \nat, 340, 687

\bibitem[{{Hernquist}(1990)}]{hernquist:profile}
---. 1990, \apj, 356, 359

\bibitem[{{Hernquist}(1992)}]{hernquist:bulgeless.mergers}
---. 1992, \apj, 400, 460

\bibitem[{{Hernquist}(1993)}]{hernquist:bulge.mergers}
---. 1993, \apj, 409, 548

\bibitem[{{Hernquist} \& {Barnes}(1991)}]{hernquist:kinematic.subsystems}
{Hernquist}, L., \& {Barnes}, J.~E. 1991, \nat, 354, 210

\bibitem[{{Hernquist} \& {Mihos}(1995)}]{hernquist.mihos:minor.mergers}
{Hernquist}, L., \& {Mihos}, J.~C. 1995, \apj, 448, 41

\bibitem[{{Hernquist} {et~al.}(1993){Hernquist}, {Spergel}, \&
  {Heyl}}]{hernquist:phasespace}
{Hernquist}, L., {Spergel}, D.~N., \& {Heyl}, J.~S. 1993, \apj, 416, 415

\bibitem[{{Hibbard} \& {Yun}(1999)}]{hibbard.yun:excess.light}
{Hibbard}, J.~E., \& {Yun}, M.~S. 1999, \apjl, 522, L93

\bibitem[{{Hoffman} {et~al.}(2009){Hoffman}, {Cox}, {Dutta}, \&
  {Hernquist}}]{hoffman:dissipation.and.gal.kinematics}
{Hoffman}, L., {Cox}, T.~J., {Dutta}, S., \& {Hernquist}, L. 2009, \apj, 705,
  920

\bibitem[{{Hoffman} {et~al.}(2010){Hoffman}, {Cox}, {Dutta}, \&
  {Hernquist}}]{hoffman:mgr.orbit.structure.vs.fgas}
---. 2010, \apj, 723, 818

\bibitem[{{Hopkins}(2013)}]{hopkins:lagrangian.pressure.sph}
{Hopkins}, P.~F. 2013, \mnras, 428, 2840

\bibitem[{{Hopkins} {et~al.}(2009{\natexlab{a}}){Hopkins}, {Cox}, {Dutta},
  {Hernquist}, {Kormendy}, \& {Lauer}}]{hopkins:cusps.ell}
{Hopkins}, P.~F., {Cox}, T.~J., {Dutta}, S.~N., {Hernquist}, L., {Kormendy},
  J., \& {Lauer}, T.~R. 2009{\natexlab{a}}, \apjs, 181, 135

\bibitem[{{Hopkins} {et~al.}(2008{\natexlab{a}}){Hopkins}, {Cox}, \&
  {Hernquist}}]{hopkins:cusps.fp}
{Hopkins}, P.~F., {Cox}, T.~J., \& {Hernquist}, L. 2008{\natexlab{a}}, \apj,
  689, 17

\bibitem[{{Hopkins} {et~al.}(2008{\natexlab{b}}){Hopkins}, {Cox}, {Kere{\v s}},
  \& {Hernquist}}]{hopkins:groups.ell}
{Hopkins}, P.~F., {Cox}, T.~J., {Kere{\v s}}, D., \& {Hernquist}, L.
  2008{\natexlab{b}}, \apjs, 175, 390

\bibitem[{{Hopkins} {et~al.}(2009{\natexlab{b}}){Hopkins}, {Cox}, {Younger}, \&
  {Hernquist}}]{hopkins:disk.survival}
{Hopkins}, P.~F., {Cox}, T.~J., {Younger}, J.~D., \& {Hernquist}, L.
  2009{\natexlab{b}}, \apj, 691, 1168

\bibitem[{{Hopkins} \& {Hernquist}(2009)}]{hopkins:seyfert.limits}
{Hopkins}, P.~F., \& {Hernquist}, L. 2009, \apj, 694, 599

\bibitem[{{Hopkins} \& {Hernquist}(2010)}]{hopkins:sb.ir.lfs}
---. 2010, \mnras, 402, 985

\bibitem[{{Hopkins} {et~al.}(2006){Hopkins}, {Hernquist}, {Cox}, {Di Matteo},
  {Robertson}, \& {Springel}}]{hopkins:qso.all}
{Hopkins}, P.~F., {Hernquist}, L., {Cox}, T.~J., {Di Matteo}, T., {Robertson},
  B., \& {Springel}, V. 2006, \apjs, 163, 1

\bibitem[{{Hopkins} {et~al.}(2008{\natexlab{c}}){Hopkins}, {Hernquist}, {Cox},
  {Dutta}, \& {Rothberg}}]{hopkins:cusps.mergers}
{Hopkins}, P.~F., {Hernquist}, L., {Cox}, T.~J., {Dutta}, S.~N., \& {Rothberg},
  B. 2008{\natexlab{c}}, \apj, 679, 156

\bibitem[{{Hopkins} {et~al.}(2008{\natexlab{d}}){Hopkins}, {Hernquist}, {Cox},
  \& {Kere{\v s}}}]{hopkins:groups.qso}
{Hopkins}, P.~F., {Hernquist}, L., {Cox}, T.~J., \& {Kere{\v s}}, D.
  2008{\natexlab{d}}, \apjs, 175, 356

\bibitem[{{Hopkins} {et~al.}(2009{\natexlab{c}}){Hopkins}, {Hernquist}, {Cox},
  {Kere{\v s}}, \& {Wuyts}}]{hopkins:cusps.evol}
{Hopkins}, P.~F., {Hernquist}, L., {Cox}, T.~J., {Kere{\v s}}, D., \& {Wuyts},
  S. 2009{\natexlab{c}}, \apj, 691, 1424

\bibitem[{{Hopkins} {et~al.}(2008{\natexlab{e}}){Hopkins}, {Hernquist}, {Cox},
  {Younger}, \& {Besla}}]{hopkins:disk.heating}
{Hopkins}, P.~F., {Hernquist}, L., {Cox}, T.~J., {Younger}, J.~D., \& {Besla},
  G. 2008{\natexlab{e}}, \apj, 688, 757

\bibitem[{{Hopkins} {et~al.}(2005){Hopkins}, {Hernquist}, {Martini}, {Cox},
  {Robertson}, {Di Matteo}, \& {Springel}}]{hopkins:lifetimes.letter}
{Hopkins}, P.~F., {Hernquist}, L., {Martini}, P., {Cox}, T.~J., {Robertson},
  B., {Di Matteo}, T., \& {Springel}, V. 2005, \apjl, 625, L71

\bibitem[{{Hopkins} {et~al.}(2009{\natexlab{d}}){Hopkins}, {Lauer}, {Cox},
  {Hernquist}, \& {Kormendy}}]{hopkins:cores}
{Hopkins}, P.~F., {Lauer}, T.~R., {Cox}, T.~J., {Hernquist}, L., \& {Kormendy},
  J. 2009{\natexlab{d}}, \apjs, 181, 486

\bibitem[{{Hopkins} \& {Quataert}(2010)}]{hopkins:zoom.sims}
{Hopkins}, P.~F., \& {Quataert}, E. 2010, \mnras, 407, 1529

\bibitem[{{Hopkins} \&
  {Quataert}(2011{\natexlab{a}})}]{hopkins:inflow.analytics}
---. 2011{\natexlab{a}}, \mnras, 415, 1027

\bibitem[{{Hopkins} \& {Quataert}(2011{\natexlab{b}})}]{hopkins:cusp.slopes}
---. 2011{\natexlab{b}}, \mnras, 411, L61

\bibitem[{{Hopkins} {et~al.}(2011{\natexlab{a}}){Hopkins}, {Quataert}, \&
  {Murray}}]{hopkins:binding.sf.prescription}
{Hopkins}, P.~F., {Quataert}, E., \& {Murray}, N. 2011{\natexlab{a}}, \mnras,
  in prep

\bibitem[{{Hopkins} {et~al.}(2011{\natexlab{b}}){Hopkins}, {Quataert}, \&
  {Murray}}]{hopkins:rad.pressure.sf.fb}
---. 2011{\natexlab{b}}, \mnras, 417, 950

\bibitem[{{Hopkins} {et~al.}(2012{\natexlab{a}}){Hopkins}, {Quataert}, \&
  {Murray}}]{hopkins:stellar.fb.winds}
---. 2012{\natexlab{a}}, \mnras, 421, 3522

\bibitem[{{Hopkins} {et~al.}(2012{\natexlab{b}}){Hopkins}, {Quataert}, \&
  {Murray}}]{hopkins:fb.ism.prop}
---. 2012{\natexlab{b}}, \mnras, 421, 3488

\bibitem[{{Hopkins} {et~al.}(2009{\natexlab{e}}){Hopkins}, {Somerville}, {Cox},
  {Hernquist}, {Jogee}, {Kere{\v s}}, {Ma}, {Robertson}, \&
  {Stewart}}]{hopkins:disk.survival.cosmo}
{Hopkins}, P.~F., {Somerville}, R.~S., {Cox}, T.~J., {Hernquist}, L., {Jogee},
  S., {Kere{\v s}}, D., {Ma}, C.-P., {Robertson}, B., \& {Stewart}, K.
  2009{\natexlab{e}}, \mnras, 397, 802

\bibitem[{{Hopkins} {et~al.}(2010){Hopkins}, {Younger}, {Hayward}, {Narayanan},
  \& {Hernquist}}]{hopkins:ir.lfs}
{Hopkins}, P.~F., {Younger}, J.~D., {Hayward}, C.~C., {Narayanan}, D., \&
  {Hernquist}, L. 2010, \mnras, 402, 1693

\bibitem[{{Ivison} {et~al.}(2010){Ivison}, {Smail}, {Papadopoulos}, {Wold},
  {Richard}, {Swinbank}, {Kneib}, \&
  {Owen}}]{ivison:smg.is.complex.merger.w.agn}
{Ivison}, R., {Smail}, I., {Papadopoulos}, P.~P., {Wold}, I., {Richard}, J.,
  {Swinbank}, A.~M., {Kneib}, J., \& {Owen}, F.~N. 2010, \mnras, 404, 198

\bibitem[{{James} {et~al.}(1999){James}, {Bate}, {Wells}, {Wright}, \&
  {Doyon}}]{James99}
{James}, P., {Bate}, C., {Wells}, M., {Wright}, G., \& {Doyon}, R. 1999,
  \mnras, 309, 585

\bibitem[{{Jesseit} {et~al.}(2009){Jesseit}, {Cappellari}, {Naab}, {Emsellem},
  \& {Burkert}}]{jesseit:merger.rem.spin.vs.gas}
{Jesseit}, R., {Cappellari}, M., {Naab}, T., {Emsellem}, E., \& {Burkert}, A.
  2009, \mnras, 397, 1202

\bibitem[{{Jogee} {et~al.}(2009)}]{jogee:merger.density.08}
{Jogee}, S., {et~al.} 2009, \apj, 697, 1971

\bibitem[{{Jonsson} {et~al.}(2006){Jonsson}, {Cox}, {Primack}, \&
  {Somerville}}]{jonsson:sunrise.attenuation}
{Jonsson}, P., {Cox}, T.~J., {Primack}, J.~R., \& {Somerville}, R.~S. 2006,
  \apj, 637, 255

\bibitem[{{Joseph} \& {Wright}(1985)}]{joseph85}
{Joseph}, R.~D., \& {Wright}, G.~S. 1985, \mnras, 214, 87

\bibitem[{{Juneau} {et~al.}(2009){Juneau}, {Narayanan}, {Moustakas}, {Shirley},
  {Bussmann}, {Kennicutt}, \& {Vanden
  Bout}}]{juneau:2009.enhanced.dense.gas.ulirgs}
{Juneau}, S., {Narayanan}, D.~T., {Moustakas}, J., {Shirley}, Y.~L.,
  {Bussmann}, R.~S., {Kennicutt}, Jr., R.~C., \& {Vanden Bout}, P.~A. 2009,
  \apj, 707, 1217

\bibitem[{{Karl} {et~al.}(2010){Karl}, {Naab}, {Johansson}, {Kotarba}, {Boily},
  {Renaud}, \& {Theis}}]{karl:2010.antennae.model}
{Karl}, S.~J., {Naab}, T., {Johansson}, P.~H., {Kotarba}, H., {Boily}, C.~M.,
  {Renaud}, F., \& {Theis}, C. 2010, \apjl, 715, L88

\bibitem[{{Kennicutt}(1998)}]{kennicutt98}
{Kennicutt}, Jr., R.~C. 1998, \apj, 498, 541

\bibitem[{{Kere{\v s}} {et~al.}(2012){Kere{\v s}}, {Vogelsberger}, {Sijacki},
  {Springel}, \& {Hernquist}}]{keres:2011.arepo.gadget.disk.angmom}
{Kere{\v s}}, D., {Vogelsberger}, M., {Sijacki}, D., {Springel}, V., \&
  {Hernquist}, L. 2012, \mnras, 425, 2027

\bibitem[{{Kewley} {et~al.}(2010){Kewley}, {Rupke}, {Zahid}, {Geller}, \&
  {Barton}}]{kewley:2010.gal.pair.metal.grad.evol}
{Kewley}, L.~J., {Rupke}, D., {Zahid}, H.~J., {Geller}, M.~J., \& {Barton},
  E.~J. 2010, \apjl, 721, L48

\bibitem[{{Khochfar} \& {Burkert}(2006)}]{khochfar:cosmo.orbits}
{Khochfar}, S., \& {Burkert}, A. 2006, \aap, 445, 403

\bibitem[{{Kitsionas}
  {et~al.}(2009)}]{kitsionas:2009.grid.sph.compare.turbulence}
{Kitsionas}, S., {et~al.} 2009, \aap, 508, 541

\bibitem[{{Kormendy} \& {Sanders}(1992)}]{kormendysanders92}
{Kormendy}, J., \& {Sanders}, D.~B. 1992, \apjl, 390, L53

\bibitem[{{Kroupa}(2002)}]{kroupa:imf}
{Kroupa}, P. 2002, Science, 295, 82

\bibitem[{{Krumholz} \& {Gnedin}(2011)}]{krumholz:2011.molecular.prescription}
{Krumholz}, M.~R., \& {Gnedin}, N.~Y. 2011, \apj, 729, 36

\bibitem[{{Lake} \& {Dressler}(1986)}]{LakeDressler86}
{Lake}, G., \& {Dressler}, A. 1986, \apj, 310, 605

\bibitem[{{Leitherer} {et~al.}(1999)}]{starburst99}
{Leitherer}, C., {et~al.} 1999, \apjs, 123, 3

\bibitem[{{Mac Low} \& {Klessen}(2004)}]{mac-low:2004.turb.sf.review}
{Mac Low}, M.-M., \& {Klessen}, R.~S. 2004, Reviews of Modern Physics, 76, 125

\bibitem[{{Mannucci} {et~al.}(2006){Mannucci}, {Della Valle}, \&
  {Panagia}}]{mannucci:2006.snIa.rates}
{Mannucci}, F., {Della Valle}, M., \& {Panagia}, N. 2006, \mnras, 370, 773

\bibitem[{{Martin} {et~al.}(2007)}]{martin:mass.flux}
{Martin}, D.~C., {et~al.} 2007, \apjs, 173, 342

\bibitem[{{McGaugh}(2005)}]{mcgaugh:tf}
{McGaugh}, S.~S. 2005, \apj, 632, 859

\bibitem[{{McKee} \& {Ostriker}(1977)}]{mckee.ostriker:ism}
{McKee}, C.~F., \& {Ostriker}, J.~P. 1977, \apj, 218, 148

\bibitem[{{Melbourne} {et~al.}(2008){Melbourne}, {Desai}, {Armus}, {Dey},
  {Brand}, {Thompson}, {Soifer}, {Matthews}, {Jannuzi}, \&
  {Houck}}]{melbourne:2008.dog.morph.smooth}
{Melbourne}, J., {Desai}, V., {Armus}, L., {Dey}, A., {Brand}, K., {Thompson},
  D., {Soifer}, B.~T., {Matthews}, K., {Jannuzi}, B.~T., \& {Houck}, J.~R.
  2008, \aj, 136, 1110

\bibitem[{{Michard}(2006)}]{michard:ssp.decomp}
{Michard}, R. 2006, \aap, 449, 519

\bibitem[{{Mihos} \& {Hernquist}(1994{\natexlab{a}})}]{mihos:cusps}
{Mihos}, J.~C., \& {Hernquist}, L. 1994{\natexlab{a}}, \apjl, 437, L47

\bibitem[{{Mihos} \& {Hernquist}(1994{\natexlab{b}})}]{mihos:starbursts.94}
---. 1994{\natexlab{b}}, \apjl, 431, L9

\bibitem[{{Mihos} \& {Hernquist}(1996)}]{mihos:starbursts.96}
---. 1996, \apj, 464, 641

\bibitem[{{Moster} {et~al.}(2010){Moster}, {Macci{\`o}}, {Somerville},
  {Johansson}, \& {Naab}}]{moster:2010.thin.disk.heating.vs.gas}
{Moster}, B.~P., {Macci{\`o}}, A.~V., {Somerville}, R.~S., {Johansson}, P.~H.,
  \& {Naab}, T. 2010, \mnras, 403, 1009

\bibitem[{{Moster} {et~al.}(2011){Moster}, {Macci{\`o}}, {Somerville}, {Naab},
  \& {Cox}}]{moster:2011.gas.halo.merger.fx}
{Moster}, B.~P., {Macci{\`o}}, A.~V., {Somerville}, R.~S., {Naab}, T., \&
  {Cox}, T.~J. 2011, \mnras, 415, 3750

\bibitem[{{Naab} \& {Burkert}(2003)}]{naab:minor.mergers}
{Naab}, T., \& {Burkert}, A. 2003, \apj, 597, 893

\bibitem[{{Naab} {et~al.}(1999){Naab}, {Burkert}, \&
  {Hernquist}}]{naab:boxy.disky.massratio}
{Naab}, T., {Burkert}, A., \& {Hernquist}, L. 1999, \apjl, 523, L133

\bibitem[{{Narayanan} {et~al.}(2009){Narayanan}, {Cox}, {Hayward}, {Younger},
  \& {Hernquist}}]{narayanan:molecular.gas.in.smgs}
{Narayanan}, D., {Cox}, T.~J., {Hayward}, C., {Younger}, J.~D., \& {Hernquist},
  L. 2009, \mnras, 400, 1919

\bibitem[{{Narayanan} {et~al.}(2010){Narayanan}, {Hayward}, {Cox}, {Hernquist},
  {Jonsson}, {Younger}, \& {Groves}}]{narayanan:smg.modeling}
{Narayanan}, D., {Hayward}, C.~C., {Cox}, T.~J., {Hernquist}, L., {Jonsson},
  P., {Younger}, J.~D., \& {Groves}, B. 2010, \mnras, 401, 1613

\bibitem[{{Narayanan} {et~al.}(2011){Narayanan}, {Krumholz}, {Ostriker}, \&
  {Hernquist}}]{narayanan:2011.xco}
{Narayanan}, D., {Krumholz}, M., {Ostriker}, E.~C., \& {Hernquist}, L. 2011,
  \mnras, 418, 664

\bibitem[{{Narayanan} {et~al.}(2012){Narayanan}, {Krumholz}, {Ostriker}, \&
  {Hernquist}}]{narayanan:2011.xco.model}
{Narayanan}, D., {Krumholz}, M.~R., {Ostriker}, E.~C., \& {Hernquist}, L. 2012,
  \mnras, 421, 3127

\bibitem[{{Papovich} {et~al.}(2005){Papovich}, {Dickinson}, {Giavalisco},
  {Conselice}, \& {Ferguson}}]{papovich:highz.sb.gal.timescales}
{Papovich}, C., {Dickinson}, M., {Giavalisco}, M., {Conselice}, C.~J., \&
  {Ferguson}, H.~C. 2005, \apj, 631, 101

\bibitem[{{Pei}(1992)}]{pei92:reddening.curves}
{Pei}, Y.~C. 1992, \apj, 395, 130

\bibitem[{{Planesas} {et~al.}(1991){Planesas}, {Scoville}, \&
  {Myers}}]{planesas:1991.1068.superclumps}
{Planesas}, P., {Scoville}, N., \& {Myers}, S.~T. 1991, \apj, 369, 364

\bibitem[{{Price} \&
  {Federrath}(2010)}]{price:2010.grid.sph.compare.turbulence}
{Price}, D.~J., \& {Federrath}, C. 2010, \mnras, 406, 1659

\bibitem[{Puech {et~al.}(2009)Puech, Hammer, Flores, Neichel, \&
  Yang}]{puech:z06.disk.postmerger}
Puech, M., Hammer, F., Flores, H., Neichel, B., \& Yang, Y. 2009, Astronomy and
  Astrophysics, 493, 899

\bibitem[{{Purcell} {et~al.}(2009){Purcell}, {Kazantzidis}, \&
  {Bullock}}]{purcell:minor.merger.thindisk.destruction}
{Purcell}, C.~W., {Kazantzidis}, S., \& {Bullock}, J.~S. 2009, \apjl, 694, L98

\bibitem[{{Rand} \& {Kulkarni}(1990)}]{rand:1990.m51.superclumps}
{Rand}, R.~J., \& {Kulkarni}, S.~R. 1990, \apjl, 349, L43

\bibitem[{{Reichardt} {et~al.}(2001){Reichardt}, {Jimenez}, \&
  {Heavens}}]{reichardt:ssp.decomp}
{Reichardt}, C., {Jimenez}, R., \& {Heavens}, A.~F. 2001, \mnras, 327, 849

\bibitem[{{Robaina} {et~al.}(2009)}]{robaina:2009.sf.fraction.from.mergers}
{Robaina}, A.~R., {et~al.} 2009, \apj, 704, 324

\bibitem[{{Robertson} {et~al.}(2006{\natexlab{a}}){Robertson}, {Bullock},
  {Cox}, {Di Matteo}, {Hernquist}, {Springel}, \&
  {Yoshida}}]{robertson:disk.formation}
{Robertson}, B., {Bullock}, J.~S., {Cox}, T.~J., {Di Matteo}, T., {Hernquist},
  L., {Springel}, V., \& {Yoshida}, N. 2006{\natexlab{a}}, \apj, 645, 986

\bibitem[{{Robertson} {et~al.}(2006{\natexlab{b}}){Robertson}, {Cox},
  {Hernquist}, {Franx}, {Hopkins}, {Martini}, \& {Springel}}]{robertson:fp}
{Robertson}, B., {Cox}, T.~J., {Hernquist}, L., {Franx}, M., {Hopkins}, P.~F.,
  {Martini}, P., \& {Springel}, V. 2006{\natexlab{b}}, \apj, 641, 21

\bibitem[{{Robertson} {et~al.}(2006{\natexlab{c}}){Robertson}, {Hernquist},
  {Cox}, {Di Matteo}, {Hopkins}, {Martini}, \&
  {Springel}}]{robertson:msigma.evolution}
{Robertson}, B., {Hernquist}, L., {Cox}, T.~J., {Di Matteo}, T., {Hopkins},
  P.~F., {Martini}, P., \& {Springel}, V. 2006{\natexlab{c}}, \apj, 641, 90

\bibitem[{{Robertson} \&
  {Bullock}(2008)}]{robertson.bullock:disk.merger.rem.vs.obs}
{Robertson}, B.~E., \& {Bullock}, J.~S. 2008, \apjl, 685, L27

\bibitem[{{Rothberg} \& {Joseph}(2004)}]{rj:profiles}
{Rothberg}, B., \& {Joseph}, R.~D. 2004, \aj, 128, 2098

\bibitem[{{Rothberg} \& {Joseph}(2006)}]{rothberg.joseph:kinematics}
---. 2006, \aj, 131, 185

\bibitem[{{Rupke} {et~al.}(2010){Rupke}, {Kewley}, \&
  {Barnes}}]{rupke:2010.metallicity.grad.merger.vs.obs}
{Rupke}, D.~S.~N., {Kewley}, L.~J., \& {Barnes}, J.~E. 2010, \apjl, 710, L156

\bibitem[{{Rupke} \& {Veilleux}(2011)}]{rupke:2011.outflow.mrk231}
{Rupke}, D.~S.~N., \& {Veilleux}, S. 2011, \apjl, 729, L27+

\bibitem[{{Saitoh} {et~al.}(2008){Saitoh}, {Daisaka}, {Kokubo}, {Makino},
  {Okamoto}, {Tomisaka}, {Wada}, \&
  {Yoshida}}]{saitoh:2008.highres.disks.high.sf.thold}
{Saitoh}, T.~R., {Daisaka}, H., {Kokubo}, E., {Makino}, J., {Okamoto}, T.,
  {Tomisaka}, K., {Wada}, K., \& {Yoshida}, N. 2008, \pasj, 60, 667

\bibitem[{{Saitoh} {et~al.}(2009){Saitoh}, {Daisaka}, {Kokubo}, {Makino},
  {Okamoto}, {Tomisaka}, {Wada}, \&
  {Yoshida}}]{saitoh:2009.firstpassage.shocks.wcooling}
---. 2009, \pasj, 61, 481

\bibitem[{{Sajina} {et~al.}(2007){Sajina}, {Yan}, {Armus}, {Choi}, {Fadda},
  {Helou}, \& {Spoon}}]{sajina:pah.qso.vs.sf}
{Sajina}, A., {Yan}, L., {Armus}, L., {Choi}, P., {Fadda}, D., {Helou}, G., \&
  {Spoon}, H. 2007, \apj, 664, 713

\bibitem[{{S{\'a}nchez-Bl{\'a}zquez} {et~al.}(2007){S{\'a}nchez-Bl{\'a}zquez},
  {Forbes}, {Strader}, {Brodie}, \& {Proctor}}]{sanchezblazquez:ssp.gradients}
{S{\'a}nchez-Bl{\'a}zquez}, P., {Forbes}, D.~A., {Strader}, J., {Brodie}, J.,
  \& {Proctor}, R. 2007, \mnras, 377, 759

\bibitem[{{Sanders} \& {Mirabel}(1996)}]{sanders96:ulirgs.mergers}
{Sanders}, D.~B., \& {Mirabel}, I.~F. 1996, \araa, 34, 749

\bibitem[{{Sanders} {et~al.}(1988){Sanders}, {Soifer}, {Elias}, {Madore},
  {Matthews}, {Neugebauer}, \& {Scoville}}]{sanders88:quasars}
{Sanders}, D.~B., {Soifer}, B.~T., {Elias}, J.~H., {Madore}, B.~F., {Matthews},
  K., {Neugebauer}, G., \& {Scoville}, N.~Z. 1988, \apj, 325, 74

\bibitem[{{Sargent} {et~al.}(1987){Sargent}, {Sanders}, {Scoville}, \&
  {Soifer}}]{sargent87}
{Sargent}, A.~I., {Sanders}, D.~B., {Scoville}, N.~Z., \& {Soifer}, B.~T. 1987,
  \apjl, 312, L35

\bibitem[{{Schinnerer}
  {et~al.}(2008)}]{schinnerer:submm.merger.w.compact.mol.gas}
{Schinnerer}, E., {et~al.} 2008, \apjl, 689, L5

\bibitem[{{Scoville} {et~al.}(1986){Scoville}, {Sanders}, {Sargent}, {Soifer},
  {Scott}, \& {Lo}}]{scoville86}
{Scoville}, N.~Z., {Sanders}, D.~B., {Sargent}, A.~I., {Soifer}, B.~T.,
  {Scott}, S.~L., \& {Lo}, K.~Y. 1986, \apjl, 311, L47

\bibitem[{{Shier} \& {Fischer}(1998)}]{ShierFischer98}
{Shier}, L.~M., \& {Fischer}, J. 1998, \apj, 497, 163

\bibitem[{{Sijacki} {et~al.}(2012){Sijacki}, {Vogelsberger}, {Keres},
  {Springel}, \& {Hernquist}}]{sijacki:2011.gadget.arepo.hydro.tests}
{Sijacki}, D., {Vogelsberger}, M., {Keres}, D., {Springel}, V., \& {Hernquist},
  L. 2012, \mnras, 424, 2999

\bibitem[{{Snyder} {et~al.}(2011){Snyder}, {Cox}, {Hayward}, {Hernquist}, \&
  {Jonsson}}]{snyder:2011.ka.gal.sims}
{Snyder}, G.~F., {Cox}, T.~J., {Hayward}, C.~C., {Hernquist}, L., \& {Jonsson},
  P. 2011, \apj, 741, 77

\bibitem[{{Soto} \& {Martin}(2010)}]{soto:ssp.grad.in.ulirgs}
{Soto}, K.~T., \& {Martin}, C.~L. 2010, \apj, 716, 332

\bibitem[{{Springel}(2005)}]{springel:gadget}
{Springel}, V. 2005, \mnras, 364, 1105

\bibitem[{Springel(2010)}]{springel:arepo}
Springel, V. 2010, \mnras, 401, 791

\bibitem[{{Springel} {et~al.}(2005){Springel}, {Di Matteo}, \&
  {Hernquist}}]{springel:models}
{Springel}, V., {Di Matteo}, T., \& {Hernquist}, L. 2005, \mnras, 361, 776

\bibitem[{{Springel} \& {Hernquist}(2002)}]{springel:entropy}
{Springel}, V., \& {Hernquist}, L. 2002, \mnras, 333, 649

\bibitem[{{Springel} \& {Hernquist}(2003)}]{springel:multiphase}
---. 2003, \mnras, 339, 289

\bibitem[{{Springel} \& {Hernquist}(2005)}]{springel:spiral.in.merger}
---. 2005, \apjl, 622, L9

\bibitem[{{Stewart} {et~al.}(2009){Stewart}, {Bullock}, {Wechsler}, \&
  {Maller}}]{stewart:disk.survival.vs.mergerrates}
{Stewart}, K.~R., {Bullock}, J.~S., {Wechsler}, R.~H., \& {Maller}, A.~H. 2009,
  \apj, 702, 307

\bibitem[{{Stewart} {et~al.}(2008){Stewart}, {Bullock}, {Wechsler}, {Maller},
  \& {Zentner}}]{stewart:mw.minor.accretion}
{Stewart}, K.~R., {Bullock}, J.~S., {Wechsler}, R.~H., {Maller}, A.~H., \&
  {Zentner}, A.~R. 2008, \apj, 683, 597

\bibitem[{{Sturm} {et~al.}(2011)}]{sturm:2011.ulirg.herschel.outflows}
{Sturm}, E., {et~al.} 2011, \apjl, 733, L16+

\bibitem[{{Swinbank} {et~al.}(2008){Swinbank}, {Lacey}, {Smail}, {Baugh},
  {Frenk}, {Blain}, {Chapman}, {Coppin}, {Ivison}, {Gonzalez}, \&
  {Hainline}}]{swinbank:smg.counts.vs.durham}
{Swinbank}, A.~M., {Lacey}, C.~G., {Smail}, I., {Baugh}, C.~M., {Frenk}, C.~S.,
  {Blain}, A.~W., {Chapman}, S.~C., {Coppin}, K.~E.~K., {Ivison}, R.~J.,
  {Gonzalez}, J.~E., \& {Hainline}, L.~J. 2008, \mnras, 391, 420

\bibitem[{{Tacconi} {et~al.}(2002){Tacconi}, {Genzel}, {Lutz}, {Rigopoulou},
  {Baker}, {Iserlohe}, \& {Tecza}}]{tacconi:ulirgs.sb.profiles}
{Tacconi}, L.~J., {Genzel}, R., {Lutz}, D., {Rigopoulou}, D., {Baker}, A.~J.,
  {Iserlohe}, C., \& {Tecza}, M. 2002, \apj, 580, 73

\bibitem[{{Tacconi} {et~al.}(2006)}]{tacconi:smg.maximal.sb.sizes}
{Tacconi}, L.~J., {et~al.} 2006, \apj, 640, 228

\bibitem[{{Tacconi} {et~al.}(2008)}]{tacconi:smg.mgr.lifetime.to.quiescent}
---. 2008, \apj, 680, 246

\bibitem[{{Tasker} \& {Tan}(2009)}]{tasker:2009.gmc.form.evol.gravalone}
{Tasker}, E.~J., \& {Tan}, J.~C. 2009, \apj, 700, 358

\bibitem[{{Teyssier} {et~al.}(2010){Teyssier}, {Chapon}, \&
  {Bournaud}}]{teyssier:2010.clumpy.sb.in.mergers}
{Teyssier}, R., {Chapon}, D., \& {Bournaud}, F. 2010, \apjl, 720, L149

\bibitem[{{Titus} {et~al.}(1997){Titus}, {Spillar}, \&
  {Johnson}}]{titus:ssp.decomp}
{Titus}, T.~N., {Spillar}, E.~J., \& {Johnson}, P. 1997, \aj, 114, 958

\bibitem[{{Torrey} {et~al.}(2012{\natexlab{a}}){Torrey}, {Cox}, {Kewley}, \&
  {Hernquist}}]{torrey:2011.metallicity.evol.merger}
{Torrey}, P., {Cox}, T.~J., {Kewley}, L., \& {Hernquist}, L.
  2012{\natexlab{a}}, \apj, 746, 108

\bibitem[{{Torrey} {et~al.}(2012{\natexlab{b}}){Torrey}, {Vogelsberger},
  {Sijacki}, {Springel}, \& {Hernquist}}]{torrey:2011.arepo.disks}
{Torrey}, P., {Vogelsberger}, M., {Sijacki}, D., {Springel}, V., \&
  {Hernquist}, L. 2012{\natexlab{b}}, \mnras, 427, 2224

\bibitem[{{Tremonti} {et~al.}(2007){Tremonti}, {Moustakas}, \&
  {Diamond-Stanic}}]{tremonti:postsb.outflows}
{Tremonti}, C.~A., {Moustakas}, J., \& {Diamond-Stanic}, A.~M. 2007, \apjl,
  663, L77

\bibitem[{{Vogelsberger} {et~al.}(2012){Vogelsberger}, {Sijacki}, {Keres},
  {Springel}, \& {Hernquist}}]{vogelsberger:2011.arepo.vs.gadget.cosmo}
{Vogelsberger}, M., {Sijacki}, D., {Keres}, D., {Springel}, V., \& {Hernquist},
  L. 2012, \mnras, 425, 3024

\bibitem[{{Walker} {et~al.}(1996){Walker}, {Mihos}, \&
  {Hernquist}}]{walker:disk.fragility.minor.merger}
{Walker}, I.~R., {Mihos}, J.~C., \& {Hernquist}, L. 1996, \apj, 460, 121

\bibitem[{{Williams} \& {McKee}(1997)}]{williams:1997.gmc.prop}
{Williams}, J.~P., \& {McKee}, C.~F. 1997, \apj, 476, 166

\bibitem[{{Wilson} {et~al.}(2006){Wilson}, {Harris}, {Longden}, \&
  {Scoville}}]{wilson:2006.arp220.superclumps}
{Wilson}, C.~D., {Harris}, W.~E., {Longden}, R., \& {Scoville}, N.~Z. 2006,
  \apj, 641, 763

\bibitem[{{Wilson} {et~al.}(2003){Wilson}, {Scoville}, {Madden}, \&
  {Charmandaris}}]{wilson:2003.supergiant.clumps}
{Wilson}, C.~D., {Scoville}, N., {Madden}, S.~C., \& {Charmandaris}, V. 2003,
  \apj, 599, 1049

\bibitem[{{Wuyts} {et~al.}(2010){Wuyts}, {Cox}, {Hayward}, {Franx},
  {Hernquist}, {Hopkins}, {Jonsson}, \& {van
  Dokkum}}]{wuyts:2010.highz.sizes.vs.models}
{Wuyts}, S., {Cox}, T.~J., {Hayward}, C.~C., {Franx}, M., {Hernquist}, L.,
  {Hopkins}, P.~F., {Jonsson}, P., \& {van Dokkum}, P.~G. 2010, \apj, 722, 1666

\bibitem[{{Wuyts} {et~al.}(2009){Wuyts}, {Franx}, {Cox}, {F{\"o}rster
  Schreiber}, {Hayward}, {Hernquist}, {Hopkins}, {Labb{\'e}}, {Marchesini},
  {Robertson}, {Toft}, \& {van Dokkum}}]{wuyts:model.numbers.and.colors.vs.obs}
{Wuyts}, S., {Franx}, M., {Cox}, T.~J., {F{\"o}rster Schreiber}, N.~M.,
  {Hayward}, C.~C., {Hernquist}, L., {Hopkins}, P.~F., {Labb{\'e}}, I.,
  {Marchesini}, D., {Robertson}, B.~E., {Toft}, S., \& {van Dokkum}, P.~G.
  2009, \apj, 700, 799

\bibitem[{{Yan} {et~al.}(2007){Yan}, {Sajina}, {Fadda}, {Choi}, {Armus},
  {Helou}, {Teplitz}, {Frayer}, \& {Surace}}]{yan:z2.sf.seds}
{Yan}, L., {Sajina}, A., {Fadda}, D., {Choi}, P., {Armus}, L., {Helou}, G.,
  {Teplitz}, H., {Frayer}, D., \& {Surace}, J. 2007, \apj, 658, 778

\bibitem[{{Yang} {et~al.}(2009){Yang}, {Hammer}, {Flores}, {Puech}, \&
  {Rodrigues}}]{yang:post.merger.disk.obs}
{Yang}, Y., {Hammer}, F., {Flores}, H., {Puech}, M., \& {Rodrigues}, M. 2009,
  \aap, 501, 437, 7 pages, 8 figures, accepted by A{\&}A

\bibitem[{{Younger} {et~al.}(2009{\natexlab{a}}){Younger}, {Hayward},
  {Narayanan}, {Cox}, {Hernquist}, \& {Jonsson}}]{younger:warm.ulirg.evol}
{Younger}, J.~D., {Hayward}, C.~C., {Narayanan}, D., {Cox}, T.~J., {Hernquist},
  L., \& {Jonsson}, P. 2009{\natexlab{a}}, \mnras, 396, L66

\bibitem[{{Younger} {et~al.}(2008{\natexlab{a}}){Younger}, {Hopkins}, {Cox}, \&
  {Hernquist}}]{younger:minor.mergers}
{Younger}, J.~D., {Hopkins}, P.~F., {Cox}, T.~J., \& {Hernquist}, L.
  2008{\natexlab{a}}, \apj, 686, 815

\bibitem[{{Younger} {et~al.}(2007)}]{younger:highz.smgs}
{Younger}, J.~D., {et~al.} 2007, \apj, 671, 1531

\bibitem[{{Younger} {et~al.}(2008{\natexlab{b}})}]{younger:smg.sizes}
---. 2008{\natexlab{b}}, \apj, 688, 59

\bibitem[{{Younger} {et~al.}(2009{\natexlab{b}})}]{younger:sma.hylirg.obs}
---. 2009{\natexlab{b}}, \apj, 704, 803

\bibitem[{{Zuckerman} \& {Evans}(1974)}]{zuckerman:1974.gmc.constraints}
{Zuckerman}, B., \& {Evans}, II, N.~J. 1974, \apjl, 192, L149

\end{thebibliography}

\begin{appendix}
\section{Simulation Images}
\label{sec:appendix:images}
\begin{figure*}
    \centering
    \scaleup
    \plotsidesize{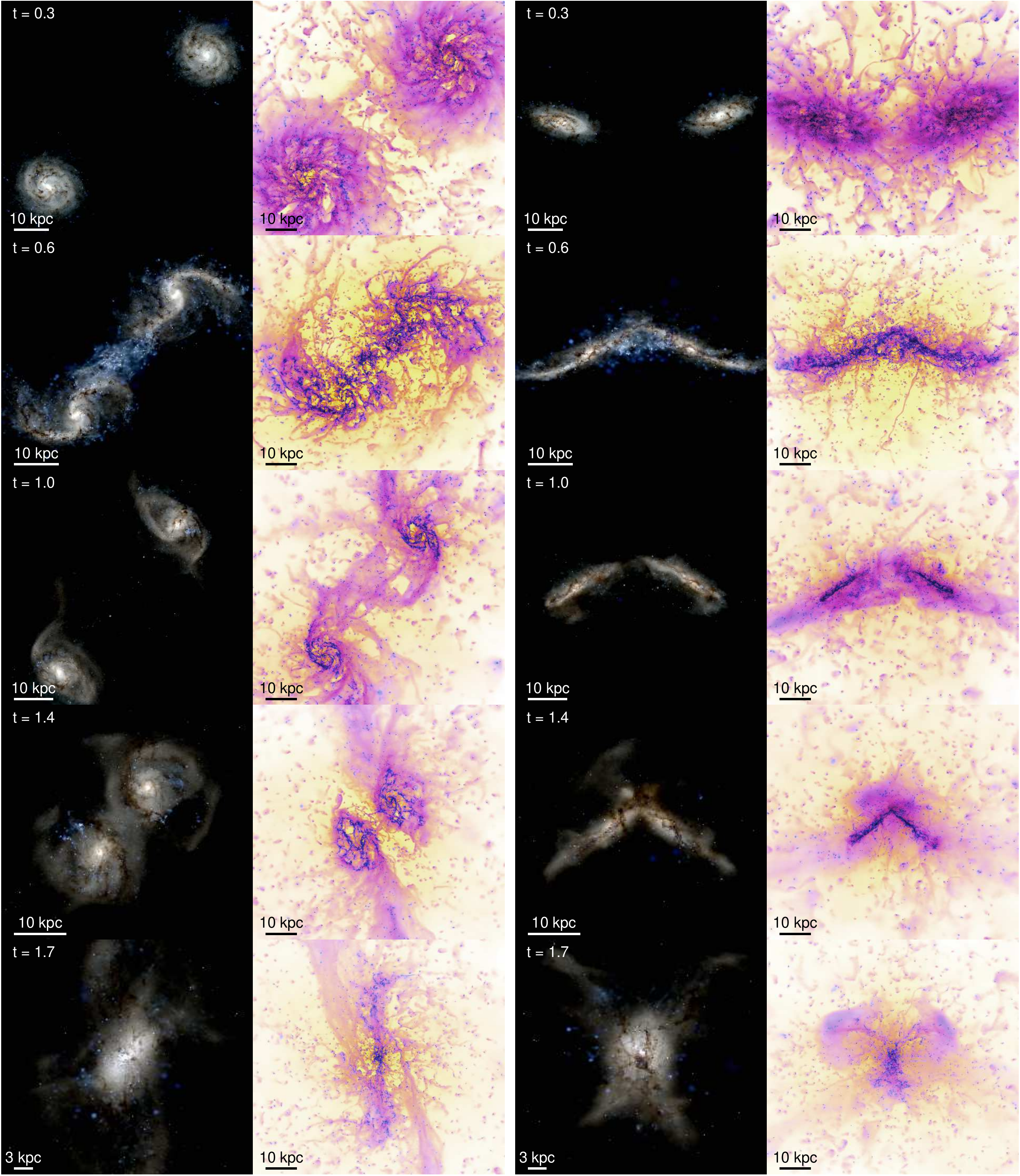}{0.97}
    \caption{Morphology of the {\bf e} (prograde) merger of the Sbc galaxy (a dwarf starburst). 
    We show two perpendicular viewing angles (left and right panel sets); each shows the optical 
    (as Fig.~\ref{fig:morph.3}; {\em left}) and gas (as Fig.~\ref{fig:morph.1}; {\em right}). 
    The time since the start of the simulation (in Gyr) is labeled.
    From top to bottom, the merger stages
    are (1) pre-merger infall, (2) just after first passage, (3) apocenter/turnaround, (4) final merger, (5) just after nuclear coalescence. 
    \label{fig:tile.sbc.e}}
\end{figure*}

\begin{figure*}
    \centering
    \scaleup
    \plotsidesize{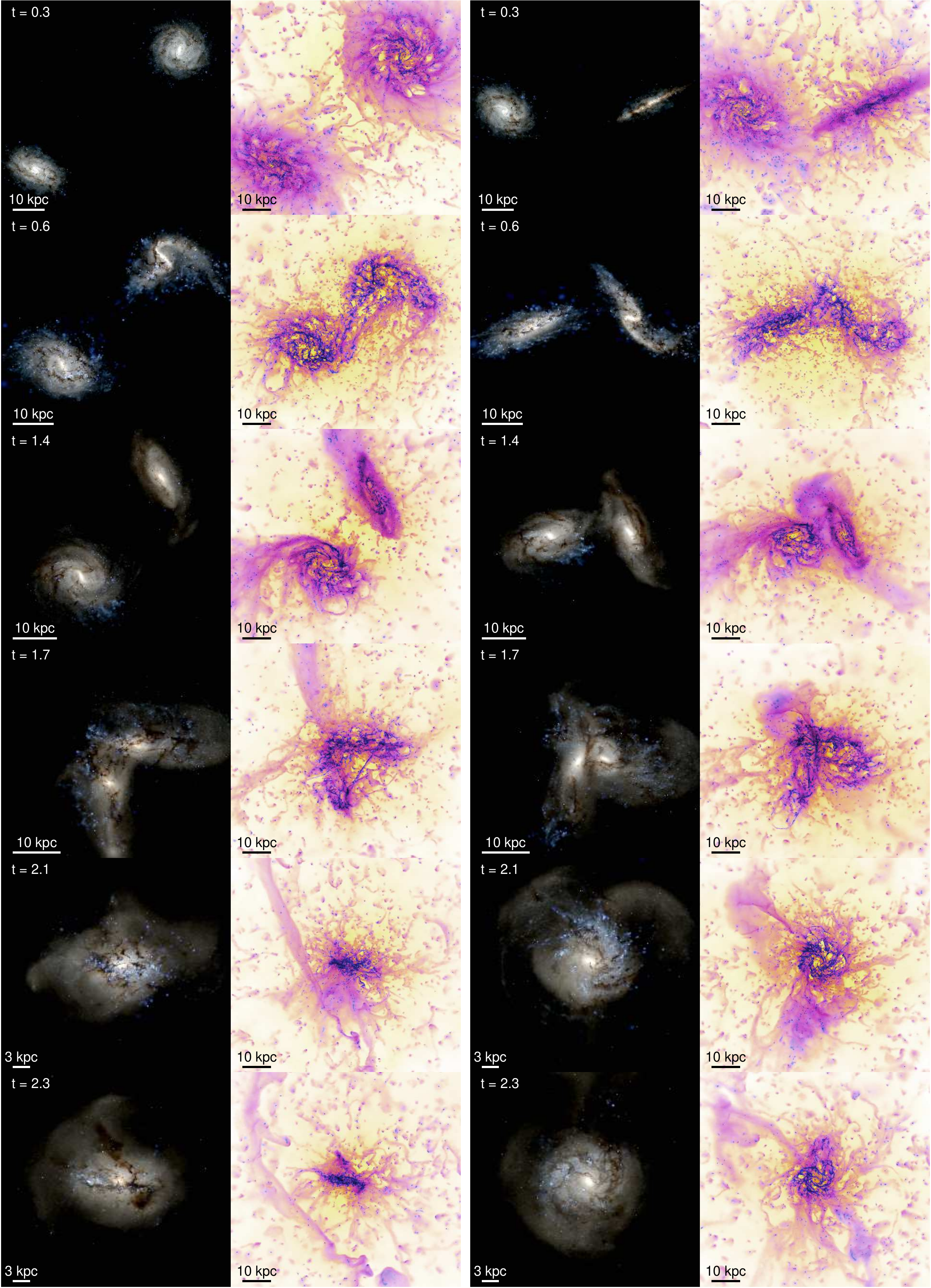}{0.95}
    \caption{As Fig.~\ref{fig:tile.sbc.e}, for the {\bf f} (retrograde) Sbc merger.
    \label{fig:tile.sbc.f}}
\end{figure*}

\begin{figure*}
    \centering
    \scaleup
    \plotsidesize{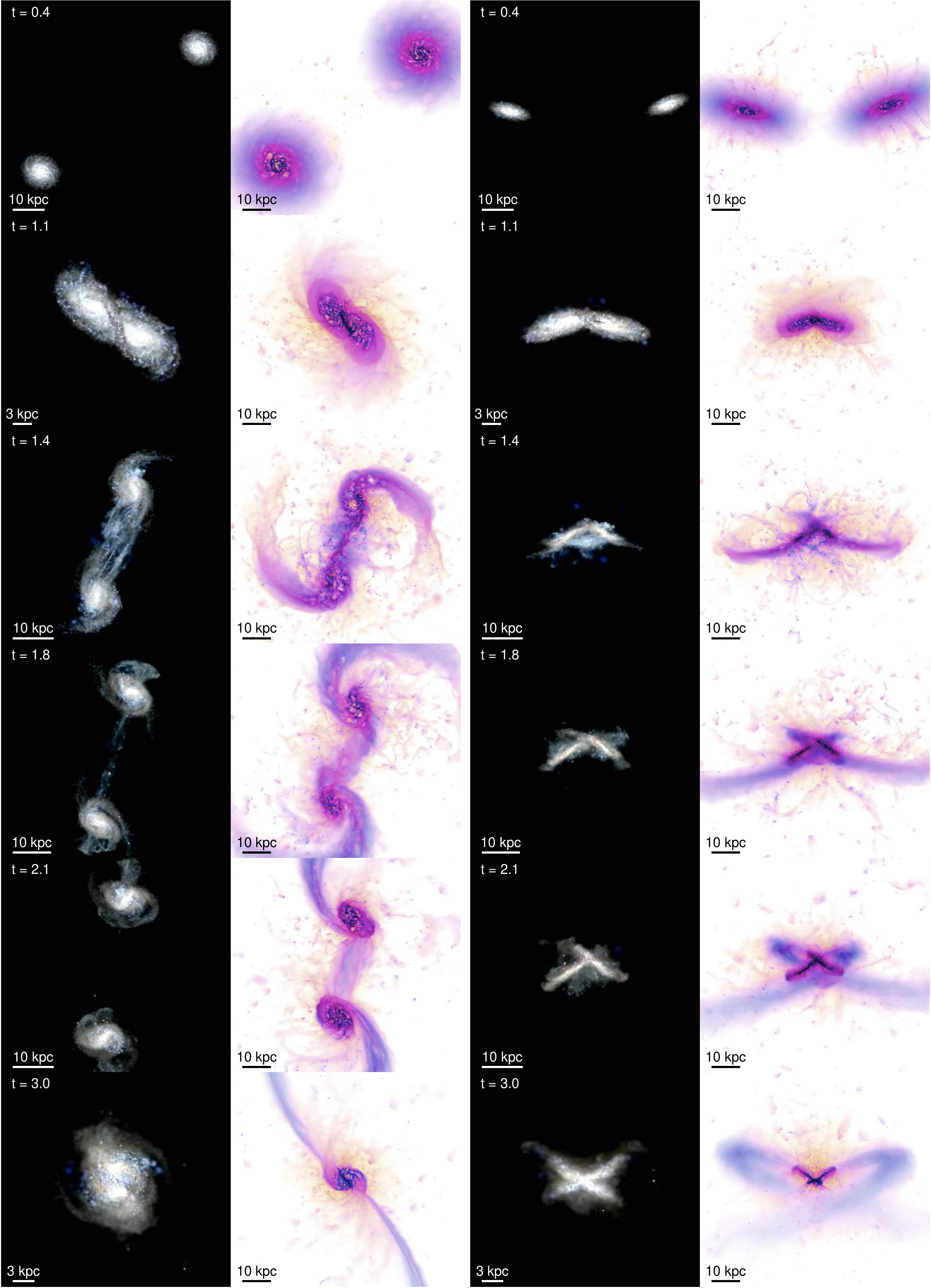}{0.83}
    \caption{As Fig.~\ref{fig:tile.sbc.e}, for the SMC (SMC-mass dwarf) {\bf e} merger.
    \label{fig:tile.smc.e}}
\end{figure*}

\begin{figure*}
    \centering
    \scaleup
    \plotsidesize{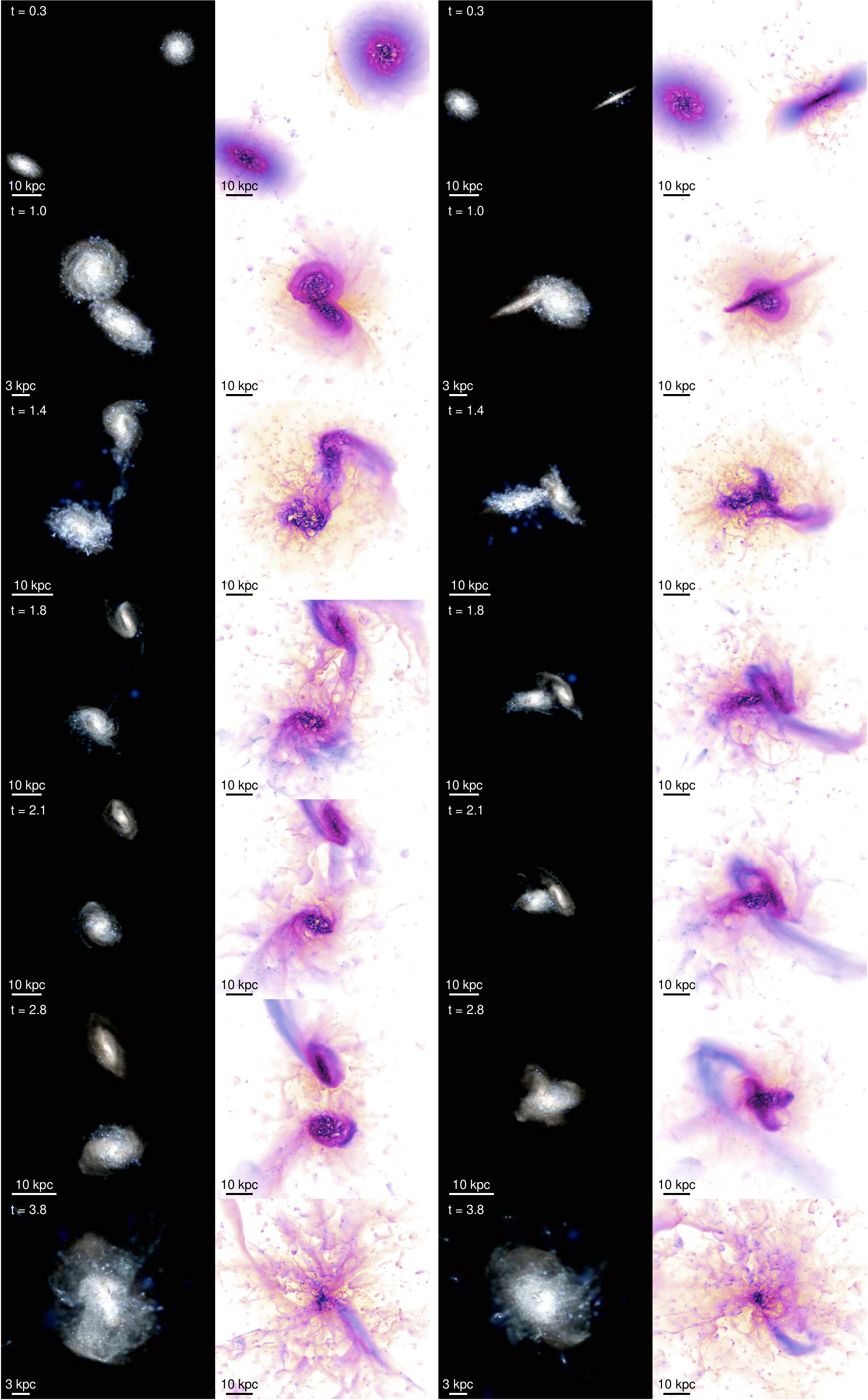}{0.83}
    \caption{As Fig.~\ref{fig:tile.sbc.e}, for the SMC {\bf f} merger.
    \label{fig:tile.smc.f}}
\end{figure*}

\begin{figure*}
    \centering
    \scaleup
    \plotsidesize{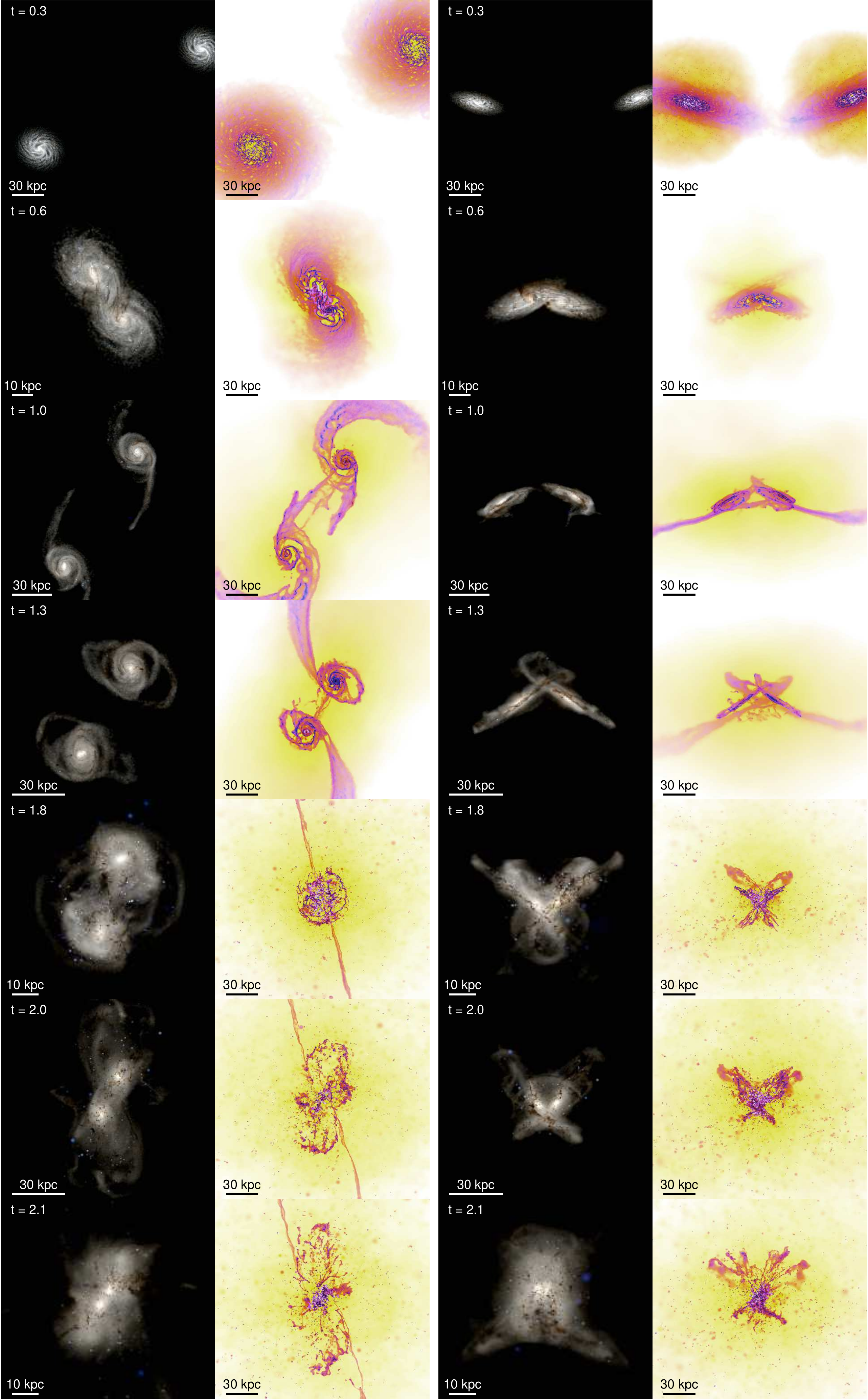}{0.83}
    \caption{As Fig.~\ref{fig:tile.sbc.e}, for the MW {\bf e} merger.
    \label{fig:tile.mw.e}}
\end{figure*}

\begin{figure*}
    \centering
    \scaleup
    \plotsidesize{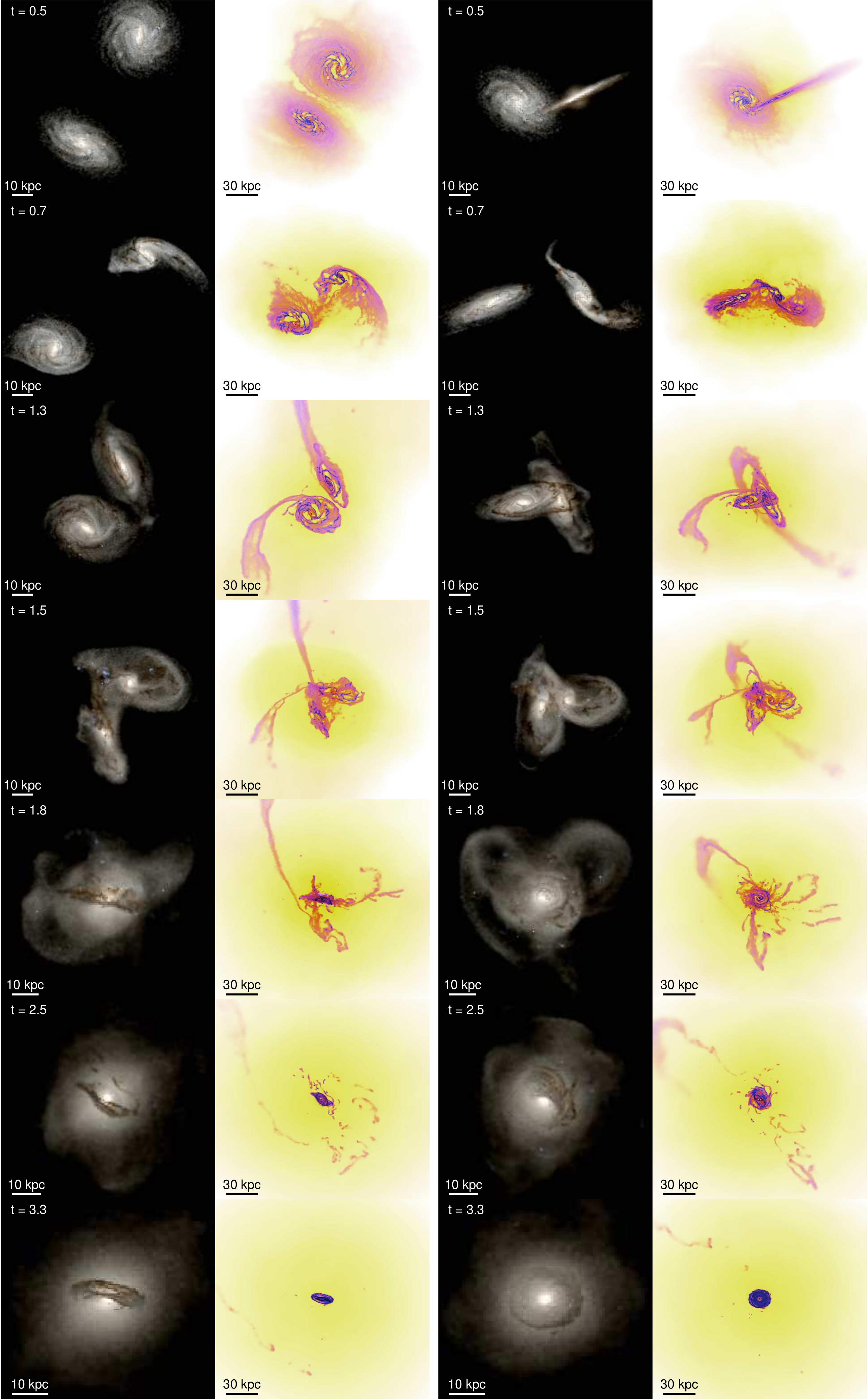}{0.83}
    \caption{As Fig.~\ref{fig:tile.sbc.e}, for the MW {\bf f} merger.
    \label{fig:tile.mw.f}}
\end{figure*}

\begin{figure*}
    \centering
    \scaleup
    \plotsidesize{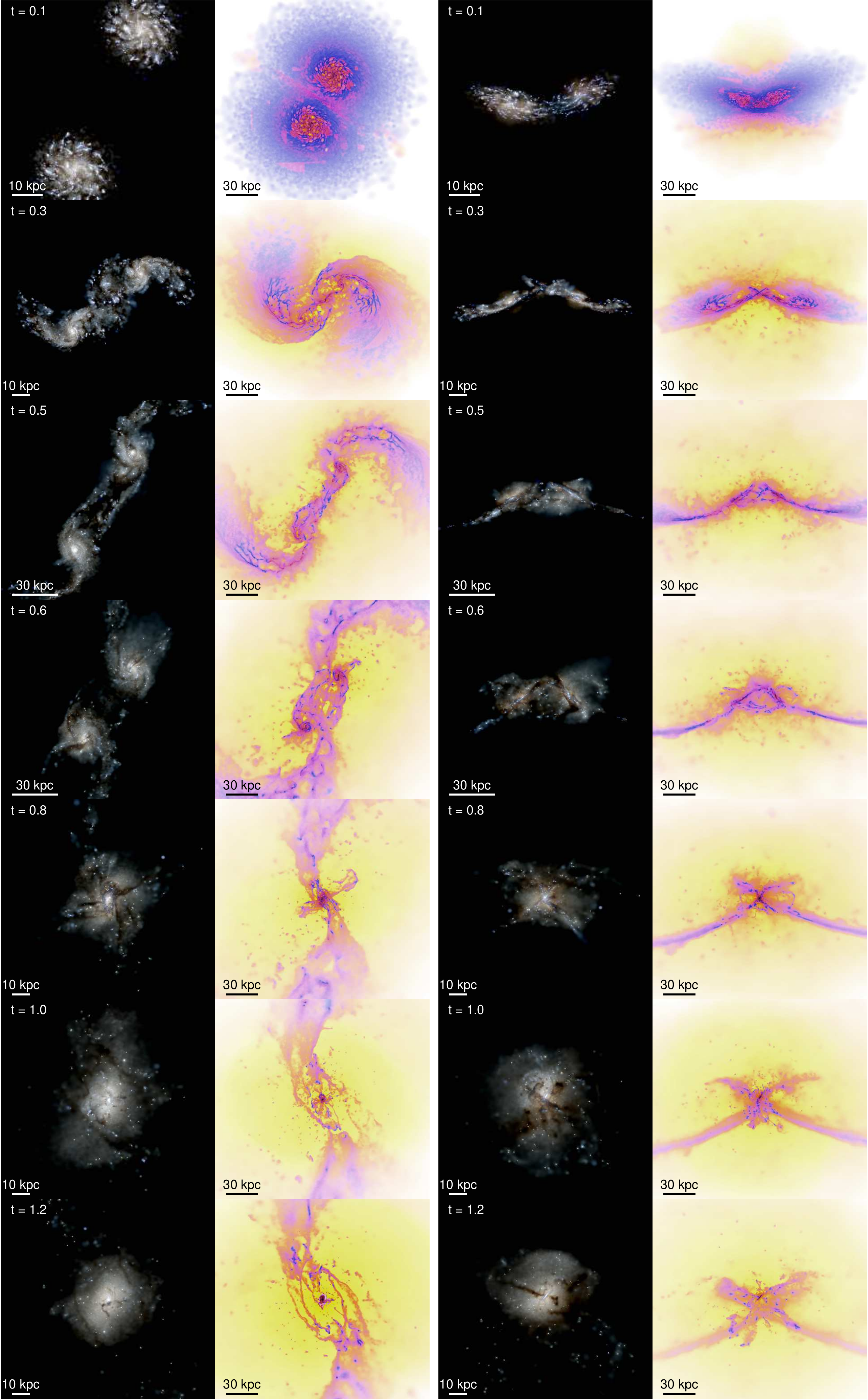}{0.83}
    \caption{As Fig.~\ref{fig:tile.sbc.e}, for the HiZ {\bf e} merger.
    \label{fig:tile.hiz.e}}
\end{figure*}

\begin{figure*}
    \centering
    \scaleup
    \plotsidesize{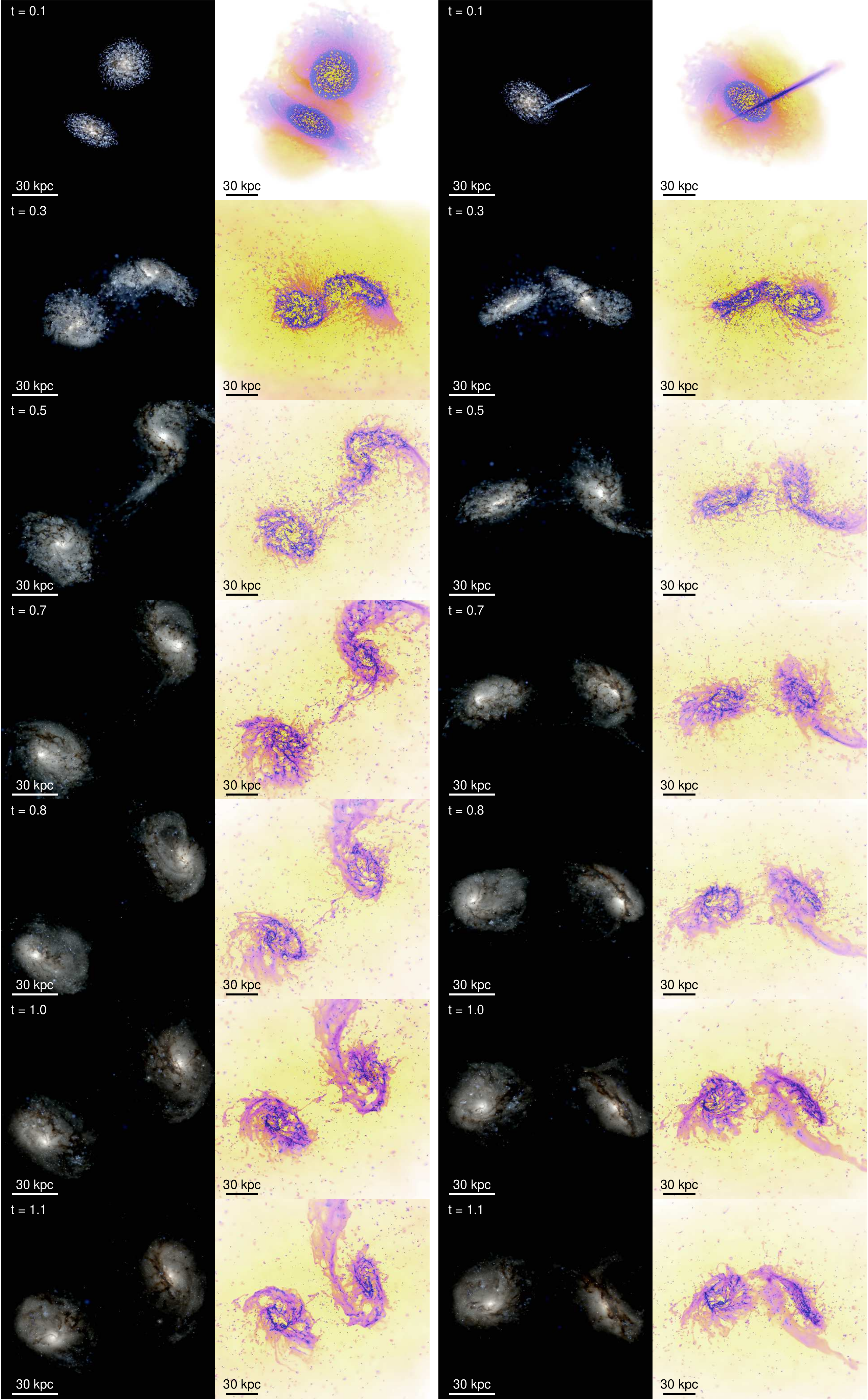}{0.83}
    \caption{As Fig.~\ref{fig:tile.sbc.e}, for the HiZ {\bf f} merger.
    \label{fig:tile.hiz.f}}
\end{figure*}

Figures~\ref{fig:tile.sbc.e}-\ref{fig:tile.hiz.f} present the images of the rest of our simulations, at different viewing angles and times during the mergers, in the style of Figures~\ref{fig:morph.1}-\ref{fig:morph.3}. Movies and additional images are available online ({\movieurl}).



\end{appendix}

\end{document}